\documentclass[useAMS,usenatbib]{mn2e}

\usepackage{graphicx}
\usepackage{amssymb}
\usepackage{natbib}
\usepackage{longtable}
\usepackage{afterpage}

\def\deg {^{\circ} }

\title[The ATLAS 5.5 GHz Survey of the Extended Chandra Deep Field South DR2]{The ATLAS 5.5 GHz Survey of the Extended Chandra Deep Field South: The Second Data Release}

\author[M. Huynh et al.] {M.T.~Huynh,$^1$\thanks{E-mail: minh.huynh@uwa.edu.au}
M.E. ~Bell,$^2$ A.M.~Hopkins,$^3$  R.P.~Norris,$^2$ N.~Seymour$^4$ \\
$^1$ International Center for Radio Astronomy Research, M468, University of Western Australia, Crawley, WA 6009, Australia \\
$^2$ CSIRO Astronomy and Space Science (CASS), PO Box 76, Epping, NSW 1710, Australia \\
$^3$ Australian Astronomical Observatory, PO Box 915, North Ryde, NSW 1670, Australia \\
$^4$ International Center for Radio Astronomy Research,  Curtin University, Perth, WA, Australia \\
}

\begin{document}

\maketitle

\label{firstpage}

\begin{abstract}

We present a new image of the 5.5 GHz radio emission from the extended Chandra Deep Field South. Deep radio observations at 5.5 GHz were obtained in 2010 and presented in the first data release.  A further 76 hours of integration has since been obtained, nearly doubling the integration time. This paper presents a new analysis of all the data. The new image reaches 8.6 $\mu$Jy rms, an improvement of about 40\% in sensitivity. We present a new catalogue of 5.5 GHz sources, identifying 212 source components, roughly 50\% more than were detected in the first data release. Source counts derived from this sample are consistent with those reported in the literature for $S_{5.5GHz} > 0.1$ mJy but significantly lower than published values in the lowest flux density bins ($S_{5.5GHz} < 0.1$ mJy), where we have more detected sources and improved statistical reliability. The 5.5 GHz radio sources were matched to 1.4 GHz sources in the literature and we find a mean spectral index of  $-0.35 \pm 0.10$ for $S_{5.5GHz} > 0.5$ mJy, consistent with the flattening of the spectral index  observed in 5 GHz sub-mJy samples. The median spectral index of the whole sample is $\alpha_{\rm med} = -0.58$, indicating that these observations may be starting to probe the star forming population. However, even at the faintest levels ($0.05 < S_{5.5GHz} < 0.1$ mJy), 39\% of the 5.5 GHz sources have flat or inverted radio spectra. Four flux density measurements from our data, across the full 4.5 to 6.5 GHz bandwidth, are combined with those from literature and we find 10\% of sources ($S_{5.5GHz} \gtrsim 0.1$ mJy) show significant curvature in their radio spectral energy distribution spanning 1.4 to 9 GHz. 

\end{abstract}

\begin{keywords}
{galaxies: evolution --- radio continuum: galaxies}
\end{keywords}

\section{Introduction}

A fundamental question in astrophysics today is how galaxies and their main constituent parts, stars and black-holes, form and evolve over cosmic time. 
A link between black holes, or active galactic nuclei (AGN), and the stellar growth of galaxies is suggested by scaling relations such as that between the black hole mass and stellar bulge mass (e.g. \citealp{magorrian1998}).
An intimate connection between AGN and star formation in galaxies is further suggested by the similar decline in AGN activity \citep{hasinger2005,aird2010} and star formation \citep{hopkins2006} from when
the Universe was half its current age to today. Additionally, this connection between galaxy and AGN evolution is reflected in the general shift of these processes from high mass galaxies in the distant Universe to low mass galaxies locally \citep{cowie1996,hasinger2005, juneau2005, mobasher2009}, commonly referred to as downsizing. Radio emission can be produced by both AGN and star-forming processes, hence radio wavelengths provide a unique dust-unbiased view of galaxy and AGN evolution. 

The first large sky-area radio surveys were conducted more than 50 years ago and the current state-of-the art surveys (e.g. NVSS, \citealp{condon1998}) now catalogue millions of sources. It is now well established
that bright radio-loud sources ($>$ 100 mJy) are associated with AGN activity (e.g. \citealp{condon1984b}). The normalised differential radio source counts, however, are observed to flatten below about 1 mJy  in a way
which cannot be explained by an extrapolation of the population of radio-loud AGNs found at higher flux densities. Star formation in strongly evolving �normal� spiral galaxies \citep{condon1984a,condon1989} and starbursting galaxies \citep{windhorst1985, rowan-robinson1993} were suggested as new populations to explain this upturn. 
The upturn in the source counts was initially explained through modelling of source populations with no need to include a substantial AGN contribution (e.g. \citealp{hopkins1998}).
However, a growing number of studies are finding that lower luminosity AGN, both radio-loud and weakly radio emitting sources (radio-loud and radio-quiet AGN respectively), make a significant contribution to the sub-mJy population \citep{jarvis2004,huynh2008,seymour2008b,smolcic2008,padovani2009,padovani2011,bonzini2013}. 

Star formation processes result in galaxies with a typical spectral index of $\alpha = -0.8$ at 1.4 GHz ($S \propto \nu^\alpha$, \citealp{condon1992}), consistent with optically-thin synchrotron emission from electrons accelerated by supernovae. The emission from the lobes of a radio jet are also synchrotron in nature, and hence also have steep spectral indices. A flat ($\alpha > -0.5$) or inverted ($\alpha > 0$) spectrum is usually attributed to the superposition of different self-absorbed components of varying sizes at the base of the radio jet of a radio-loud AGN. Thermal Bremsstrahlung (free-free) emission found in HII regions usually has a flatter spectral index but this becomes significant in normal galaxies only for rest-frame frequencies $>$10 GHz (e.g. \citealp{murphy2009}). The radio spectral index and radio spectral energy distribution can therefore provide important information on the nature of radio sources. 

The spectral index of radio sources has been studied for a few decades.  For the brightest sources ($\sim$1 Jy), the majority of 1.4 GHz-selected sources were found to be steep with a spectral index of $\alpha = -0.8$  \citep{condon1984b}, however a 5 GHz selected sample at similar flux densities shows a broad flat spectrum population of sources with $\alpha \sim 0$ \citep{witzel1979}. This bright flat spectrum population is compact (unresolved) and more likely to be quasars than steep spectrum sources \citep{peacock1981}. The fraction of flat spectrum sources decreases with decreasing flux density such that the average spectral index is steep at the tens of mJy level (e.g. \citealp{condon1981,owen1983}).  There is now emerging evidence that the spectral index flattens again at sub-mJy levels, but the nature and properties of these faint radio sources is still unclear. 
The flattening of the average spectral index at sub-mJy levels has been observed in faint 5 GHz selected samples \citep{prandoni2006,huynh2012a} and recently confirmed in sub-mJy samples selected at even higher frequencies ($>$10 GHz, \citealp{whittam2013,franzen2014}). However sub-mJy sources selected at 1.4 GHz or 610 MHz do not appear to exhibit a flattening in their average spectral index \citep{ibar2009}.
The observed flattening of the spectral index in higher frequency samples is not easily reproduced from extrapolations of the 1.4 GHz population, indicating that either there is a new population of faint, flat spectrum sources which are missing from 1.4 GHz selected samples, or the higher frequency radio emission of the known 1.4 GHz population is not well-modelled. 

In order to study the faint 5.5 GHz population we observed the extended Chandra Deep Field South with the Compact Array Broadband Backend (CABB; \citealp{wilson2011}) on the Australia Telescope Compact Array. Our observing run in 2010 consisted of 144 hours of observations, and this was supplemented by initial pilot observations of 20 hours from August 2009. A total of 42 pointings was used to uniformly sample the full 30 $\times$ 30 arcmin eCDFS region at 6cm, achieving $\sim$12 $\mu$Jy rms over roughly 0.25 deg$^2$ with a restoring beam of 4.9 $\times$ 2.0 arcsec. The survey description, image reduction and catalogue were presented in \cite{huynh2012a} (hereafter H12). Further 6cm observations of the extended Chandra Deep Field South were obtained in 2012 in a program to detect faint variable radio sources, nearly doubling the effective integration time. This paper presents a new and more sensitive 6cm image from a reduction of all the data. This new image covers 0.34 deg$^2$ with a typical sensitivity in the inner region of $\sim$9 $\mu$Jy rms, making it the largest mosaic ever made at 6cm to these depths.  We describe the survey and wide-field wide-band imaging techniques in Section 2. In Section 3 we discuss the extraction and characterisation of sources and present the source catalogue. Source counts from the new data and an analysis of the radio spectral energy distribution of the sources are presented in Sections 4 and 5, respectively. 

\section{The Observations}

\subsection{Observing Strategy}

The extended Chandra Deep Field South (eCDFS) was observed with the Compact Array Broadband Backend (CABB; Wilson et al. 2011) on the Australia Telescope Compact Array (ATCA) with the full 2048 MHz bandwidth centred at 5.5 GHz. We chose a 42 pointing hexagonal mosaic with spacings of 5 arcmin (approximately 0.5 FWHM of the primary beam) to uniformly sample the full 30 $\times$ 30 arcmin eCDFS region, centered approximately at RA = 3h32m22s and Dec = $-27\deg48\arcmin37\arcsec$ (J2000). The 20 hours in 2009 and 144 hours in 2010 were allocated under ATCA observing program C2028. This data resulted in a rms sensitivity of 11.9 $\mu$Jy and synthesised beam size of 4.9 $\times$ 2.0 arcsec (H12), under hereafter Epoch 1 and Data Release 1. 

Further observations were obtained in 2012 via ATCA program C2670. The C2670 program was conceived as a blind search for sub-mJy level sources that are variable on time scales of months to roughly a year, with a secondary goal of testing the Variables and Slow Transients (VAST, \citealp{murphy2013}) data pipeline. A total of 54 hours in May--June 2012 (Epoch 2) and 47 hours in August 2012 (Epoch 3) was allocated to C2670, and the data was taken using the mosaicing strategy of H12.  The three epochs are summarised in Table \ref{tab:obssumm}. An analysis of the variable sources is presented in \cite{bell2015}. Here we present a reduction of the full dataset, i.e. all three epochs, to obtain the most sensitive image possible. 

\begin{table}\caption{Summary of the ATCA observations used in this data release.}  
 \begin{tabular}{llcc}  
\hline
Program ID & Epoch and Date & Array & Net Integration \\ 
                   &                        &           &         Time (h)     \\ \hline
C2028      &     1,  2009 Aug 12, 14 & 6D & 13.8 \\
C2028      &     1,  2010 Jan   5 -- 15 & 6A & 91.0 \\
C2670      &     2,  2012 May 31 -- June 4 & 6A & 41.7 \\
C2670      &     3,  2012 Aug 14 -- 18 & 6D & 34.3 \\
\hline
\end{tabular}
\label{tab:obssumm}
\end{table}

\subsection{Wide-field Wide-band Imaging}

The new generation of wide-band receivers on radio interferometers such as ATCA and the Very Large Array (VLA) have led to new challenges in radio imaging. The 2 GHz bandwidth is a significant fraction of the central frequency of the observations. The primary beam response, the synthesized beam and the flux density of most sources vary significantly with frequency. One way to mitigate the issues with a large bandwidth is to divide {\em uv} data into sub-bands and then force nearly identical beam-sizes with an appropriate ``robustness" parameter \citep{briggs1995}. This sub-division approach was used to image VLA data spanning 2 -- 4 GHz \citep{condon2012}. While the fractional bandwidth is less for our ATCA data centred at 5.5 GHz, we tested two imaging schemes: one where the {\em uv} data is not divided into sub-bands (hereafter full-band reduction), and a second scheme where the 2048 MHz CABB band is divided into 512 MHz sub-bands (hereafter sub-band reduction). 

We used the Multichannel Image Reconstruction, Image Analysis and Display (MIRIAD) software package to reduce the CABB data. This is the standard package used for ATCA data and has undergone several enhancements since H12 to better handle the wide-band of CABB. These include an option to allow calibration task {\em gpcal} to solve for gain variation across the band and an option for {\em linmos} to apply several frequency-averaged primary beams instead of one primary beam across the full 2 GHz band. The full-band and sub-band reductions use the same calibration scheme. In the calibration step we set the number of frequency bins in {\em gpcal} to four (i.e. 512 MHz bins) and for the primary beam correction we set {\em linmos} to apply ten frequency bins.  The 42 pointings were individually reduced and imaged. Automated flagging was performed using the MIRIAD task {\em pgflag}. {\em pgflag} is based on AOFLAGGER \citep{offringa2010} which was developed for LOFAR but now used at many telescopes. 

The steps for the full-band reduction are similar to that in H12, but with some improvements to the cleaning and self calibration steps. We performed multi-frequency synthesis imaging with {\em invert} using the same robust weighting as H12, robust = 1, and set the image size to 2500 $\times$ 2500 pixels with 0.5 arcsec pixels. This is larger than H12 because the frequency varying primary beam response means a larger image is needed to capture the larger field of view at the low frequency end.  Multi-frequency cleaning was performed with the task {\em mfclean} with the clean region set to about 9.6 arcmin. This extends to just beyond the 10\% response level at 4.5 GHz, the lower end of the band, therefore encompassing the full region of interest. We found two iterations of phase self-calibration produced good results. The first iteration was performed with a model set from 100 {\em mfclean} iterations (i.e. bright sources only), and the second with the model set by cleaning to 4$\sigma$. The individual pointings were restored with the same beam, the average beam of the 42 pointings, 5.0 $\times$ 2.0 arcsec. The individual pointings were then mosaiced together using the task {\em linmos}, which applies the ten frequency-varying primary beams. The edges of each pointing beyond 9.5 arcmin were removed before the combination, to discard the uncleaned areas with a very low primary beam response from the final mosaic. 

In the sub-band reduction the calibrated data was split into 4 sub-bands of 512 MHz, resulting in fractional bandwidths of 0.09 -- 0.11, much less than 1. Each sub-band was imaged with a different robust weighting that resulted in similar beam-sizes. Multi-frequency cleaning and self-calibration was then performed for each pointing and each sub-band using the same strategy as for the full-band reduction. The individual images were restored with the same beam, the average beam of the 4 $\times$ 42 images. Finally, as for the full-band reduction, the 4 $\times$ 42 images were mosaiced together using {\em linmos}.

\begin{figure*}
\includegraphics[width=1.5\columnwidth]{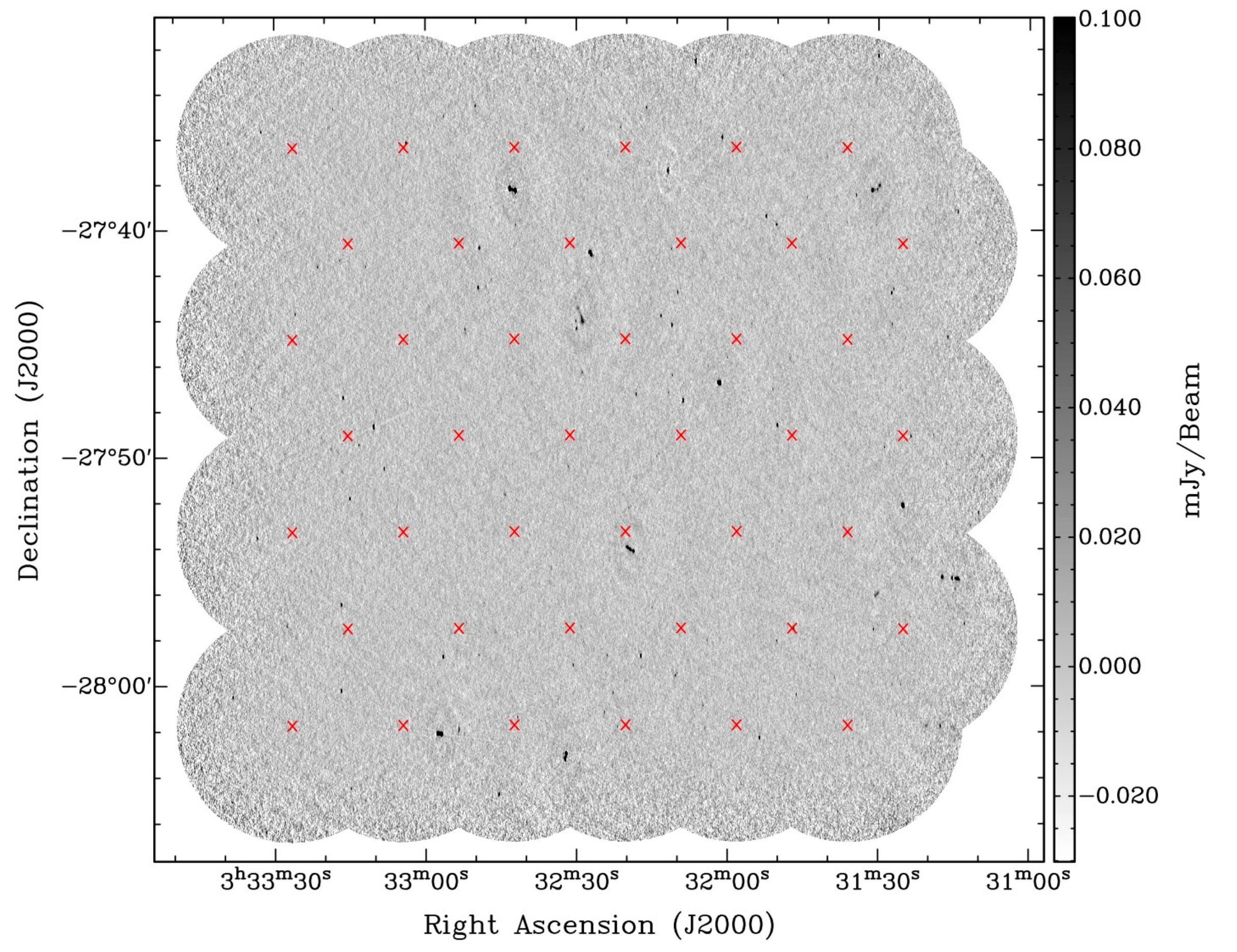}
\caption{The eCDFS 5.5 GHz full-band mosaic with the greyscale set to the range $-0.03$ to 0.1 mJy. The red crosses mark the 42 individual pointings of the mosaic. The total area covered by this mosaic is 0.34 deg$^2$.} 
\label{fig:ecdfs_grey}
\end{figure*}

\subsection{Image Analysis: Sensitivity and Clean Bias}

We used the MIRIAD task {\em sigest}, iteratively clipping the pixels, to estimate the noise in the inner 20 $\times$ 20 arcmin region of the full-band and sub-band mosaic.  We find the noise in the full-band reduced mosaic is 8.6 $\mu$Jy/beam, and 8.7 $\mu$Jy/beam for the sub-band reduced mosaic. The full-band reduced mosaic therefore has slightly lower noise than the sub-band reduced mosaic, at the 1\% level. On visual inspection of the two mosaics the sidelobes around bright sources appear to be marginally more prominent in the sub-band reduction compared to the full-band reduction. This may be due to the better self-calibration from {\em mfclean} models produced in the full-band reduction, which goes deeper than the sub-band imaging and has better {\em uv}-coverage. We use the full-band reduced mosaic in the production of the catalogue. 

\begin{figure}
\includegraphics[width=0.99\columnwidth]{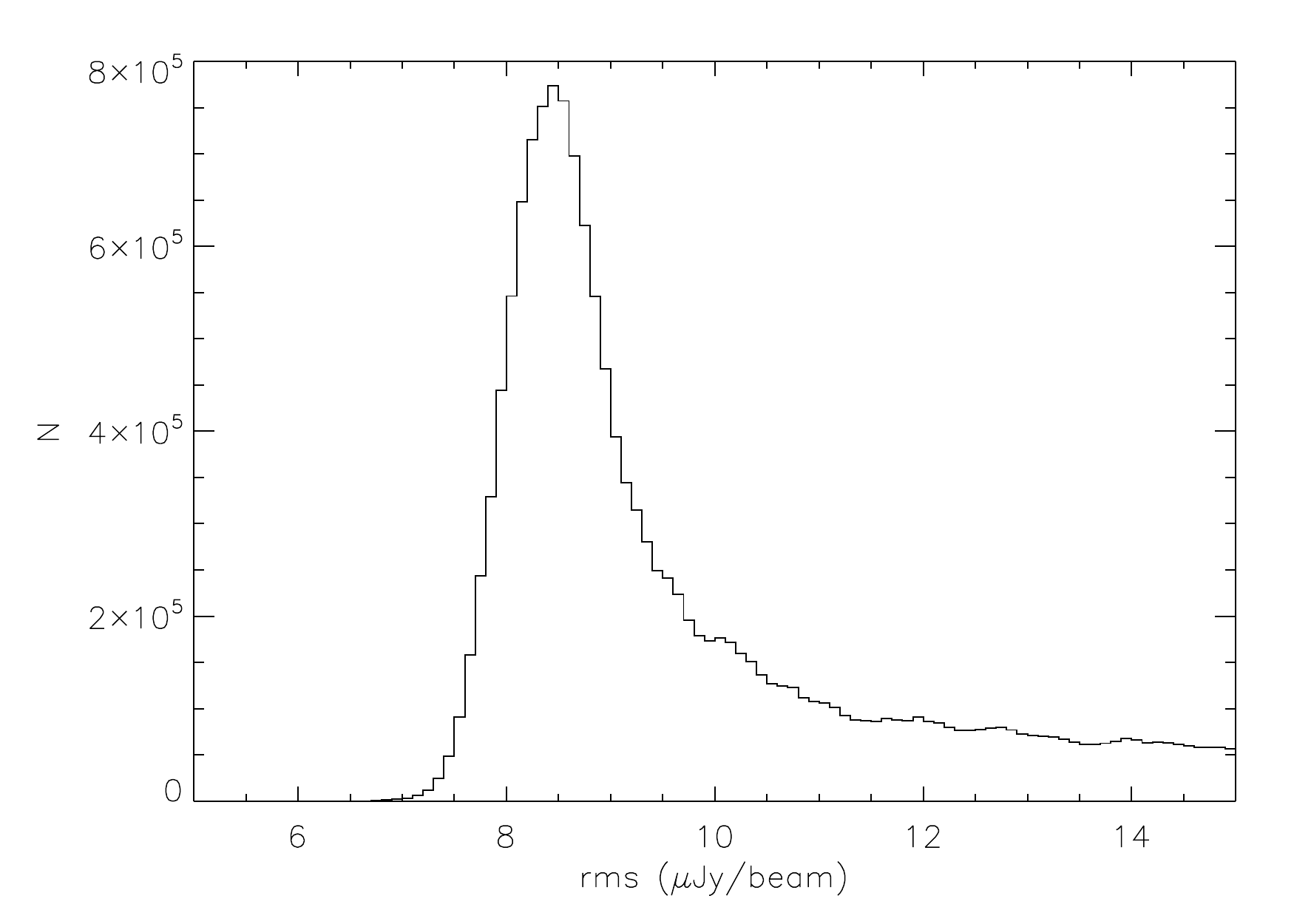}
\caption{The noise distribution, as determined from the noise image generated by SExtractor.} 
\label{fig:rmsdist}
\end{figure}

The full-band mosaic is shown in Figure \ref{fig:ecdfs_grey}, where regions greater than 5 arcmin ($\sim$ 0.5 FWHM of the primary beam) of the outer pointings have been removed to minimise primary beam affects and avoid high levels of non-Gaussian noise which may affect the source extraction. The noise properties of this full-band mosaic were investigated using SExtractor \citep{bertin1996}. Briefly,  SExtractor calculates the background and rms for a region (or `mesh') around each pixel using a combination of clipping and mode estimation. SExtractor with mesh-sizes of 8 -- 12 times the synthesized beam is known to produce good noise estimates of radio images \citep{schinnerer2007,schinnerer2010,huynh2005,huynh2012b}.
 Figure \ref{fig:rmsdist} shows the histogram of the pixels in the noise image generated by SExtractor, using a mesh-size of 10 times the synthesized beam. The peak in the distribution is 8.4 $\mu$Jy/beam, broadly consistent with the {\em sigest} result of 8.6 $\mu$Jy/beam for the inner part of the mosaic. The median of the noise distribution is 9.3 $\mu$Jy/beam, so half of the mosaic has an rms noise level lower than this. The tail at high noise levels ($>$11$\mu$Jy/beam) is due to the higher levels of noise at the edge of the mosaic from the primary beam response and increased noise around bright sources. 

When {\em uv} coverage is incomplete the cleaning process can redistribute flux from real sources on to noise peaks. This clean bias is generally only a problem for snapshot observations where {\em uv} coverage is poor. Although our {\em uv} coverage is good from the 180 hours split between 6A and 6D configurations, we performed tests to check the extent of the clean bias in the full-band mosaic. Point sources were injected into the {\em uv} data at random positions to avoid being confused with real sources. The {\em uv} data was then imaged with the same cleaning depth as the production images, and the source output flux densities  compared to the input values. The fake sources were injected one at a time to avoid source confusion, and the process repeated 4000 times to obtain a large sampling. We find the median clean bias is $\sim$5\% for the faintest sources at 50 $\mu$Jy and it is negligible for brighter sources ($> 150$ $\mu$Jy). We therefore conclude clean bias is not a significant issue. 

\section{Source Extraction}

There are many radio source extraction tools available, including AIPS and MIRIAD Gaussian fitting routines {\em sad}, {\em vsad} and {\em imsad}, the false discovery rate algorithm {\em sfind} \citep{hopkins2002}, and newer codes such as Duchamp \citep{whiting2012},  BLOBCAT \citep{hales2012}, and AEGEAN \citep{hancock2012}. Most of these source finding algorithms use a simple S/N thresholding technique whereby a source is deemed a true source if it has a peak flux density, or pixel value, above a set threshold. Following our previous work in H12, we use the MIRIAD task {\em sfind} to search for sources. The {\em sfind} task implements a false-discovery rate algorithm \citep{miller2001}, which compares the distribution of image pixels to that of an image containing only noise to return a list of source detections. The user set threshold is the fraction of sources which are allowed to be false, not a S/N. 

\begin{figure*}
\includegraphics[width=4cm]{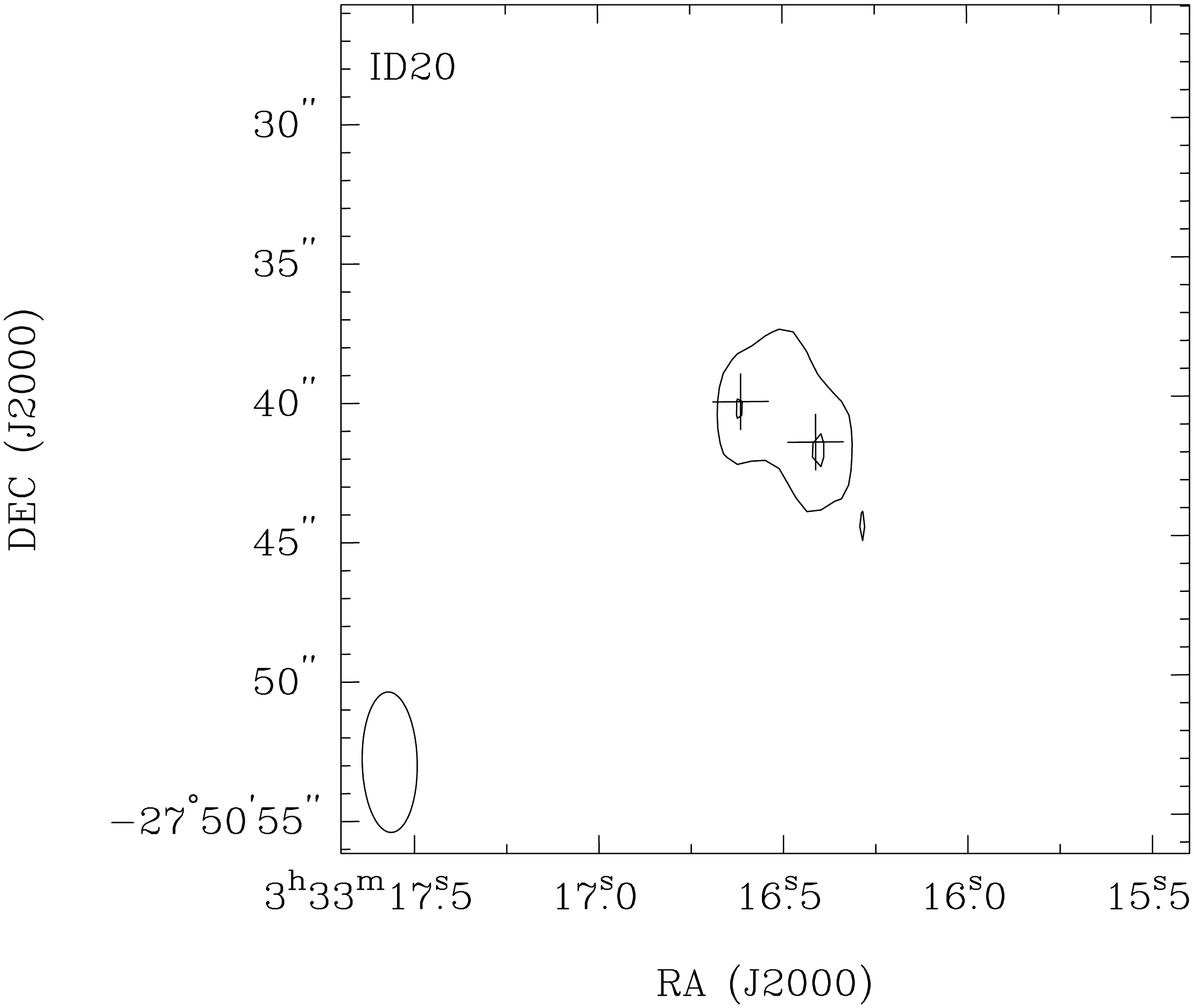}
\includegraphics[width=4cm]{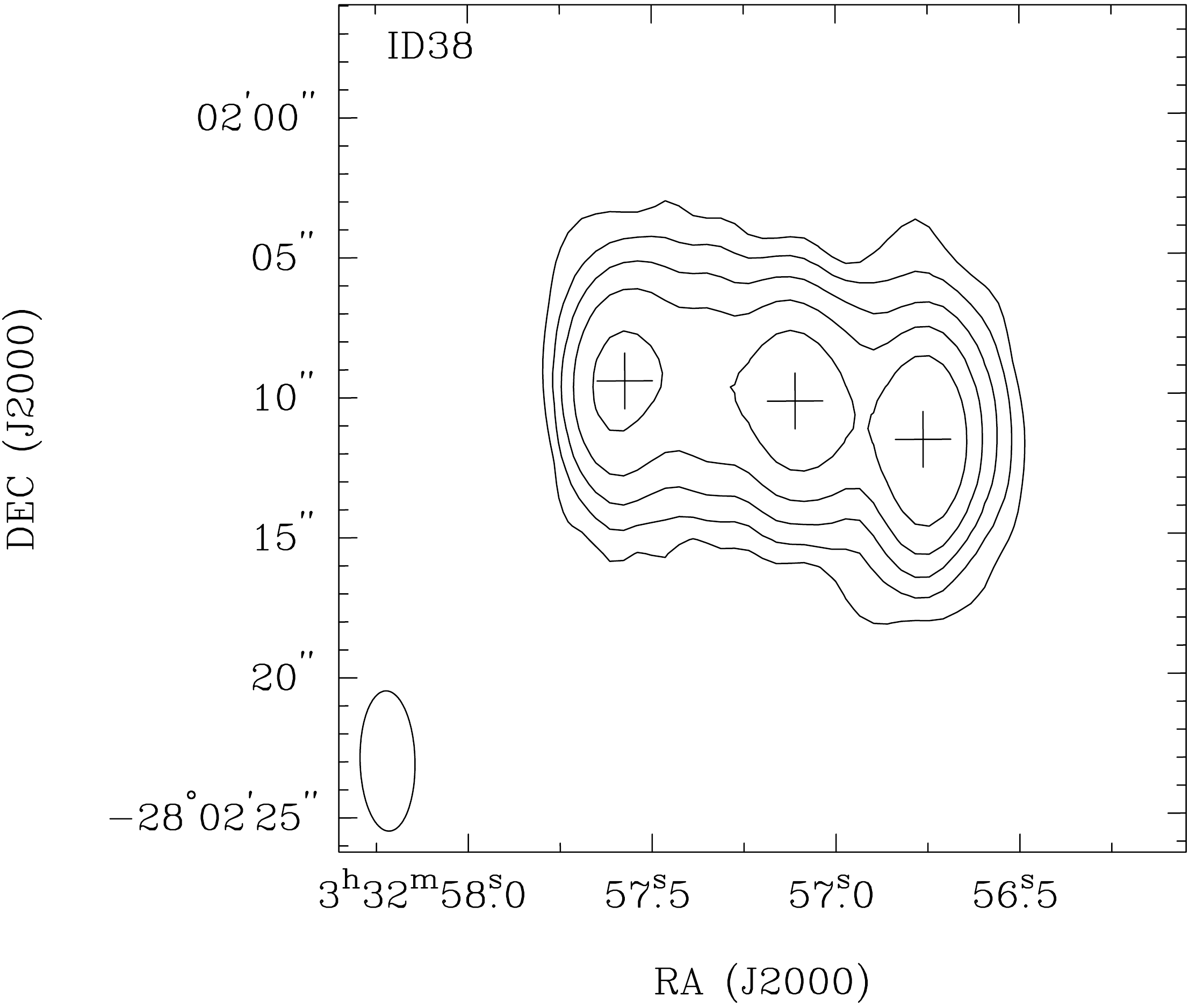}
\includegraphics[width=4cm]{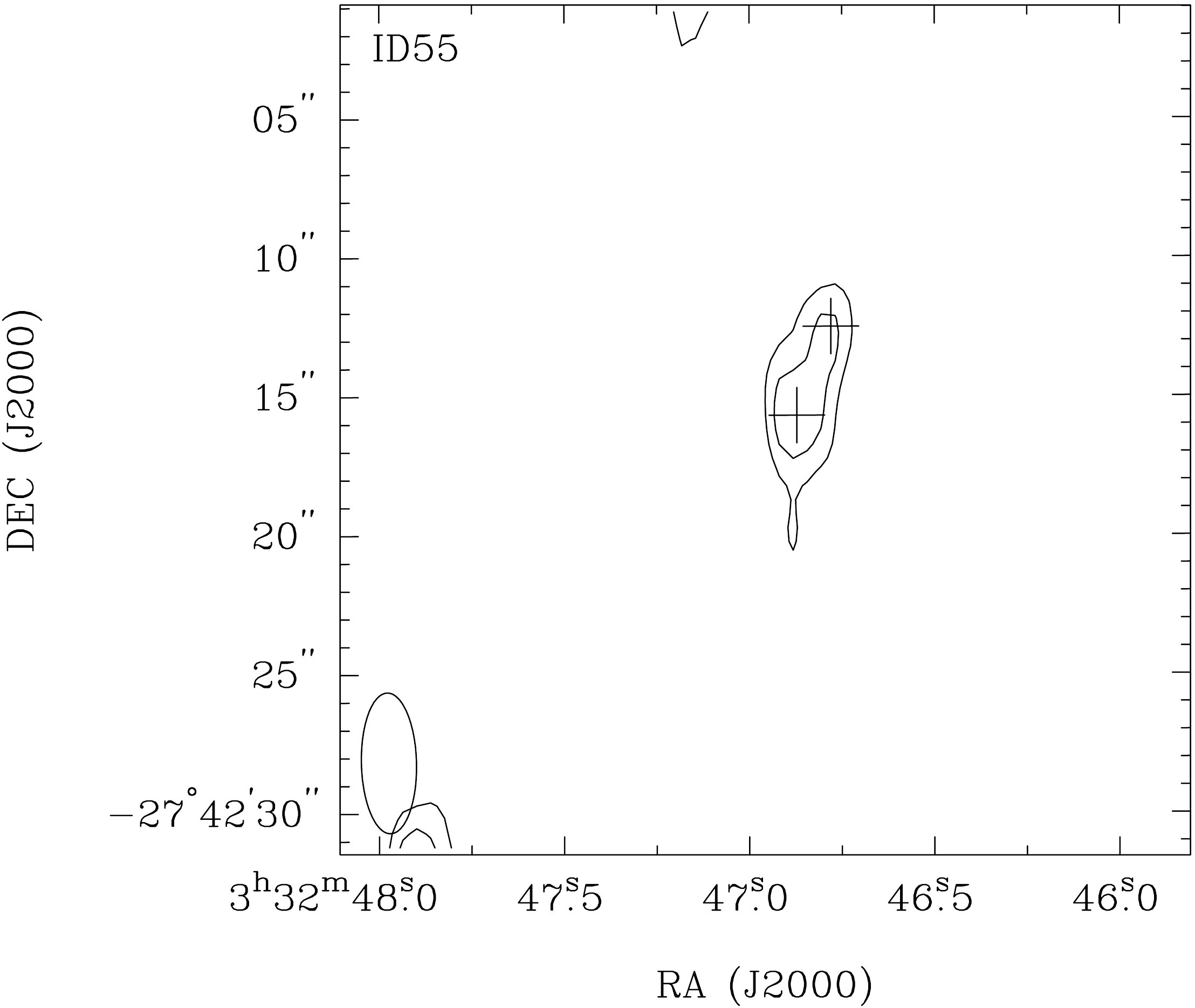}
\includegraphics[width=4cm]{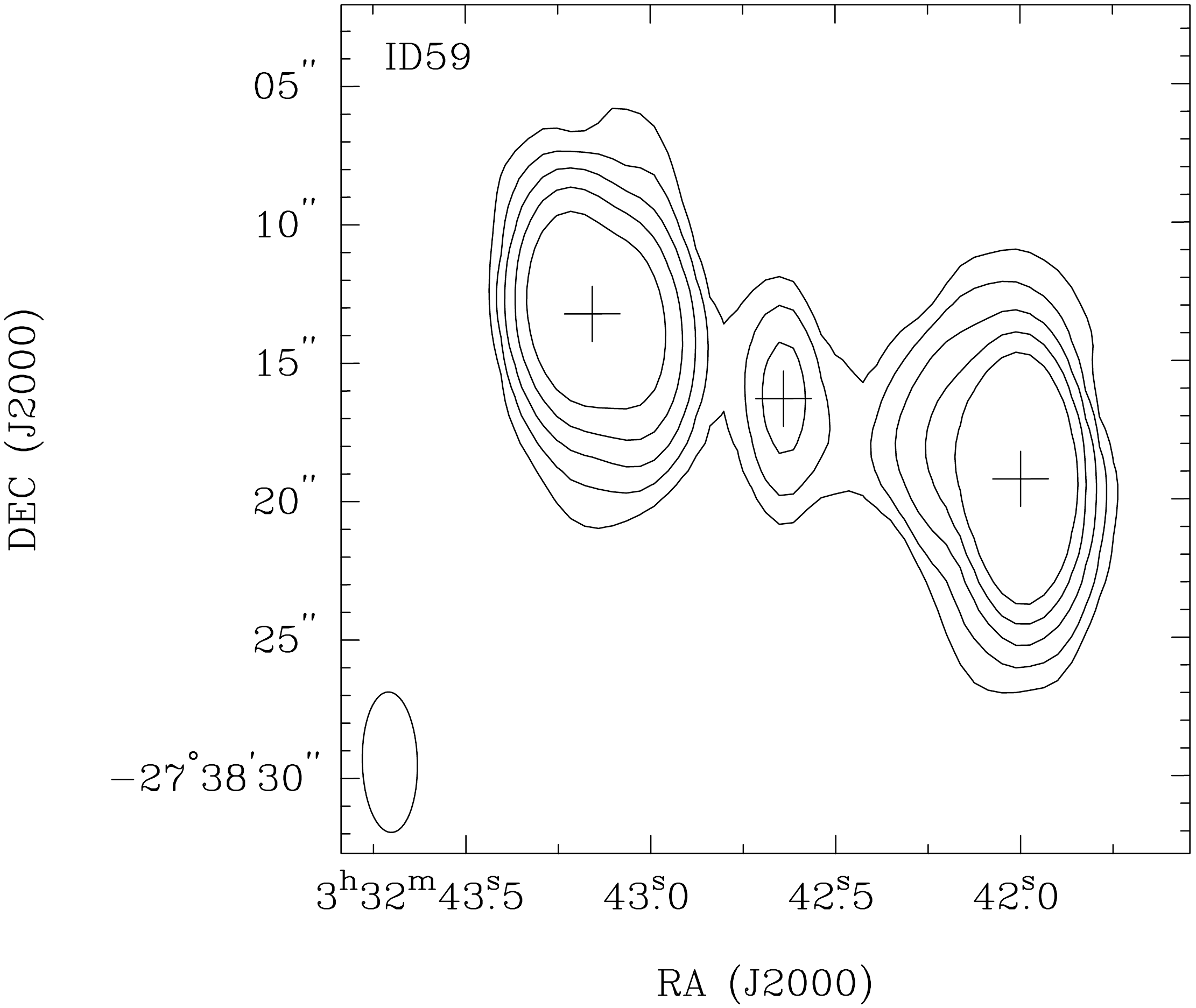}
\includegraphics[width=4cm]{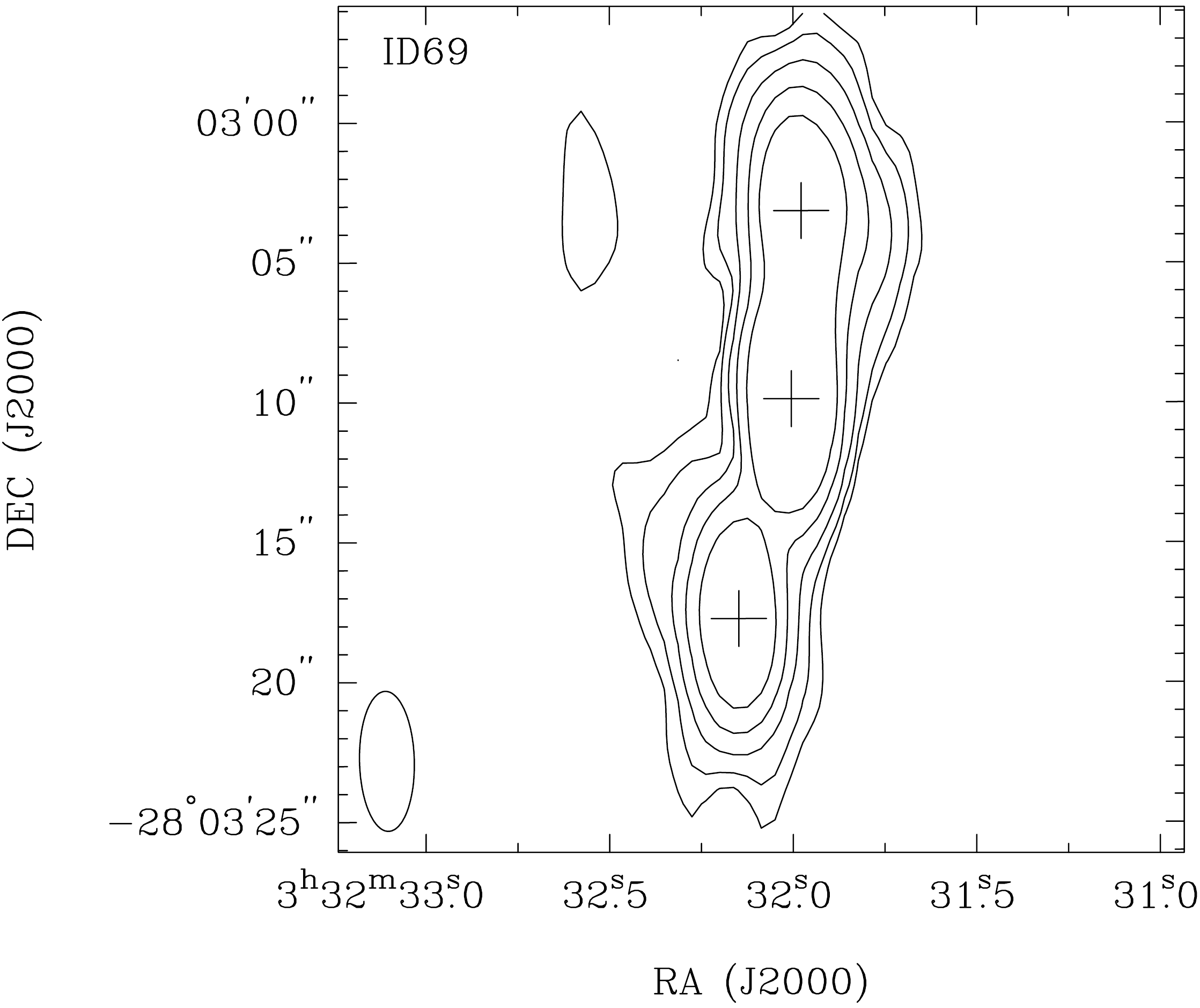}
\includegraphics[width=4cm]{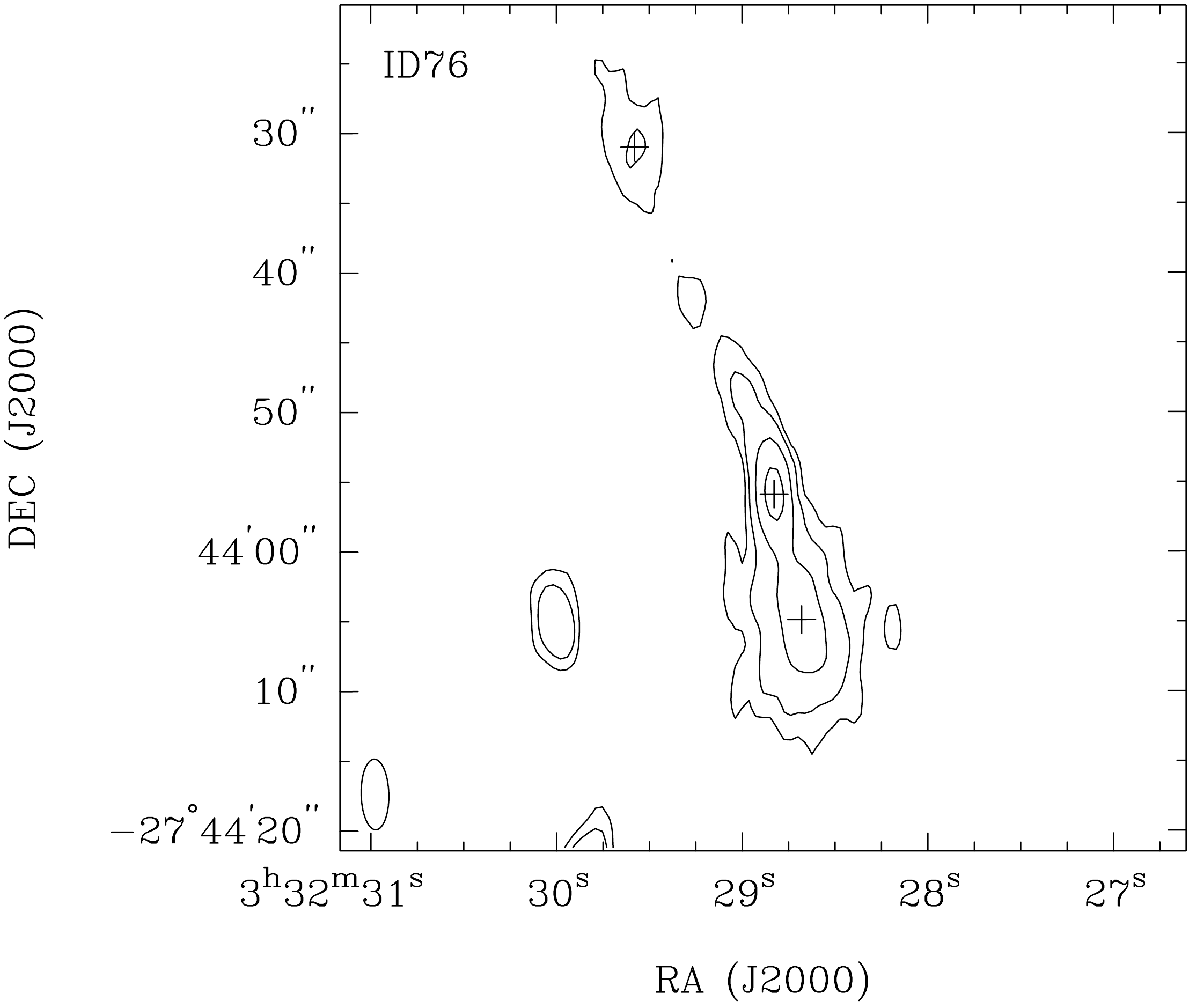}
\includegraphics[width=4cm]{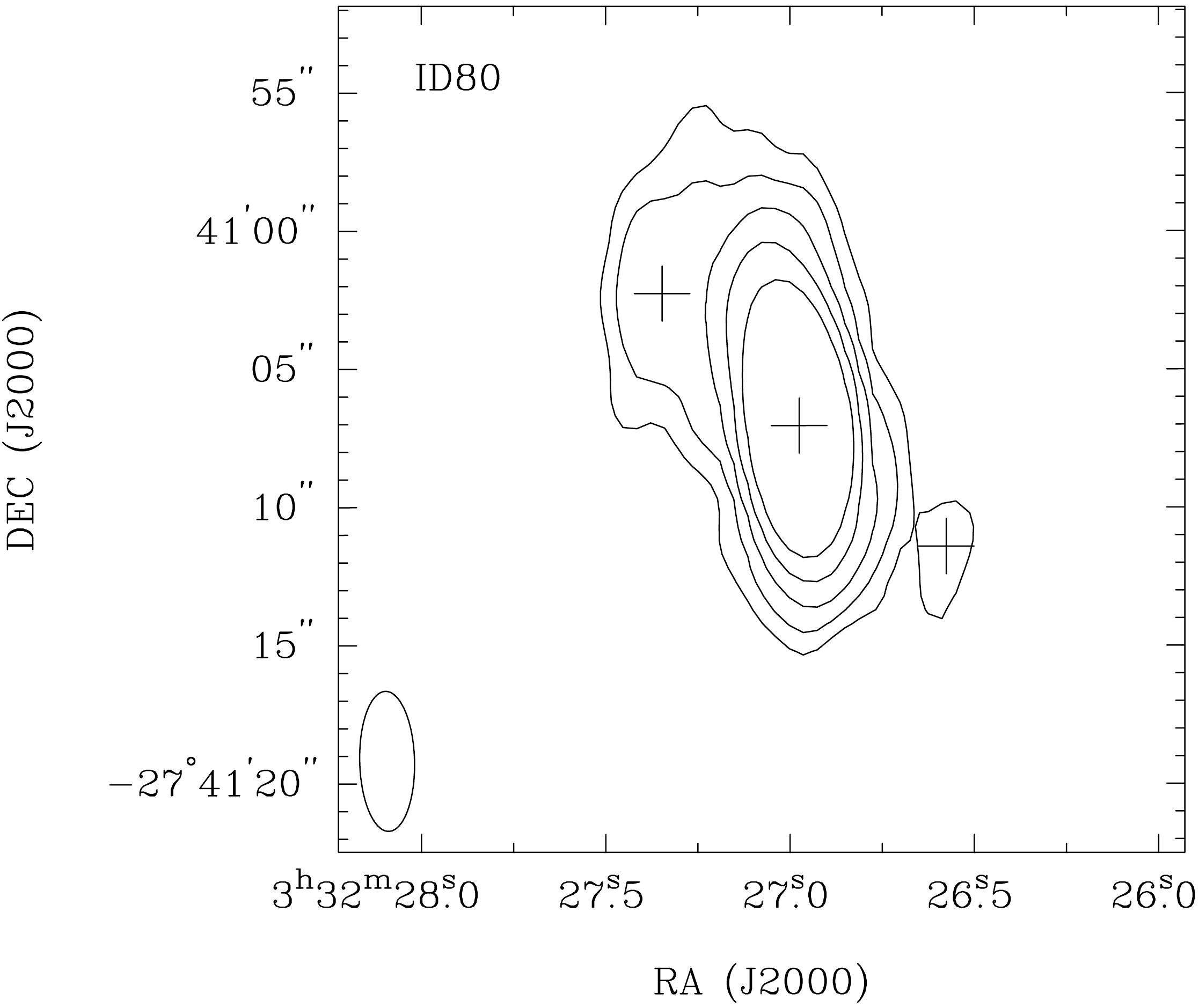}
\includegraphics[width=4cm]{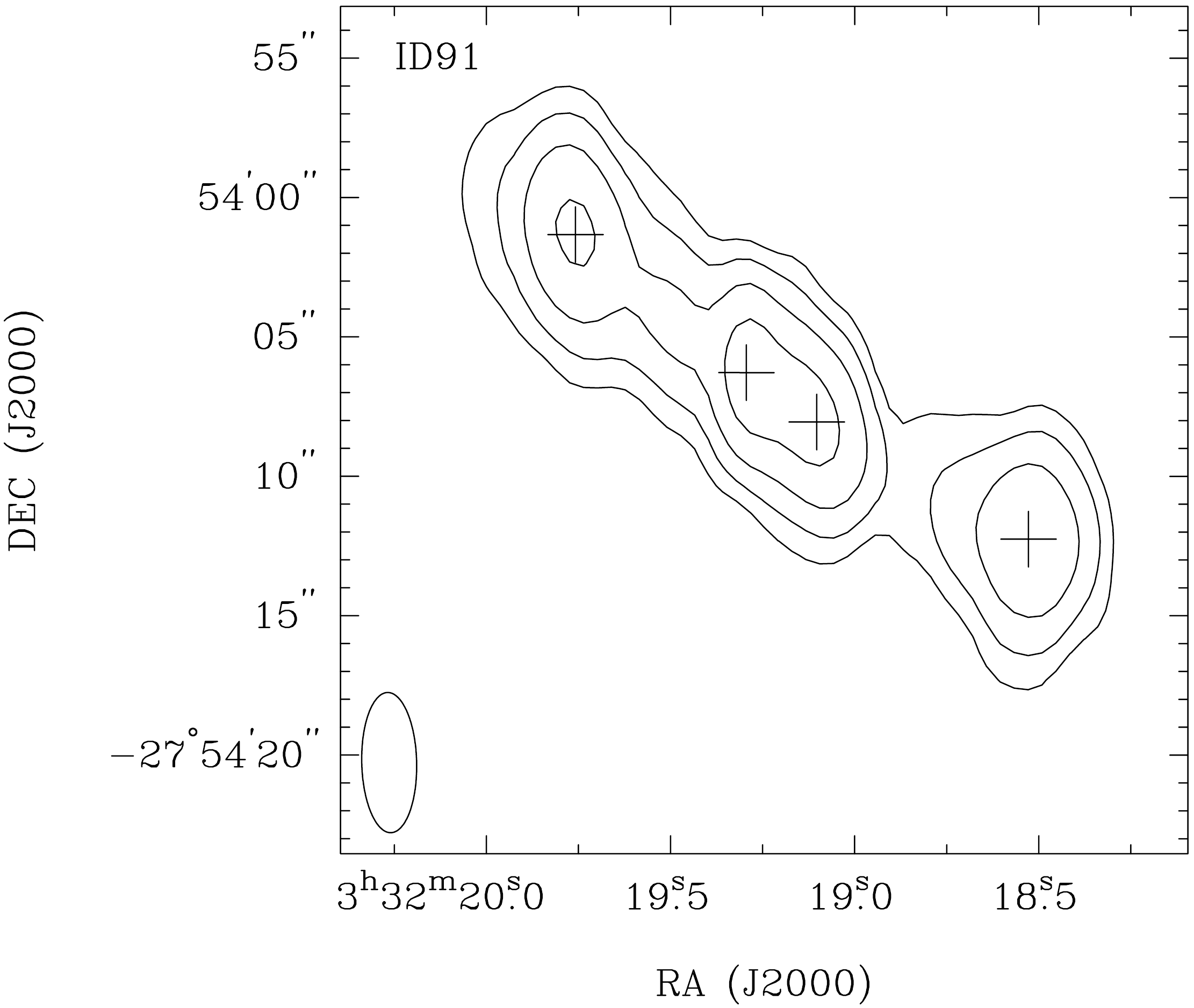}
\includegraphics[width=4cm]{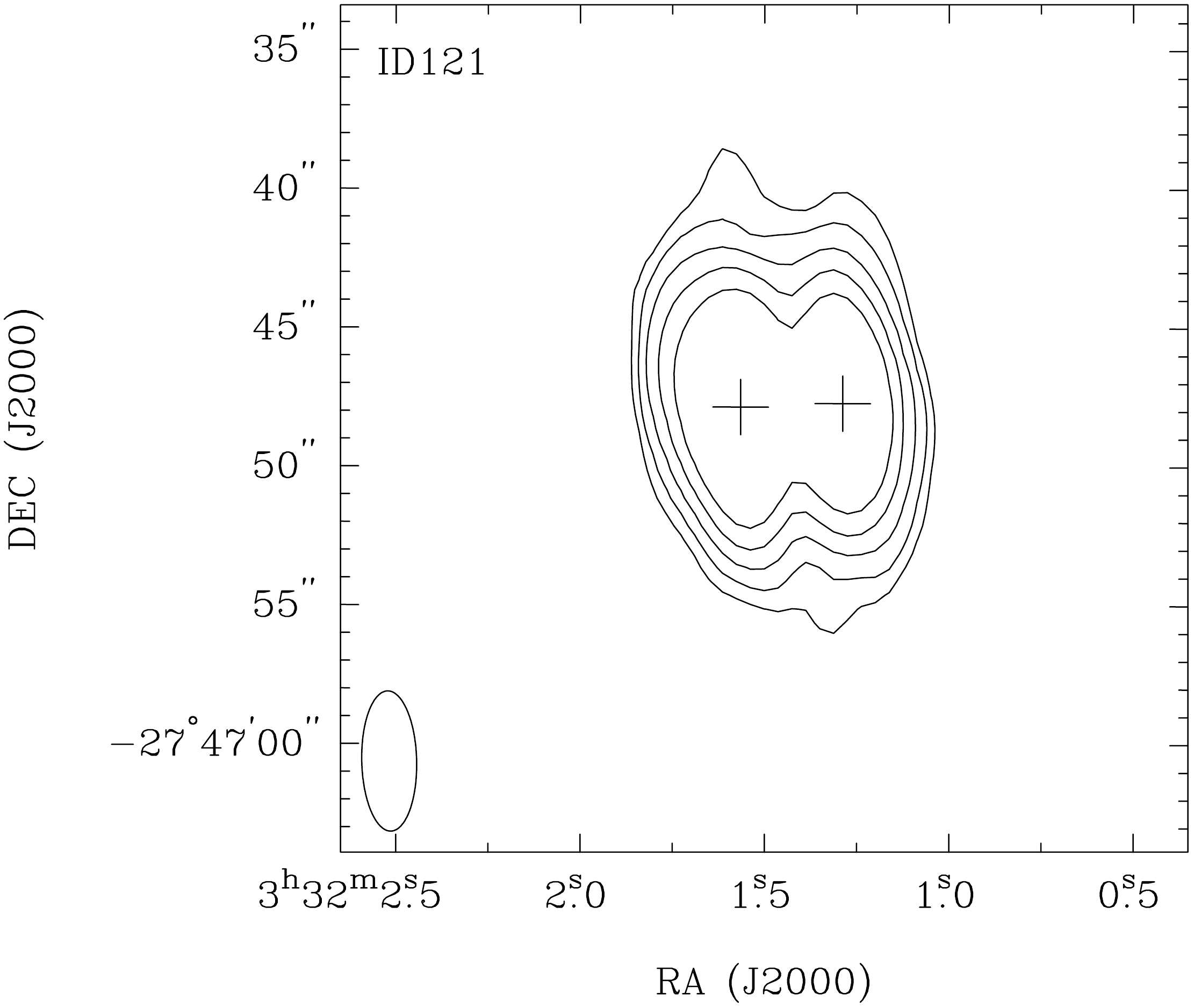} 
\includegraphics[width=4cm]{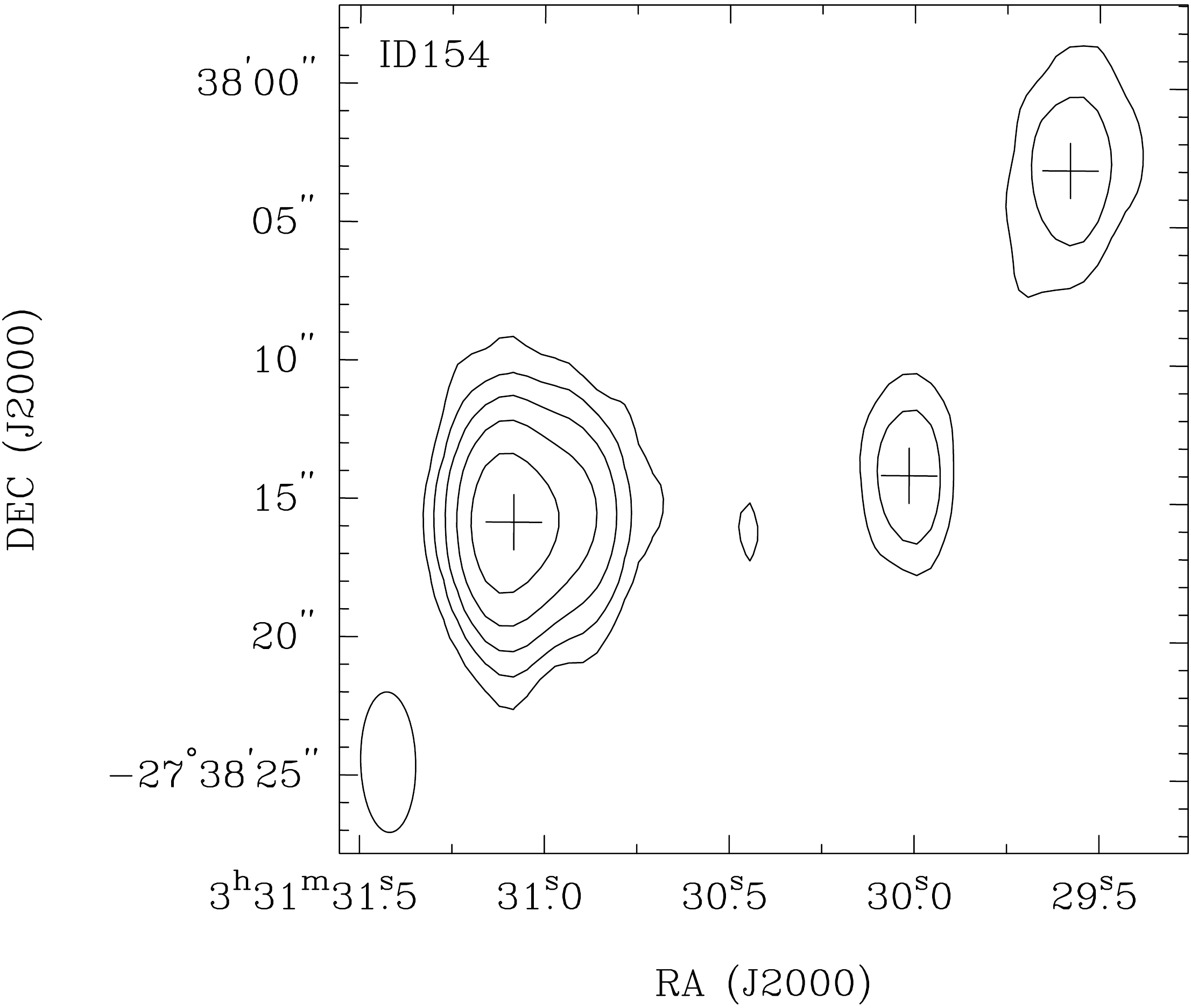}
\includegraphics[width=4cm]{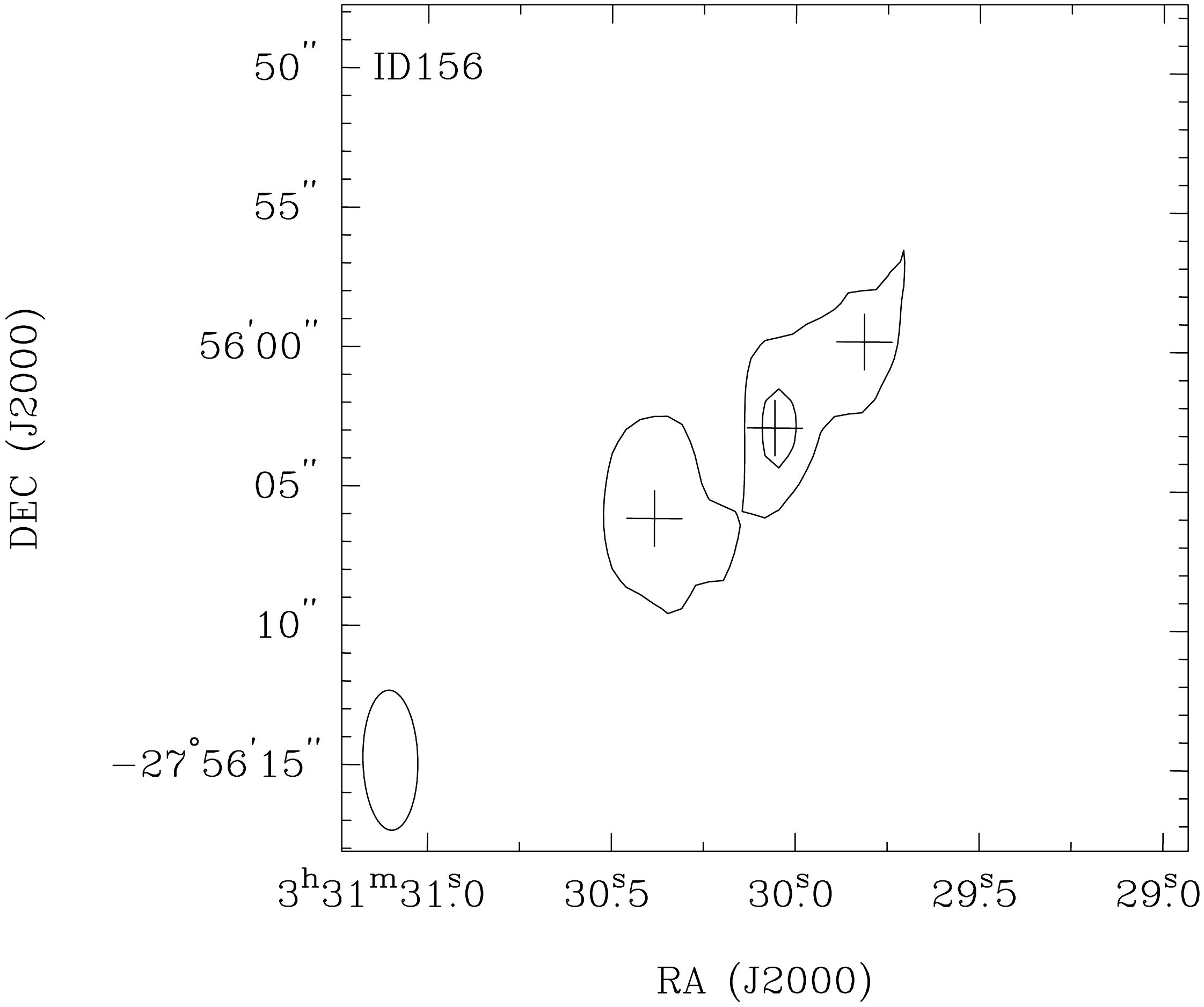}
\includegraphics[width=4cm]{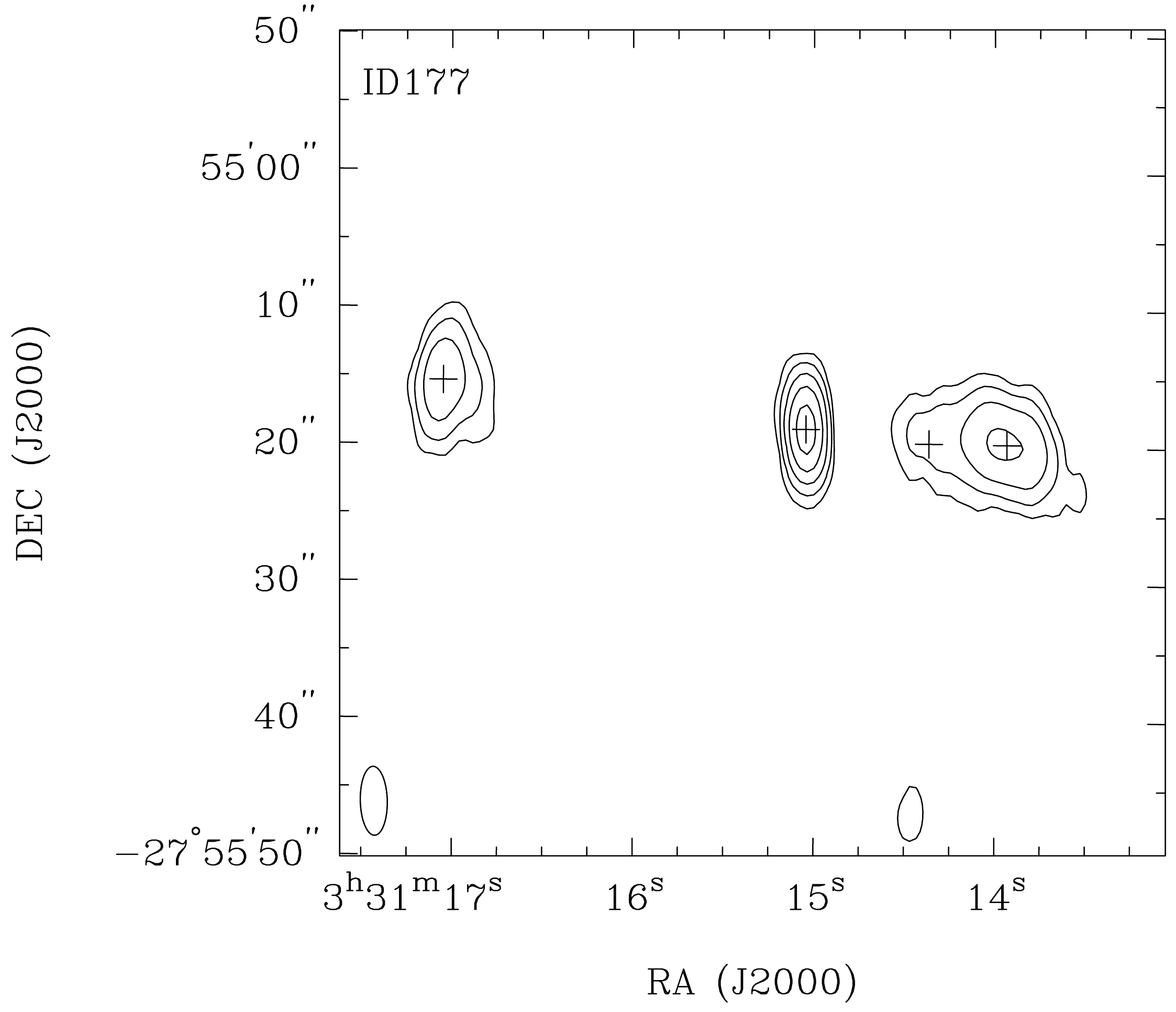}
\caption{Contour images of the multiple sources in the catalogue. The images are 30 $\times$ 30 arcsec in size, except for IDs 76 and 177, which are 1 $\times$ 1 arcmin. 
The contour levels are set at 5, 10, 20, 40 and 80 times the local noise level. However IDs 20, 55 and 76 also have a 3 sigma contour to highlight more detail in the source morphology. 
The synthesized beam is shown in the bottom left corner. Crosses mark the positions of the catalogued components.}
 \label{fig:multicomp}
\end{figure*}

We searched the full-band mosaic shown in Figure \ref{fig:ecdfs_grey}, which has a total area of 0.34 deg$^2$.
As in H12 we ran {\em sfind} with `rmsbox' set to 10 synthesized beamwidths and `alpha' set to 1. If the noise is perfectly Gaussian then setting `alpha' to 1 returns a list of sources which is 99\% reliable. 
Each {\em sfind} source was then individually fit as a point source and a Gaussian with MIRIAD task {\em imfit}. We identified 12 multiple component sources via visual inspection (see Figure \ref{fig:multicomp}). These sources exhibit classical core-lobe or lobe-lobe radio AGN morphology and are components of a single source. They were fitted as multiple Gaussians with {\em imfit} where necessary and the components listed individually in the final catalogue. There are 212 source components and 189 sources in the final catalogue.

\subsection{Deconvolution}
\label{sec:deconv}

\begin{figure} 
\centering   
\includegraphics[width=0.95\columnwidth]{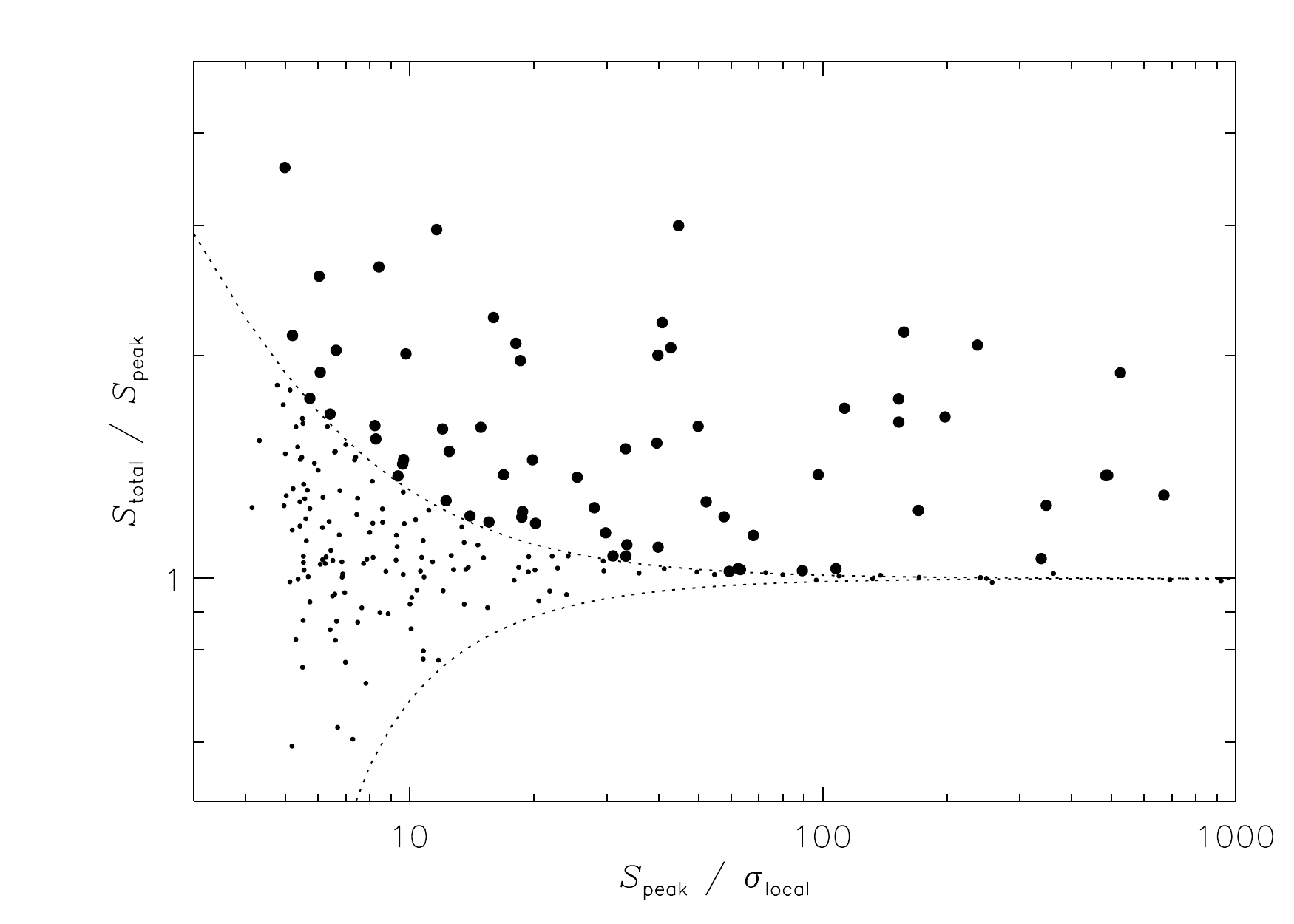}   
\caption{The ratio of integrated ($S_{\rm tot}$) flux density to peak flux density ($S_{\rm peak}$) as a function of source signal to noise ($S_{\rm peak}/\sigma$). The dotted line shows the upper and lower envelopes of the flux ratio distribution that contains 90\% of the unresolved sources. The large dots indicate sources which are deconvolved successfully and considered resolved.}
\label{fig:deconv}
\end{figure}

The ratio of integrated to peak flux density gives a direct measure of the extension of a source.  We performed the same analysis as in H12 to determine if a source is resolved, using the ratio of integrated flux to the peak flux (see Equation 1 of H12), where the peak flux is the peak of the fitted Gaussian. Whether a source is successfully deconvolved depends on the S/N ratio of the source and not just the synthesised beam-size. Using the Gaussian fits from {\em imfit}, we show the integrated flux density to peak flux density as a function of S/N in Figure \ref{fig:deconv}. 

Assuming the sources with $S_{\rm tot}/S_{\rm peak} < 1$ are due to noise then an envelope can be defined as: 
\begin{equation}
 S_{\rm tot}/S_{\rm peak} = 1 + a / (S_{\rm peak}/\sigma)^{b} .
 \end{equation}
In H12 we defined this envelope with a = 10 and b = 1.5. Figure 4 shows the lower curve, Equation 1 mirrored across $S_{\rm tot}/S_{\rm peak} = 1$, sufficiently encompasses all the $S_{\rm tot}/S_{\rm peak} < 1$. 
Sources which lie above the envelope, Equation 1, are considered successfully deconvolved. We add the extra criterion that $S_{\rm tot}/S_{\rm peak} > 1.02$ to account for the uncertainty in Gaussian fitting, which would otherwise push compact bright sources over the deconvolved line.  We find that 66/212 (31\%) source components lie above the upper envelope and have $S_{\rm tot}/S_{\rm peak} > 1.02$, and we consider these to be successfully deconvolved (i.e. resolved). 

\subsection{The Source Catalogue}

The source catalogue is reported in Table \ref{tab:cat}.  Point-source measurements are given for sources which are not successfully deconvolved. The integrated source flux density and deconvolved source sizes from the Gaussian fits are given for the resolved, or successfully deconvolved, sources.
Absolute calibration errors dominate for high S/N sources, but internal fitting errors shown in Table \ref{tab:cat} dominate for the majority of sources, which are low S/N. 

Column (1) - ID. A letter, such as `a', `b', etc., indicates a component of a multiple source. 

Column (2) - Source IAU name

Columns (3) and (4) - Source position: Right Ascension and Declination (J2000)

Column (5) - Point source flux density ($\mu$Jy). (Peak flux density for deconvolved sources.)

Column (6) - Uncertainty in point source flux density ($\mu$Jy). (Uncertainty in peak flux density for deconvolved sources .) 

Column (7) - Integrated flux density ($\mu$Jy). Zero indicates source is not successfully deconvolved and hence no integrated flux density is given. 

Column (8) - Uncertainty in integrated flux density ($\mu$Jy). Zero indicates source is not successfully deconvolved. 

Column (9) - Deconvolved major axis (arcsec). Zero indicates source is not successfully deconvolved. 

Column (10) - Deconvolved minor axis (arcsec). Zero indicates source is not successfully deconvolved. 

Column (11) - Deconvolved position angle (degrees), measured from north through east. Zero indicates source is not successfully deconvolved. 

Column(12) - Local noise level, rms, in $\mu$Jy. 

\begin{table*}\caption{The ATLAS 5.5 GHz Data Release 2 Catalogue}  
{\scriptsize\begin{tabular}{llllrrrrrrrr}  
\hline
ID & IAU name & RA & Dec  & $S_{pnt}$ & d$S_{pnt}$ & $S_{int}$ & d$S_{int}$ & Decon& Decon & Decon & $\sigma_{local}$ \\  
     &            &   (J2000) &  (J2000)   & ($\mu$Jy)  & ($\mu$Jy)  & ($\mu$Jy) & ($\mu$Jy) & Bmajor & Bminor & PA & \\ \hline

      1 &   ATCDFS5 J0033348.75-280233.1 & 03:33:48.75 & -28:02:33.1 &        283 &        30 &          0 & 0 &   0.00 &   0.00 &    0.0 &       30.5 \\
      2 &   ATCDFS5 J0033341.31-273809.0 & 03:33:41.31 & -27:38:09.0 &        306 &        17 &          0 & 0 &   0.00 &   0.00 &    0.0 &       22.4 \\
      3 &   ATCDFS5 J0033338.35-280030.9 & 03:33:38.35 & -28:00:30.9 &        544 &        27 &          0 & 0 &   0.00 &   0.00 &    0.0 &       16.9 \\
      4 &   ATCDFS5 J0033334.58-274751.3 & 03:33:34.58 & -27:47:51.3 &        155 &        14 &          0 & 0 &   0.00 &   0.00 &    0.0 &       15.4 \\
      5 &   ATCDFS5 J0033333.43-275332.9 & 03:33:33.43 & -27:53:32.9 &        505 &        12 &          0 & 0 &   0.00 &   0.00 &    0.0 &       14.0 \\
      6 &   ATCDFS5 J0033333.14-273932.7 & 03:33:33.14 & -27:39:32.7 &         96 &        12 &          0 & 0 &   0.00 &   0.00 &    0.0 &       14.4 \\
      7 &   ATCDFS5 J0033333.14-274602.1 & 03:33:33.14 & -27:46:02.1 &         95 &        14 &          0 & 0 &   0.00 &   0.00 &    0.0 &       14.2 \\
      8 &   ATCDFS5 J0033332.56-273538.9 & 03:33:32.56 & -27:35:38.9 &        421 &        13 &          0 & 0 &   0.00 &   0.00 &    0.0 &       14.1 \\
      9 &   ATCDFS5 J0033327.54-275726.1 & 03:33:27.54 & -27:57:26.1 &        113 &        12 &        155 & 38 &   2.59 &   1.31 &   16.3 &       12.0 \\
     10 &   ATCDFS5 J0033325.85-274343.0 & 03:33:25.85 & -27:43:43.0 &        231 &        12 &          0 & 0 &   0.00 &   0.00 &    0.0 &       10.2 \\
     11 &   ATCDFS5 J0033322.74-275459.9 & 03:33:22.74 & -27:54:59.9 &         93 &        12 &          0 & 0 &   0.00 &   0.00 &    0.0 &       10.5 \\
     12 &   ATCDFS5 J0033321.31-274138.6 & 03:33:21.31 & -27:41:38.6 &        265 &        10 &          0 & 0 &   0.00 &   0.00 &    0.0 &       10.7 \\
     13 &   ATCDFS5 J0033320.60-274910.0 & 03:33:20.60 & -27:49:10.0 &         56 &         9 &          0 & 0 &   0.00 &   0.00 &    0.0 &       10.1 \\
     14 &   ATCDFS5 J0033319.05-273530.6 & 03:33:19.05 & -27:35:30.6 &         72 &        18 &          0 & 0 &   0.00 &   0.00 &    0.0 &       10.2 \\
     15 &   ATCDFS5 J0033318.71-274940.2 & 03:33:18.71 & -27:49:40.2 &         76 &        10 &          0 & 0 &   0.00 &   0.00 &    0.0 &       10.1 \\
     16 &   ATCDFS5 J0033318.29-273440.0 & 03:33:18.29 & -27:34:40.0 &        108 &        15 &          0 & 0 &   0.00 &   0.00 &    0.0 &       10.4 \\
     17 &   ATCDFS5 J0033316.94-274121.9 & 03:33:16.94 & -27:41:21.9 &         74 &        11 &        120 & 40 &   2.70 &   1.74 &  -34.1 &        9.0 \\
     18 &   ATCDFS5 J0033316.76-280016.1 & 03:33:16.76 & -28:00:16.1 &       1286 &        15 &          0 & 0 &   0.00 &   0.00 &    0.0 &        9.3 \\
     19 &   ATCDFS5 J0033316.73-275630.4 & 03:33:16.73 & -27:56:30.4 &        697 &         9 &          0 & 0 &   0.00 &   0.00 &    0.0 &        9.5 \\
    20A &   ATCDFS5 J0033316.61-275040.0 & 03:33:16.61 & -27:50:40.0 &         55 &        14 &          0 & 0 &   0.00 &   0.00 &    0.0 &        9.6 \\
    20B &   ATCDFS5 J0033316.41-275041.5 & 03:33:16.41 & -27:50:41.5 &         55 &        15 &          0 & 0 &   0.00 &   0.00 &    0.0 &        9.6 \\
     21 &   ATCDFS5 J0033316.35-274725.1 & 03:33:16.35 & -27:47:25.1 &       1298 &        15 &          0 & 0 &   0.00 &   0.00 &    0.0 &        9.8 \\
     22 &   ATCDFS5 J0033314.98-275151.4 & 03:33:14.98 & -27:51:51.4 &        704 &        13 &          0 & 0 &   0.00 &   0.00 &    0.0 &        8.8 \\
     23 &   ATCDFS5 J0033314.84-280432.1 & 03:33:14.84 & -28:04:32.1 &        246 &        14 &          0 & 0 &   0.00 &   0.00 &    0.0 &       13.9 \\
     24 &   ATCDFS5 J0033313.13-274930.5 & 03:33:13.13 & -27:49:30.5 &        137 &        11 &          0 & 0 &   0.00 &   0.00 &    0.0 &        8.9 \\
     25 &   ATCDFS5 J0033312.63-275231.8 & 03:33:12.63 & -27:52:31.8 &         67 &        12 &          0 & 0 &   0.00 &   0.00 &    0.0 &        8.3 \\
     26 &   ATCDFS5 J0033311.80-274138.7 & 03:33:11.80 & -27:41:38.7 &        100 &         7 &          0 & 0 &   0.00 &   0.00 &    0.0 &        8.6 \\
     27 &   ATCDFS5 J0033310.19-274842.2 & 03:33:10.19 & -27:48:42.2 &      10114 &        54 &          0 & 0 &   0.00 &   0.00 &    0.0 &       11.0 \\
     28 &   ATCDFS5 J0033309.73-274802.0 & 03:33:09.73 & -27:48:02.0 &         89 &        12 &        127 & 44 &   3.73 &   0.72 &   19.6 &        9.2 \\
     29 &   ATCDFS5 J0033308.17-275033.3 & 03:33:08.17 & -27:50:33.3 &        499 &         8 &          0 & 0 &   0.00 &   0.00 &    0.0 &        9.1 \\
     30 &   ATCDFS5 J0033305.11-274028.6 & 03:33:05.11 & -27:40:28.6 &         51 &        12 &          0 & 0 &   0.00 &   0.00 &    0.0 &        8.5 \\
     31 &   ATCDFS5 J0033304.45-273802.1 & 03:33:04.45 & -27:38:02.1 &         63 &        10 &          0 & 0 &   0.00 &   0.00 &    0.0 &        8.5 \\
     32 &   ATCDFS5 J0033303.73-273611.1 & 03:33:03.73 & -27:36:11.1 &        300 &        14 &        333 & 28 &   1.96 &   0.51 &   -3.9 &        9.0 \\
     33 &   ATCDFS5 J0033302.68-275642.7 & 03:33:02.68 & -27:56:42.7 &         61 &        10 &          0 & 0 &   0.00 &   0.00 &    0.0 &        8.0 \\
     34 &   ATCDFS5 J0033301.82-273637.2 & 03:33:01.82 & -27:36:37.2 &         65 &         8 &          0 & 0 &   0.00 &   0.00 &    0.0 &        9.2 \\
     35 &   ATCDFS5 J0033301.83-274540.4 & 03:33:01.83 & -27:45:40.4 &         49 &         8 &          0 & 0 &   0.00 &   0.00 &    0.0 &        8.8 \\
     36 &   ATCDFS5 J0033259.30-273534.5 & 03:32:59.30 & -27:35:34.5 &         60 &        10 &          0 & 0 &   0.00 &   0.00 &    0.0 &        9.7 \\
     37 &   ATCDFS5 J0033259.21-274325.4 & 03:32:59.21 & -27:43:25.4 &         63 &        14 &          0 & 0 &   0.00 &   0.00 &    0.0 &        9.4 \\
    38A &   ATCDFS5 J0033257.57-280209.4 & 03:32:57.57 & -28:02:09.4 &       1428 &        48 &       2426 & 166 &   2.74 &   1.94 &  -39.0 &       12.7 \\
    38B &   ATCDFS5 J0033257.11-280210.2 & 03:32:57.11 & -28:02:10.2 &       1961 &        49 &       4222 & 226 &   3.71 &   1.17 &   78.7 &       12.5 \\
    38C &   ATCDFS5 J0033256.76-280211.6 & 03:32:56.76 & -28:02:11.6 &       2413 &        60 &       3990 & 184 &   2.32 &   2.12 &   56.5 &       12.2 \\
     39 &   ATCDFS5 J0033256.47-275848.3 & 03:32:56.47 & -27:58:48.3 &        921 &        14 &        949 & 26 &   1.18 &   0.16 &    3.9 &        8.6 \\
     40 &   ATCDFS5 J0033256.26-273500.7 & 03:32:56.26 & -27:35:00.7 &        122 &        10 &          0 & 0 &   0.00 &   0.00 &    0.0 &       10.3 \\
     41 &   ATCDFS5 J0033253.34-280159.3 & 03:32:53.34 & -28:01:59.3 &        564 &        18 &        683 & 40 &   3.31 &   0.31 &   -0.9 &        9.8 \\
     42 &   ATCDFS5 J0033252.89-273838.5 & 03:32:52.89 & -27:38:38.5 &         52 &         7 &          0 & 0 &   0.00 &   0.00 &    0.0 &        8.3 \\
     43 &   ATCDFS5 J0033252.24-280209.7 & 03:32:52.24 & -28:02:09.7 &         65 &        10 &          0 & 0 &   0.00 &   0.00 &    0.0 &       10.5 \\
     44 &   ATCDFS5 J0033252.06-274425.6 & 03:32:52.06 & -27:44:25.6 &        203 &        14 &          0 & 0 &   0.00 &   0.00 &    0.0 &        8.8 \\
     45 &   ATCDFS5 J0033251.82-274436.7 & 03:32:51.82 & -27:44:36.7 &         70 &        13 &          0 & 0 &   0.00 &   0.00 &    0.0 &        9.0 \\
     46 &   ATCDFS5 J0033251.83-275717.4 & 03:32:51.83 & -27:57:17.4 &         51 &         9 &          0 & 0 &   0.00 &   0.00 &    0.0 &        8.4 \\
     47 &   ATCDFS5 J0033249.95-273432.9 & 03:32:49.95 & -27:34:32.9 &        139 &        16 &        206 & 59 &   3.63 &   0.94 &  -19.6 &       11.1 \\
     48 &   ATCDFS5 J0033249.93-273446.2 & 03:32:49.93 & -27:34:46.2 &         59 &        11 &          0 & 0 &   0.00 &   0.00 &    0.0 &       10.7 \\
     49 &   ATCDFS5 J0033249.43-274235.4 & 03:32:49.43 & -27:42:35.4 &        846 &        20 &        867 & 36 &   0.92 &   0.24 &    4.8 &        9.5 \\
     50 &   ATCDFS5 J0033249.20-274050.8 & 03:32:49.20 & -27:40:50.8 &       2366 &        28 &          0 & 0 &   0.00 &   0.00 &    0.0 &        9.5 \\
     51 &   ATCDFS5 J0033249.32-275844.1 & 03:32:49.32 & -27:58:44.1 &         70 &         9 &        109 & 27 &   3.41 &   1.45 &   16.8 &        8.5 \\
     52 &   ATCDFS5 J0033248.54-274934.0 & 03:32:48.54 & -27:49:34.0 &         44 &        10 &          0 & 0 &   0.00 &   0.00 &    0.0 &        8.4 \\
     53 &   ATCDFS5 J0033247.89-274232.7 & 03:32:47.89 & -27:42:32.7 &         76 &        12 &          0 & 0 &   0.00 &   0.00 &    0.0 &       10.1 \\
     54 &   ATCDFS5 J0033246.95-273903.3 & 03:32:46.95 & -27:39:03.3 &         50 &        10 &          0 & 0 &   0.00 &   0.00 &    0.0 &        9.1 \\
    55A &   ATCDFS5 J0033246.87-274215.6 & 03:32:46.87 & -27:42:15.6 &         72 &        14 &          0 & 0 &   0.00 &   0.00 &    0.0 &        9.1 \\
    55B &   ATCDFS5 J0033246.78-274212.4 & 03:32:46.78 & -27:42:12.4 &         59 &        17 &          0 & 0 &   0.00 &   0.00 &    0.0 &        8.9 \\
     56 &   ATCDFS5 J0033245.37-280449.9 & 03:32:45.37 & -28:04:49.9 &        612 &        31 &        933 & 106 &   3.11 &   0.38 &  -39.6 &       15.4 \\
     57 &   ATCDFS5 J0033244.26-275141.0 & 03:32:44.26 & -27:51:41.0 &        126 &        16 &          0 & 0 &   0.00 &   0.00 &    0.0 &        8.1 \\
     58 &   ATCDFS5 J0033244.05-275144.0 & 03:32:44.05 & -27:51:44.0 &         88 &        15 &          0 & 0 &   0.00 &   0.00 &    0.0 &        8.1 \\
    59A &   ATCDFS5 J0033243.15-273813.2 & 03:32:43.15 & -27:38:13.2 &       4612 &       257 &       9538 & 792 &   3.58 &   1.62 &   64.5 &       19.5 \\
    59B &   ATCDFS5 J0033242.64-273816.3 & 03:32:42.64 & -27:38:16.3 &        519 &        29 &        647 & 71 &   2.44 &   1.00 &    4.8 &       18.6 \\
    59C &   ATCDFS5 J0033241.99-273819.2 & 03:32:41.99 & -27:38:19.2 &      10668 &       441 &      13820 & 826 &   1.82 &   1.32 &   23.1 &       15.9 \\
     60 &   ATCDFS5 J0033242.62-273825.7 & 03:32:42.62 & -27:38:25.7 &         74 &        13 &          0 & 0 &   0.00 &   0.00 &    0.0 &       12.5 \\
     61 &   ATCDFS5 J0033241.99-273949.4 & 03:32:41.99 & -27:39:49.4 &        129 &         9 &          0 & 0 &   0.00 &   0.00 &    0.0 &        9.0 \\
     62 &   ATCDFS5 J0033241.62-280127.9 & 03:32:41.62 & -28:01:27.9 &        124 &         9 &          0 & 0 &   0.00 &   0.00 &    0.0 &        8.7 \\
     63 &   ATCDFS5 J0033240.82-275547.4 & 03:32:40.82 & -27:55:47.4 &         53 &         5 &          0 & 0 &   0.00 &   0.00 &    0.0 &        7.8 \\
     64 &   ATCDFS5 J0033239.47-275301.5 & 03:32:39.47 & -27:53:01.5 &         52 &         9 &          0 & 0 &   0.00 &   0.00 &    0.0 &        8.5 \\
     65 &   ATCDFS5 J0033237.73-275000.9 & 03:32:37.73 & -27:50:00.9 &         56 &        13 &          0 & 0 &   0.00 &   0.00 &    0.0 &        8.6 \\
     66 &   ATCDFS5 J0033237.23-275748.2 & 03:32:37.23 & -27:57:48.2 &         56 &         7 &          0 & 0 &   0.00 &   0.00 &    0.0 &        9.0 \\
     67 &   ATCDFS5 J0033234.93-275455.9 & 03:32:34.93 & -27:54:55.9 &         54 &        10 &          0 & 0 &   0.00 &   0.00 &    0.0 &        8.6 \\
     68 &   ATCDFS5 J0033232.55-280303.0 & 03:32:32.55 & -28:03:03.0 &        105 &        12 &          0 & 0 &   0.00 &   0.00 &    0.0 &       12.9 \\
    69A &   ATCDFS5 J0033232.14-280317.7 & 03:32:32.14 & -28:03:17.7 &       2254 &        77 &       2785 & 196 &   1.98 &   1.12 &    0.2 &       13.2 \\
    69B &   ATCDFS5 J0033232.00-280309.8 & 03:32:32.00 & -28:03:09.8 &       4525 &       177 &       4813 & 401 &   1.8 &   0.08 &    -17.3 &       13.4 \\
    69C &   ATCDFS5 J0033231.97-280303.1 & 03:32:31.97 & -28:03:03.1 &       2042 &       109 &       3323 & 389 &   3.28 &   1.83 &    1.5 &       13.4 \\
     70 &   ATCDFS5 J0033231.67-273415.5 & 03:32:31.67 & -27:34:15.5 &         67 &        13 &          0 & 0 &   0.00 &   0.00 &    0.0 &       11.0 \\
 
\hline

\end{tabular}
}
\label{tab:cat}
\end{table*}

\setcounter{table}{1}
\begin{table*}
\caption{continued}  
{\scriptsize
\begin{tabular}{llllrrrrrrrr}  
\hline
ID & IAU name & RA & Dec & $S_{pnt}$ & d$S_{pnt}$ & $S_{int}$ & d$S_{int}$ & Decon& Decon & Decon & $\sigma_{local}$ \\  
     &                   &    (J2000)    &       (J2000)                 & ($\mu$Jy)  & ($\mu$Jy)  & ($\mu$Jy) & ($\mu$Jy) & Bmajor & Bminor & PA & \\ \hline

     71 &   ATCDFS5 J0033231.54-275029.0 & 03:32:31.54 & -27:50:29.0 &        103 &        10 &          0 & 0 &   0.00 &   0.00 &    0.0 &        9.2 \\
     72 &   ATCDFS5 J0033230.56-275911.2 & 03:32:30.56 & -27:59:11.2 &        117 &         8 &        187 & 33 &   5.04 &   0.89 &   11.2 &        7.9 \\
     73 &   ATCDFS5 J0033230.00-274405.0 & 03:32:30.00 & -27:44:05.0 &        109 &         9 &        174 & 21 &   2.73 &   1.51 &   42.4 &        9.0 \\
     74 &   ATCDFS5 J0033229.86-274424.6 & 03:32:29.86 & -27:44:24.6 &        193 &        10 &        381 & 43 &   4.44 &   1.85 &  -21.9 &       10.4 \\
     75 &   ATCDFS5 J0033229.99-274302.3 & 03:32:29.99 & -27:43:02.3 &         47 &         8 &          0 & 0 &   0.00 &   0.00 &    0.0 &        8.7 \\
    76A &   ATCDFS5 J0033229.57-274331.0 & 03:32:29.57 & -27:43:31.0 &         63 &        13 &        228 & 85 &   5.43 &   4.41 &    4.8 &       12.7 \\
    76B &   ATCDFS5 J0033228.82-274355.8 & 03:32:28.82 & -27:43:55.8 &        311 &        17 &        450 & 73 &   3.86 &   1.05 &    9.5 &       15.7 \\
    76C &   ATCDFS5 J0033228.68-274404.8 & 03:32:28.68 & -27:44:04.8 &        164 &        15 &       1024 & 224 &  12.45 &   4.17 &    8.6 &       14.8 \\
     77 &   ATCDFS5 J0033228.73-274620.6 & 03:32:28.73 & -27:46:20.6 &        166 &        11 &        204 & 24 &   2.26 &   0.97 &   11.4 &        8.8 \\
     78 &   ATCDFS5 J0033228.58-273536.6 & 03:32:28.58 & -27:35:36.6 &         67 &        13 &        111 & 76 &   3.96 &   1.66 &    7.2 &       10.4 \\
     79 &   ATCDFS5 J0033228.35-273841.8 & 03:32:28.35 & -27:38:41.8 &         57 &        12 &          0 & 0 &   0.00 &   0.00 &    0.0 &        9.2 \\
    80A &   ATCDFS5 J0033227.34-274102.2 & 03:32:27.34 & -27:41:02.2 &        191 &        12 &        397 & 69 &   4.31 &   2.42 &    2.9 &       10.6 \\
    80B &   ATCDFS5 J0033226.97-274107.0 & 03:32:26.97 & -27:41:07.0 &       5114 &       150 &       7049 & 325 &   3.35 &   0.88 &   17.3 &       10.5 \\
    80C &   ATCDFS5 J0033226.57-274111.4 & 03:32:26.57 & -27:41:11.4 &         68 &        10 &        138 & 79 &   4.48 &   1.99 &  -21.5 &       10.2 \\
     81 &   ATCDFS5 J0033226.75-280454.9 & 03:32:26.75 & -28:04:54.9 &         84 &        18 &          0 & 0 &   0.00 &   0.00 &    0.0 &       15.0 \\
     82 &   ATCDFS5 J0033224.30-280114.5 & 03:32:24.30 & -28:01:14.5 &        147 &         9 &          0 & 0 &   0.00 &   0.00 &    0.0 &       10.0 \\
     83 &   ATCDFS5 J0033223.81-275845.1 & 03:32:23.81 & -27:58:45.1 &        104 &        12 &          0 & 0 &   0.00 &   0.00 &    0.0 &        8.6 \\
     84 &   ATCDFS5 J0033223.69-273648.3 & 03:32:23.69 & -27:36:48.3 &         83 &        10 &          0 & 0 &   0.00 &   0.00 &    0.0 &        9.0 \\
     85 &   ATCDFS5 J0033222.70-274127.2 & 03:32:22.70 & -27:41:27.2 &         53 &         8 &          0 & 0 &   0.00 &   0.00 &    0.0 &        8.5 \\
     86 &   ATCDFS5 J0033222.61-280023.9 & 03:32:22.61 & -28:00:23.9 &         90 &         9 &        181 & 51 &   4.49 &   2.18 &   -7.1 &        9.2 \\
     87 &   ATCDFS5 J0033222.52-274804.4 & 03:32:22.52 & -27:48:04.4 &         55 &         6 &          0 & 0 &   0.00 &   0.00 &    0.0 &        8.3 \\
     88 &   ATCDFS5 J0033221.72-280153.2 & 03:32:21.72 & -28:01:53.2 &         93 &         6 &          0 & 0 &   0.00 &   0.00 &    0.0 &        9.9 \\
     89 &   ATCDFS5 J0033221.28-274436.1 & 03:32:21.28 & -27:44:36.1 &         87 &        10 &          0 & 0 &   0.00 &   0.00 &    0.0 &        9.3 \\
     90 &   ATCDFS5 J0033221.07-273530.6 & 03:32:21.07 & -27:35:30.6 &        102 &        10 &          0 & 0 &   0.00 &   0.00 &    0.0 &       10.5 \\
    91A &   ATCDFS5 J0033219.75-275401.3 & 03:32:19.75 & -27:54:01.3 &        489 &        16 &       1003 & 72 &   3.64 &   2.22 &   42.2 &       11.4 \\
    91B &   ATCDFS5 J0033219.29-275406.2 & 03:32:19.29 & -27:54:06.2 &        581 &        77 &        738 & 81 &   1.85 &   0.25 &   54.6 &       11.1 \\
    91C &   ATCDFS5 J0033219.10-275408.0 & 03:32:19.10 & -27:54:08.0 &        545 &        41 &        875 & 102 &   2.98 &   0.83 &   49.6 &       10.9 \\
    91D &   ATCDFS5 J0033218.52-275412.2 & 03:32:18.52 & -27:54:12.2 &        411 &        19 &        911 & 104 &   3.37 &   2.74 &   67.5 &       10.1 \\
     92 &   ATCDFS5 J0033219.80-274123.2 & 03:32:19.80 & -27:41:23.2 &         81 &         8 &          0 & 0 &   0.00 &   0.00 &    0.0 &        8.4 \\
     93 &   ATCDFS5 J0033219.50-275218.1 & 03:32:19.50 & -27:52:18.1 &         77 &        12 &          0 & 0 &   0.00 &   0.00 &    0.0 &        9.6 \\
     94 &   ATCDFS5 J0033218.02-274718.6 & 03:32:18.02 & -27:47:18.6 &        422 &        15 &          0 & 0 &   0.00 &   0.00 &    0.0 &        8.5 \\
     95 &   ATCDFS5 J0033217.05-275846.6 & 03:32:17.05 & -27:58:46.6 &       1718 &        13 &          0 & 0 &   0.00 &   0.00 &    0.0 &       10.0 \\
     96 &   ATCDFS5 J0033217.04-275916.7 & 03:32:17.04 & -27:59:16.7 &         50 &        11 &         88 & 31 &   3.41 &   2.05 &   10.8 &        8.8 \\
     97 &   ATCDFS5 J0033215.95-273438.5 & 03:32:15.95 & -27:34:38.5 &        217 &        15 &        258 & 38 &   2.03 &   0.91 &    2.3 &       10.8 \\
     98 &   ATCDFS5 J0033215.39-273706.9 & 03:32:15.39 & -27:37:06.9 &         58 &         8 &          0 & 0 &   0.00 &   0.00 &    0.0 &        9.3 \\
     99 &   ATCDFS5 J0033214.83-275640.3 & 03:32:14.83 & -27:56:40.3 &         82 &        10 &          0 & 0 &   0.00 &   0.00 &    0.0 &        8.8 \\
    100 &   ATCDFS5 J0033213.89-275001.0 & 03:32:13.89 & -27:50:01.0 &         92 &        11 &          0 & 0 &   0.00 &   0.00 &    0.0 &        8.4 \\
    101 &   ATCDFS5 J0033213.48-274953.5 & 03:32:13.48 & -27:49:53.5 &         90 &        12 &          0 & 0 &   0.00 &   0.00 &    0.0 &        8.7 \\
    102 &   ATCDFS5 J0033213.23-274241.2 & 03:32:13.23 & -27:42:41.2 &         44 &        12 &          0 & 0 &   0.00 &   0.00 &    0.0 &        8.2 \\
    103 &   ATCDFS5 J0033213.08-274350.9 & 03:32:13.08 & -27:43:50.9 &        283 &        11 &        424 & 29 &   2.42 &   1.79 &   -2.7 &        8.5 \\
    104 &   ATCDFS5 J0033211.65-273726.2 & 03:32:11.65 & -27:37:26.2 &      11886 &        70 &          0 & 0 &   0.00 &   0.00 &    0.0 &       13.0 \\
    105 &   ATCDFS5 J0033211.53-274713.3 & 03:32:11.53 & -27:47:13.3 &         90 &         8 &          0 & 0 &   0.00 &   0.00 &    0.0 &        8.5 \\
    106 &   ATCDFS5 J0033211.50-274816.2 & 03:32:11.50 & -27:48:16.2 &         50 &        12 &          0 & 0 &   0.00 &   0.00 &    0.0 &        8.8 \\
    107 &   ATCDFS5 J0033210.92-274415.2 & 03:32:10.92 & -27:44:15.2 &       2052 &        13 &          0 & 0 &   0.00 &   0.00 &    0.0 &        8.5 \\
    108 &   ATCDFS5 J0033210.99-274053.8 & 03:32:10.99 & -27:40:53.8 &        183 &         9 &          0 & 0 &   0.00 &   0.00 &    0.0 &        9.2 \\
    109 &   ATCDFS5 J0033210.79-274628.1 & 03:32:10.79 & -27:46:28.1 &        111 &         9 &          0 & 0 &   0.00 &   0.00 &    0.0 &        8.6 \\
    110 &   ATCDFS5 J0033210.16-275938.4 & 03:32:10.16 & -27:59:38.4 &        154 &        15 &        183 & 38 &   1.87 &   0.48 &  -32.1 &        9.9 \\
    111 &   ATCDFS5 J0033209.81-275932.3 & 03:32:09.81 & -27:59:32.3 &         67 &        10 &          0 & 0 &   0.00 &   0.00 &    0.0 &        9.9 \\
    112 &   ATCDFS5 J0033209.71-274248.4 & 03:32:09.71 & -27:42:48.4 &        517 &        11 &          0 & 0 &   0.00 &   0.00 &    0.0 &        8.8 \\
    113 &   ATCDFS5 J0033208.67-274734.6 & 03:32:08.67 & -27:47:34.6 &       3533 &        36 &          0 & 0 &   0.00 &   0.00 &    0.0 &        9.7 \\
    114 &   ATCDFS5 J0033208.53-274649.0 & 03:32:08.53 & -27:46:49.0 &         63 &         7 &          0 & 0 &   0.00 &   0.00 &    0.0 &        9.2 \\
    115 &   ATCDFS5 J0033206.10-273235.7 & 03:32:06.10 & -27:32:35.7 &      13803 &       114 &          0 & 0 &   0.00 &   0.00 &    0.0 &       20.0 \\
    116 &   ATCDFS5 J0033204.68-280057.5 & 03:32:04.68 & -28:00:57.5 &         73 &        15 &          0 & 0 &   0.00 &   0.00 &    0.0 &        9.3 \\
    117 &   ATCDFS5 J0033204.31-280157.0 & 03:32:04.31 & -28:01:57.0 &         61 &        10 &          0 & 0 &   0.00 &   0.00 &    0.0 &        9.8 \\
    118 &   ATCDFS5 J0033203.88-275805.1 & 03:32:03.88 & -27:58:05.1 &        111 &         9 &          0 & 0 &   0.00 &   0.00 &    0.0 &        8.5 \\
    119 &   ATCDFS5 J0033203.67-274603.9 & 03:32:03.67 & -27:46:03.9 &         60 &         9 &          0 & 0 &   0.00 &   0.00 &    0.0 &        8.7 \\
    120 &   ATCDFS5 J0033202.84-275613.2 & 03:32:02.84 & -27:56:13.2 &         63 &         8 &          0 & 0 &   0.00 &   0.00 &    0.0 &        7.5 \\
   121A &   ATCDFS5 J0033201.56-274647.8 & 03:32:01.56 & -27:46:47.8 &       4910 &        178 &       6763 & 437 &   2.12 &   0.93 &   51.3 &       10.1 \\
   121B &   ATCDFS5 J0033201.28-274647.7 & 03:32:01.28 & -27:46:47.7 &       3576 &        192 &       4489 & 477 &   2.25 &   0.66 &   30.9 &       10.3 \\
    122 &   ATCDFS5 J0033200.84-273557.0 & 03:32:00.84 & -27:35:57.0 &       2417 &        22 &          0 & 0 &   0.00 &   0.00 &    0.0 &        9.5 \\
    123 &   ATCDFS5 J0033159.83-274540.7 & 03:31:59.83 & -27:45:40.7 &         81 &         9 &          0 & 0 &   0.00 &   0.00 &    0.0 &        8.3 \\
    124 &   ATCDFS5 J0033158.93-274359.4 & 03:31:58.93 & -27:43:59.4 &         51 &         7 &          0 & 0 &   0.00 &   0.00 &    0.0 &        8.4 \\
    125 &   ATCDFS5 J0033158.33-273747.9 & 03:31:58.33 & -27:37:47.9 &         49 &         6 &          0 & 0 &   0.00 &   0.00 &    0.0 &        8.8 \\
    126 &   ATCDFS5 J0033157.75-274208.9 & 03:31:57.75 & -27:42:08.9 &         54 &         6 &          0 & 0 &   0.00 &   0.00 &    0.0 &        8.0 \\
    127 &   ATCDFS5 J0033155.00-274410.7 & 03:31:55.00 & -27:44:10.7 &         75 &         9 &          0 & 0 &   0.00 &   0.00 &    0.0 &        8.8 \\
    128 &   ATCDFS5 J0033154.88-275341.0 & 03:31:54.88 & -27:53:41.0 &         51 &        10 &          0 & 0 &   0.00 &   0.00 &    0.0 &        8.0 \\
    129 &   ATCDFS5 J0033153.42-280221.3 & 03:31:53.42 & -28:02:21.3 &        665 &        12 &          0 & 0 &   0.00 &   0.00 &    0.0 &       10.5 \\
    130 &   ATCDFS5 J0033152.12-273926.5 & 03:31:52.12 & -27:39:26.5 &        558 &        12 &          0 & 0 &   0.00 &   0.00 &    0.0 &        8.7 \\
    131 &   ATCDFS5 J0033151.31-275056.0 & 03:31:51.31 & -27:50:56.0 &         52 &         8 &          0 & 0 &   0.00 &   0.00 &    0.0 &        8.0 \\
    132 &   ATCDFS5 J0033150.78-274703.9 & 03:31:50.78 & -27:47:03.9 &        110 &         6 &          0 & 0 &   0.00 &   0.00 &    0.0 &        8.4 \\
    133 &   ATCDFS5 J0033150.13-273948.3 & 03:31:50.13 & -27:39:48.3 &        243 &        18 &        333 & 83 &   3.32 &   1.03 &   10.3 &        9.6 \\
    134 &   ATCDFS5 J0033150.02-275806.3 & 03:31:50.02 & -27:58:06.3 &        173 &        10 &          0 & 0 &   0.00 &   0.00 &    0.0 &        8.6 \\
    135 &   ATCDFS5 J0033149.88-274839.0 & 03:31:49.88 & -27:48:39.0 &        850 &        35 &       1173 & 82 &   1.81 &   1.15 &   77.3 &        8.7 \\
    136 &   ATCDFS5 J0033148.74-273311.9 & 03:31:48.74 & -27:33:11.9 &         90 &        10 &          0 & 0 &   0.00 &   0.00 &    0.0 &       12.5 \\
    137 &   ATCDFS5 J0033147.38-274542.2 & 03:31:47.38 & -27:45:42.2 &        121 &         9 &        147 & 24 &   2.99 &   0.49 &   -7.0 &        8.6 \\
    138 &   ATCDFS5 J0033146.58-275734.6 & 03:31:46.58 & -27:57:34.6 &        155 &        17 &        188 & 47 &   2.53 &   0.53 &   19.3 &        8.3 \\
    139 &   ATCDFS5 J0033146.09-280026.5 & 03:31:46.09 & -28:00:26.5 &        186 &        10 &          0 & 0 &   0.00 &   0.00 &    0.0 &        8.7 \\
    140 &   ATCDFS5 J0033145.91-274539.1 & 03:31:45.91 & -27:45:39.1 &         55 &        10 &          0 & 0 &   0.00 &   0.00 &    0.0 &        9.3 \\

\hline

\end{tabular}
}
\label{tab:cat2}
\end{table*}

\setcounter{table}{1}
\begin{table*}
\caption{continued}  
{\scriptsize
\begin{tabular}{llllrrrrrrrr}  
\hline
ID & IAU name & RA & Dec & $S_{pnt}$ & d$S_{pnt}$ & $S_{int}$ & d$S_{int}$ & Decon& Decon & Decon & $\sigma_{local}$ \\  
     &                   &    (J2000)    &       (J2000)                 & ($\mu$Jy)  & ($\mu$Jy)  & ($\mu$Jy) & ($\mu$Jy) & Bmajor & Bminor & PA & \\ \hline

    141 &   ATCDFS5 J0033144.02-273836.2 & 03:31:44.02 & -27:38:36.2 &         79 &         9 &          0 & 0 &   0.00 &   0.00 &    0.0 &        8.2 \\
    142 &   ATCDFS5 J0033143.34-275102.6 & 03:31:43.34 & -27:51:02.6 &         54 &        12 &          0 & 0 &   0.00 &   0.00 &    0.0 &        8.4 \\
    143 &   ATCDFS5 J0033143.42-274248.7 & 03:31:43.42 & -27:42:48.7 &         38 &         7 &          0 & 0 &   0.00 &   0.00 &    0.0 &        7.9 \\
    144 &   ATCDFS5 J0033143.22-275405.5 & 03:31:43.22 & -27:54:05.5 &         52 &         5 &          0 & 0 &   0.00 &   0.00 &    0.0 &        8.5 \\
    145 &   ATCDFS5 J0033140.05-273648.1 & 03:31:40.05 & -27:36:48.1 &         91 &        16 &          0 & 0 &   0.00 &   0.00 &    0.0 &        9.2 \\
    146 &   ATCDFS5 J0033139.54-274120.1 & 03:31:39.54 & -27:41:20.1 &         71 &         9 &          0 & 0 &   0.00 &   0.00 &    0.0 &        8.4 \\
    147 &   ATCDFS5 J0033139.04-275259.1 & 03:31:39.04 & -27:52:59.1 &         53 &         7 &          0 & 0 &   0.00 &   0.00 &    0.0 &        8.6 \\
    148 &   ATCDFS5 J0033138.47-275942.1 & 03:31:38.47 & -27:59:42.1 &         71 &         8 &          0 & 0 &   0.00 &   0.00 &    0.0 &        8.7 \\
    149 &   ATCDFS5 J0033137.79-280533.6 & 03:31:37.79 & -28:05:33.6 &        109 &        17 &          0 & 0 &   0.00 &   0.00 &    0.0 &       17.2 \\
    150 &   ATCDFS5 J0033136.09-273940.8 & 03:31:36.09 & -27:39:40.8 &         58 &        10 &          0 & 0 &   0.00 &   0.00 &    0.0 &        8.5 \\
    151 &   ATCDFS5 J0033135.20-273508.9 & 03:31:35.20 & -27:35:08.9 &         53 &         6 &          0 & 0 &   0.00 &   0.00 &    0.0 &        9.3 \\
    152 &   ATCDFS5 J0033134.22-273828.7 & 03:31:34.22 & -27:38:28.7 &        268 &        16 &          0 & 0 &   0.00 &   0.00 &    0.0 &        9.0 \\
    153 &   ATCDFS5 J0033132.81-280116.2 & 03:31:32.81 & -28:01:16.2 &         58 &         9 &          0 & 0 &   0.00 &   0.00 &    0.0 &        9.5 \\
   154A &   ATCDFS5 J0033131.08-273815.8 & 03:31:31.08 & -27:38:15.8 &       1792 &        83 &       3133 & 234 &   2.63 &   1.72 &  -77.0 &       11.7 \\
   154B &   ATCDFS5 J0033130.01-273814.0 & 03:31:30.01 & -27:38:14.0 &        219 &        24 &        303 & 63 &   1.72 &   1.60 &  -55.1 &       13.0 \\
   154C &   ATCDFS5 J0033129.58-273802.9 & 03:31:29.58 & -27:38:02.9 &        200 &        13 &        450 & 62 &   4.24 &   2.64 &  -20.1 &       12.5 \\
    155 &   ATCDFS5 J0033130.74-275734.9 & 03:31:30.74 & -27:57:34.9 &        196 &         9 &          0 & 0 &   0.00 &   0.00 &    0.0 &        8.4 \\
   156A &   ATCDFS5 J0033130.38-275606.0 & 03:31:30.38 & -27:56:06.0 &         90 &        11 &        237 & 52 &   4.94 &   3.14 &    7.9 &       10.7 \\
   156B &   ATCDFS5 J0033130.05-275602.8 & 03:31:30.05 & -27:56:02.8 &        105 &        11 &        152 & 44 &   3.58 &   1.00 &  -15.1 &       10.9 \\
   156C &   ATCDFS5 J0033129.81-275559.7 & 03:31:29.81 & -27:55:59.7 &         65 &        11 &        167 & 98 &   5.47 &   2.81 &    0.0 &       10.8 \\
    157 &   ATCDFS5 J0033129.90-275722.7 & 03:31:29.90 & -27:57:22.7 &         56 &        10 &          0 & 0 &   0.00 &   0.00 &    0.0 &        8.8 \\
    158 &   ATCDFS5 J0033129.77-273218.4 & 03:31:29.77 & -27:32:18.4 &       1735 &        23 &          0 & 0 &   0.00 &   0.00 &    0.0 &       18.1 \\
    159 &   ATCDFS5 J0033128.59-274935.0 & 03:31:28.59 & -27:49:35.0 &        180 &         8 &          0 & 0 &   0.00 &   0.00 &    0.0 &        8.9 \\
    160 &   ATCDFS5 J0033127.57-274439.1 & 03:31:27.57 & -27:44:39.1 &         56 &        10 &          0 & 0 &   0.00 &   0.00 &    0.0 &        9.5 \\
    161 &   ATCDFS5 J0033127.20-274247.2 & 03:31:27.20 & -27:42:47.2 &        584 &        13 &        667 & 26 &   2.23 &   0.52 &   -8.2 &        8.6 \\
    162 &   ATCDFS5 J0033127.04-275958.2 & 03:31:27.04 & -27:59:58.2 &        135 &        13 &          0 & 0 &   0.00 &   0.00 &    0.0 &        9.7 \\
    163 &   ATCDFS5 J0033127.06-274409.7 & 03:31:27.06 & -27:44:09.7 &        173 &         8 &          0 & 0 &   0.00 &   0.00 &    0.0 &        9.3 \\
    164 &   ATCDFS5 J0033126.78-274237.1 & 03:31:26.78 & -27:42:37.1 &        108 &        13 &        137 & 37 &   2.94 &   0.90 &   -2.6 &        8.8 \\
    165 &   ATCDFS5 J0033125.27-275958.6 & 03:31:25.27 & -27:59:58.6 &         85 &         9 &          0 & 0 &   0.00 &   0.00 &    0.0 &        9.8 \\
    166 &   ATCDFS5 J0033124.90-275208.0 & 03:31:24.90 & -27:52:08.0 &       6454 &       205 &      12243 & 648 &   3.48 &   1.01 &   59.6 &       12.3 \\
    167 &   ATCDFS5 J0033124.63-280454.3 & 03:31:24.63 & -28:04:54.3 &        103 &        14 &          0 & 0 &   0.00 &   0.00 &    0.0 &       18.4 \\
    168 &   ATCDFS5 J0033123.30-274905.8 & 03:31:23.30 & -27:49:05.8 &        547 &        14 &        559 & 29 &   0.99 &   0.15 &    2.5 &        9.2 \\
    169 &   ATCDFS5 J0033123.07-275430.0 & 03:31:23.07 & -27:54:30.0 &         58 &        14 &          0 & 0 &   0.00 &   0.00 &    0.0 &       10.1 \\
    170 &   ATCDFS5 J0033122.21-275755.1 & 03:31:22.21 & -27:57:55.1 &         44 &         8 &          0 & 0 &   0.00 &   0.00 &    0.0 &        9.0 \\
    171 &   ATCDFS5 J0033121.85-275445.4 & 03:31:21.85 & -27:54:45.4 &         78 &        10 &          0 & 0 &   0.00 &   0.00 &    0.0 &        9.9 \\
    172 &   ATCDFS5 J0033120.21-280146.7 & 03:31:20.21 & -28:01:46.7 &         81 &        15 &        154 & 69 &   2.95 &   2.57 &    6.0 &       13.3 \\
    173 &   ATCDFS5 J0033120.15-273901.1 & 03:31:20.15 & -27:39:01.1 &        112 &        11 &          0 & 0 &   0.00 &   0.00 &    0.0 &       10.2 \\
    174 &   ATCDFS5 J0033119.90-273549.9 & 03:31:19.90 & -27:35:49.9 &         67 &        10 &          0 & 0 &   0.00 &   0.00 &    0.0 &       12.2 \\
    175 &   ATCDFS5 J0033118.73-274902.2 & 03:31:18.73 & -27:49:02.2 &        117 &        11 &          0 & 0 &   0.00 &   0.00 &    0.0 &       10.7 \\
    176 &   ATCDFS5 J0033117.34-280147.3 & 03:31:17.34 & -28:01:47.3 &        458 &        15 &        491 & 28 &   1.17 &   0.59 &    5.5 &       13.7 \\
   177A &   ATCDFS5 J0033117.04-275515.3 & 03:31:17.04 & -27:55:15.3 &        479 &        26 &        959 & 91 &   4.49 &   2.03 &  -14.0 &       12.0 \\
   177B &   ATCDFS5 J0033115.04-275518.7 & 03:31:15.04 & -27:55:18.7 &       1551 &        20 &          0 & 0 &   0.00 &   0.00 &    0.0 &       14.1 \\
   177C &   ATCDFS5 J0033114.36-275519.7 & 03:31:14.36 & -27:55:19.7 &        163 &        14 &        483 & 89 &   5.36 &   2.74 &   49.6 &       14.0 \\
   177D &   ATCDFS5 J0033113.93-275519.7 & 03:31:13.93 & -27:55:19.7 &        619 &        17 &       1857 & 149 &   5.42 &   1.99 &   68.2 &       13.8 \\
    178 &   ATCDFS5 J0033115.99-274443.1 & 03:31:15.99 & -27:44:43.1 &        357 &        17 &        412 & 42 &   2.21 &   0.67 &    2.4 &       12.0 \\
    179 &   ATCDFS5 J0033114.46-275546.6 & 03:31:14.46 & -27:55:46.6 &        117 &        18 &          0 & 0 &   0.00 &   0.00 &    0.0 &       12.7 \\
    180 &   ATCDFS5 J0033114.51-273906.6 & 03:31:14.51 & -27:39:06.6 &         71 &        14 &          0 & 0 &   0.00 &   0.00 &    0.0 &       12.1 \\
    181 &   ATCDFS5 J0033113.95-273910.4 & 03:31:13.95 & -27:39:10.4 &        533 &        24 &          0 & 0 &   0.00 &   0.00 &    0.0 &       12.7 \\
    182 &   ATCDFS5 J0033112.58-275717.9 & 03:31:12.58 & -27:57:17.9 &        235 &        10 &          0 & 0 &   0.00 &   0.00 &    0.0 &       12.0 \\
    183 &   ATCDFS5 J0033111.80-275817.3 & 03:31:11.80 & -27:58:17.3 &         69 &        17 &          0 & 0 &   0.00 &   0.00 &    0.0 &       13.6 \\
    184 &   ATCDFS5 J0033111.50-275258.5 & 03:31:11.50 & -27:52:58.5 &         91 &        16 &          0 & 0 &   0.00 &   0.00 &    0.0 &       14.5 \\
    185 &   ATCDFS5 J0033109.81-275225.3 & 03:31:09.81 & -27:52:25.3 &        652 &        21 &          0 & 0 &   0.00 &   0.00 &    0.0 &       15.6 \\
    186 &   ATCDFS5 J0033109.94-274915.8 & 03:31:09.94 & -27:49:15.8 &         68 &        14 &        144 & 25 &   6.02 &   1.84 &   -1.3 &       13.0 \\
    187 &   ATCDFS5 J0033109.18-274954.5 & 03:31:09.18 & -27:49:54.5 &        140 &        18 &          0 & 0 &   0.00 &   0.00 &    0.0 &       14.3 \\
    188 &   ATCDFS5 J0033107.97-275047.6 & 03:31:07.97 & -27:50:47.6 &         78 &        15 &          0 & 0 &   0.00 &   0.00 &    0.0 &       15.2 \\
    189 &   ATCDFS5 J0033106.15-273837.7 & 03:31:06.15 & -27:38:37.7 &        142 &        14 &          0 & 0 &   0.00 &   0.00 &    0.0 &       16.8 \\

\hline

\end{tabular}
}
\label{tab:cat3}
\end{table*}

\subsection{Flux comparison with Data Release 1 and VLA survey}

\begin{figure*}
\includegraphics[width=0.9\columnwidth]{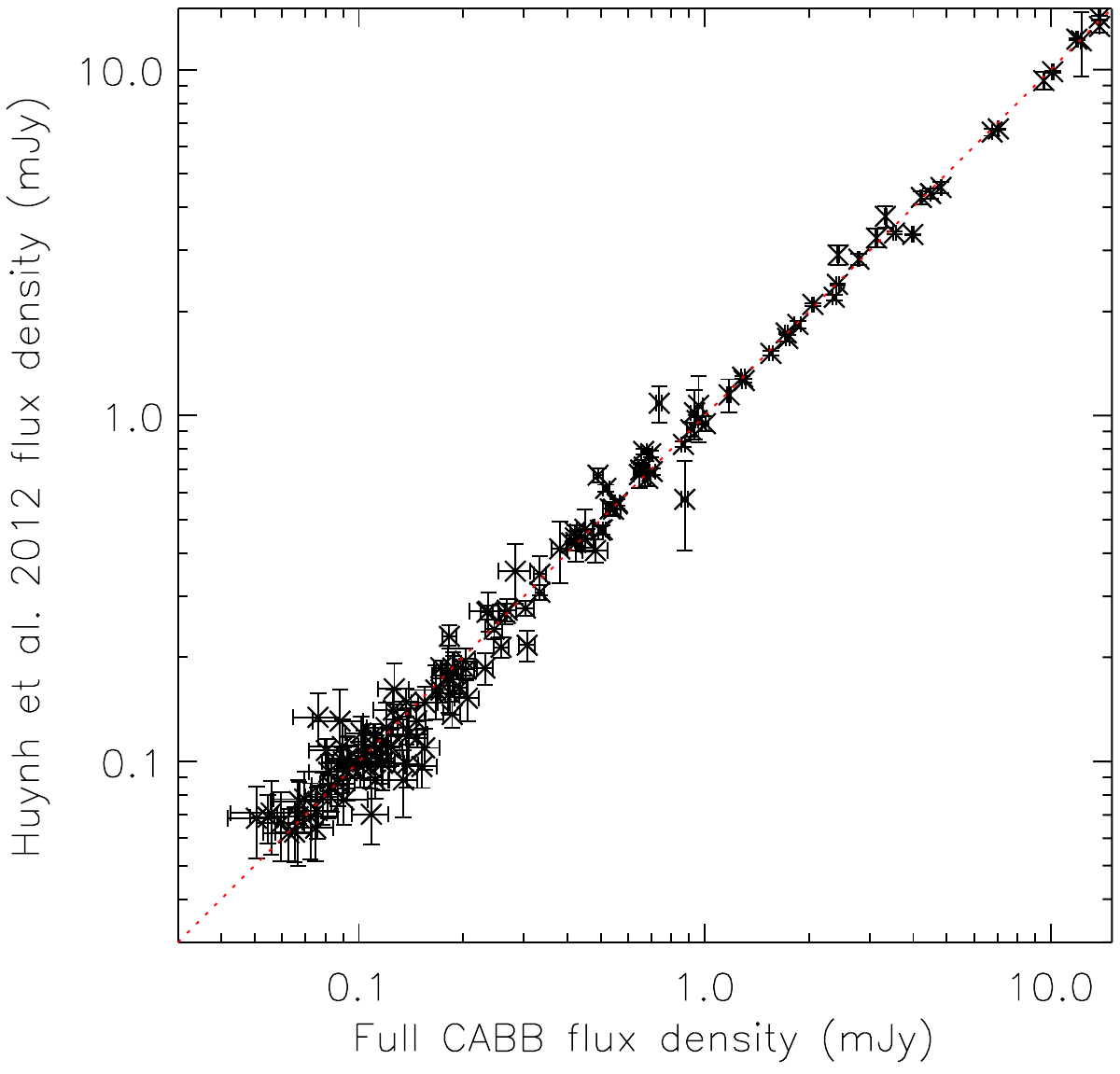}
\hspace{5mm}
\includegraphics[width=0.9\columnwidth]{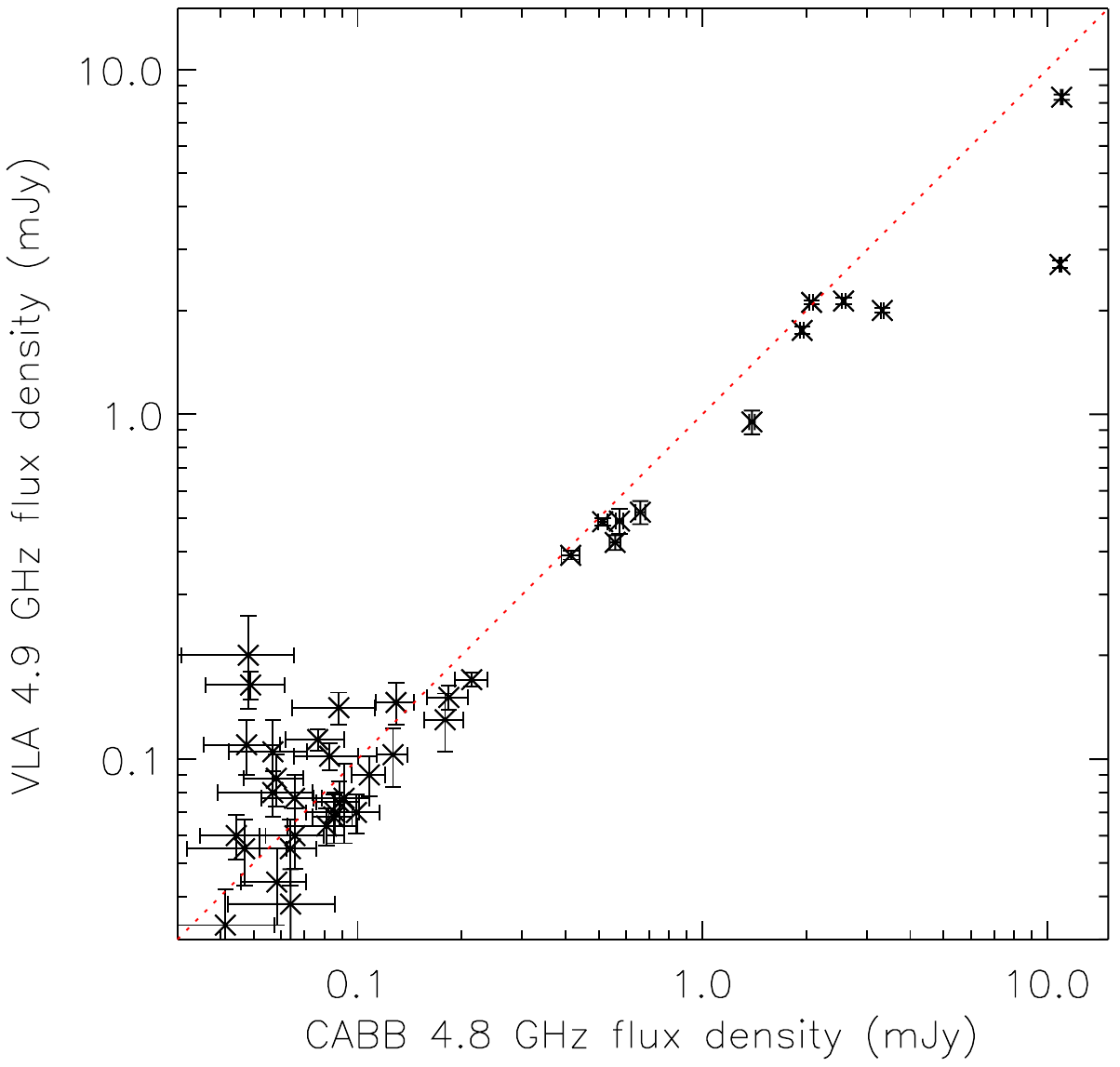}
\caption{LEFT: Comparison of the flux densities in this data release using the full band reduction (i.e. flux densities from Table 2) compared to Data Release 1 (H12). The sources lie close to the dotted line, which shows that flux densities in this data release are consistent with those measured in Data Release 1 (H12). RIGHT: The 4.8 GHz sub-band flux density versus the 4.9 GHz VLA flux density, for sources with a VLA measurement. }
\label{fig:fluxcomp}
\end{figure*}

Transients and sources that are variable on timescales of months and years are discussed in a separate paper \citep{bell2015}, but as a consistency check we compared the flux densities of the sources in this release with Data Release 1 (H12) flux densities. The flux densities for sources detected in both data releases are shown in Figure \ref{fig:fluxcomp}. We find no significant difference in the average flux densities of sources between the data releases. The ratio of DR2 (this work) to DR1 (H12) flux densities has a mean of $1.02 \pm 0.01$ and median of 1.01.

We also compare our flux densities with that from the VLA. Four VLA pointings were used to cover a region of approximately 20 $\times$ 20 arcmin in the eCDFS at 4.9 GHz. The sensitivity of the VLA observations ranged from 7 $\mu$Jy/beam rms at the pointing centers to 50 $\mu$Jy/beam rms at the edges \citep{kellermann2008}.
The resolution of the VLA 4.9 GHz image is about 3.5 arcsec, which is similar to the synthesized beam of our ATCA imaging, but to minimise resolution effects we compared the single component sources only. We compared the VLA 4.9 GHz flux densities with our 4.8 GHz sub-band flux densities to minimise spectral index effects (Figure \ref{fig:fluxcomp}).  
We find ATCA/VLA flux density ratio has a mean of 1.13 $\pm$ 0.09 and median of 1.09. For a spectral index of $\alpha = -0.8$ ($S \propto \nu^\alpha$) we expect the ATCA flux densities to be about a few percent greater than the VLA measurements, if the VLA and ATCA are calibrated on the same scale. The ATCA flux densities therefore appear to be $\sim$10\% greater than VLA flux densities for this frequency, which is generally consistent with our earlier estimate of ATCA flux densities being $\sim$20\% greater (H12). Our earlier estimate included faint ($<3\sigma$) VLA 6cm sources which are excluded in this analysis. 
The ATCA and VLA flux density scales both claim to be tied to within a few percent of the \cite{Baars1977} scale, so the source of this discrepancy is unclear. 

\subsection{Completeness and Flux Boosting}
\label{sec:sims}

\begin{figure*}
\includegraphics[width=0.9\columnwidth]{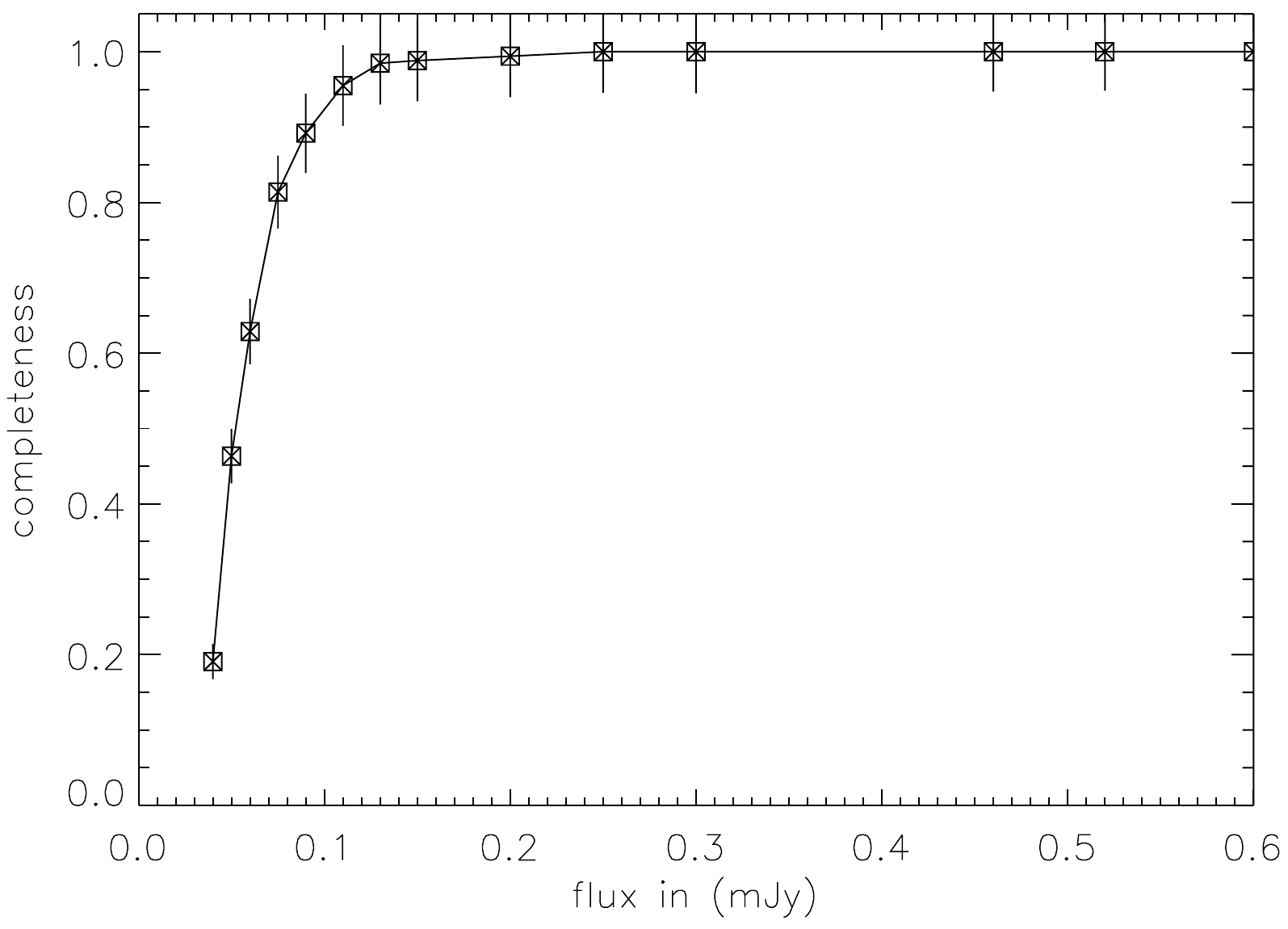}
\hspace{5mm}
\includegraphics[width=0.9\columnwidth]{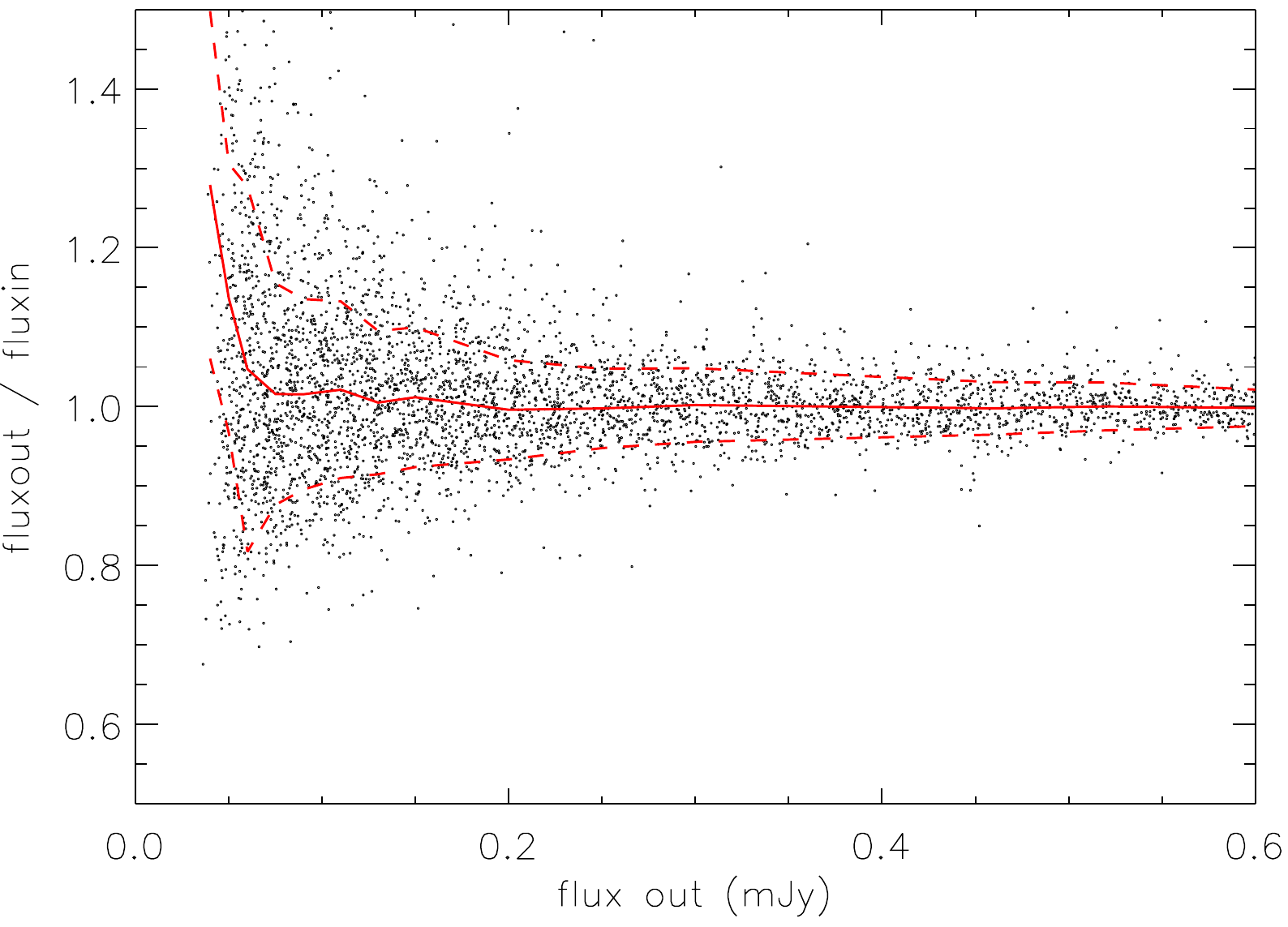}
\caption{LEFT: Completeness as a function of input flux density, as derived from the Monte-Carlo simulations. Completeness is the number of extracted sources divided by number of input sources. RIGHT: The distribution of output/input flux density as a function of output flux density for the simulated sources. The solid red line is the median of the simulation and the dashed lines mark the 1 sigma upper and lower bounds. The effect of flux boosting at the faint end is dramatically illustrated by the rapid upturn below about 0.075 mJy.}
\label{fig:comp-boost}
\end{figure*}

As in H12, we performed Monte-Carlo simulations to estimate the completeness of the source catalogue.
Artificial point sources were injected on to random locations of the mosaic and then extracted using the same method that produced the catalogue. 
Although the hexagonal mosaicing pattern results in fairly uniform noise across most of the image, the edges of the mosaic have increased noise levels due to the primary beam response and therefore lower completeness. 
We recovered the overall completeness level of the generated catalogue by injecting sources over the full area of the mosaic from which sources are extracted for the catalogue. 
We injected 8000 artificial sources for reliable statistics, and injected a single source at a time, to avoid confusion effects. 
 The input flux density varied from 20 to 2000 $\mu$Jy to sample the full range of interest. The completeness
as a function of flux density is shown in Figure \ref{fig:comp-boost}.
The completeness rises steeply from about 20\% at 40 $\mu$Jy to approximately 90\% at 100 $\mu$Jy 
 The 50\% completeness level occurs at approximately 52 $\mu$Jy (cf. the 50\% completeness level of 75 $\mu$Jy for Data Release 1; H12). 

Sources that lie on a noise peak have increased flux densities and therefore have a higher probability
of being detected, while sources which lie on a noise trough have decreased flux densities and may be excluded altogether.
This can lead to a flux boosting of sources, and this effect is strongest in the faintest flux density bins.
 The degree of flux boosting can be estimated from the ratio of output to input flux density of
the simulations (Figure \ref{fig:comp-boost}). In the faintest bins we find that flux densities are
boosted by about 14\% at 50 $\mu$Jy and 28\% at 40 $\mu$Jy, on average. The flux boosting is negligible for sources with flux densities brighter than about 75 $\mu$Jy.

\begin{figure*}
\includegraphics[width=0.85\columnwidth]{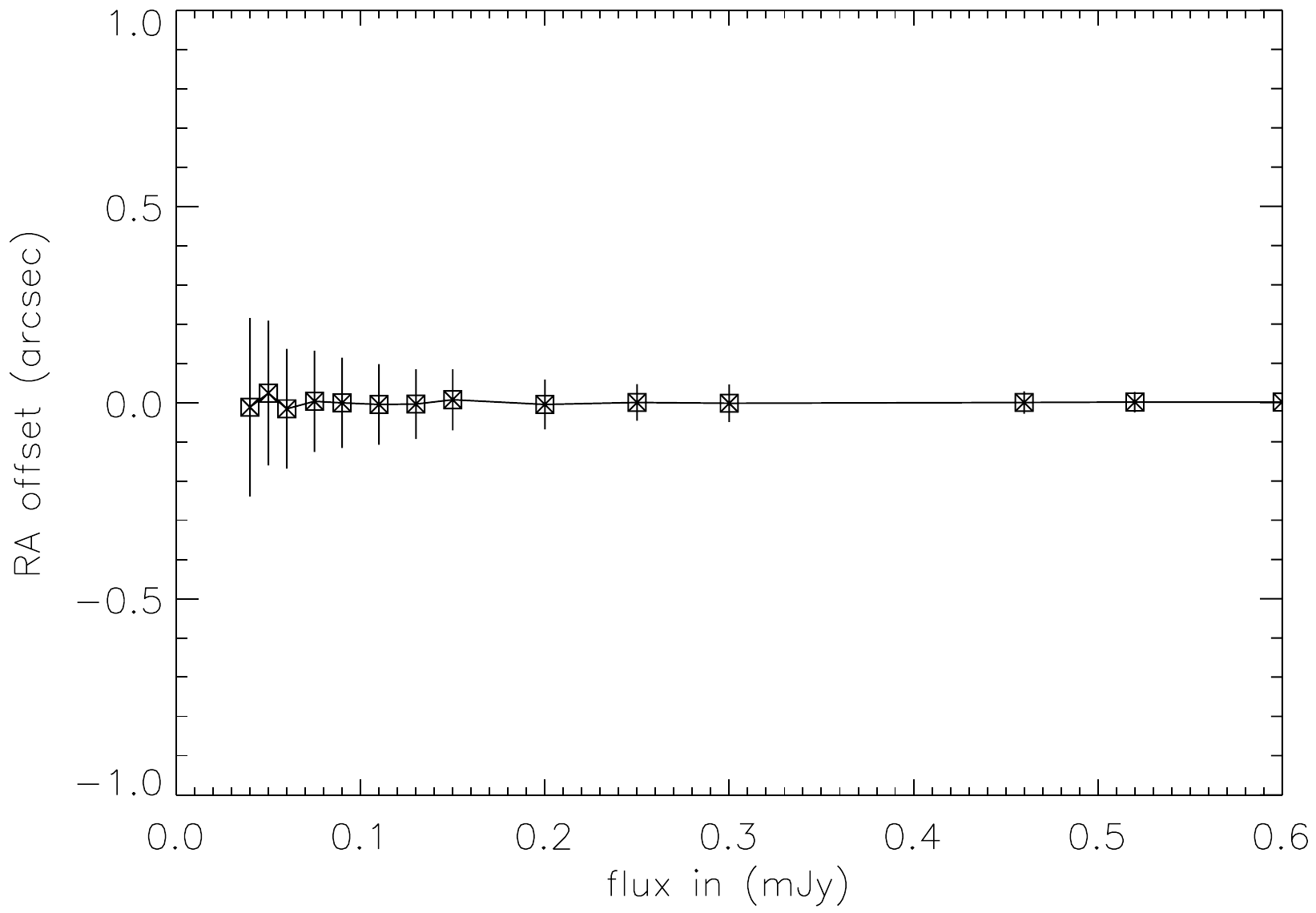}
\hspace{15mm}
\includegraphics[width=0.85\columnwidth]{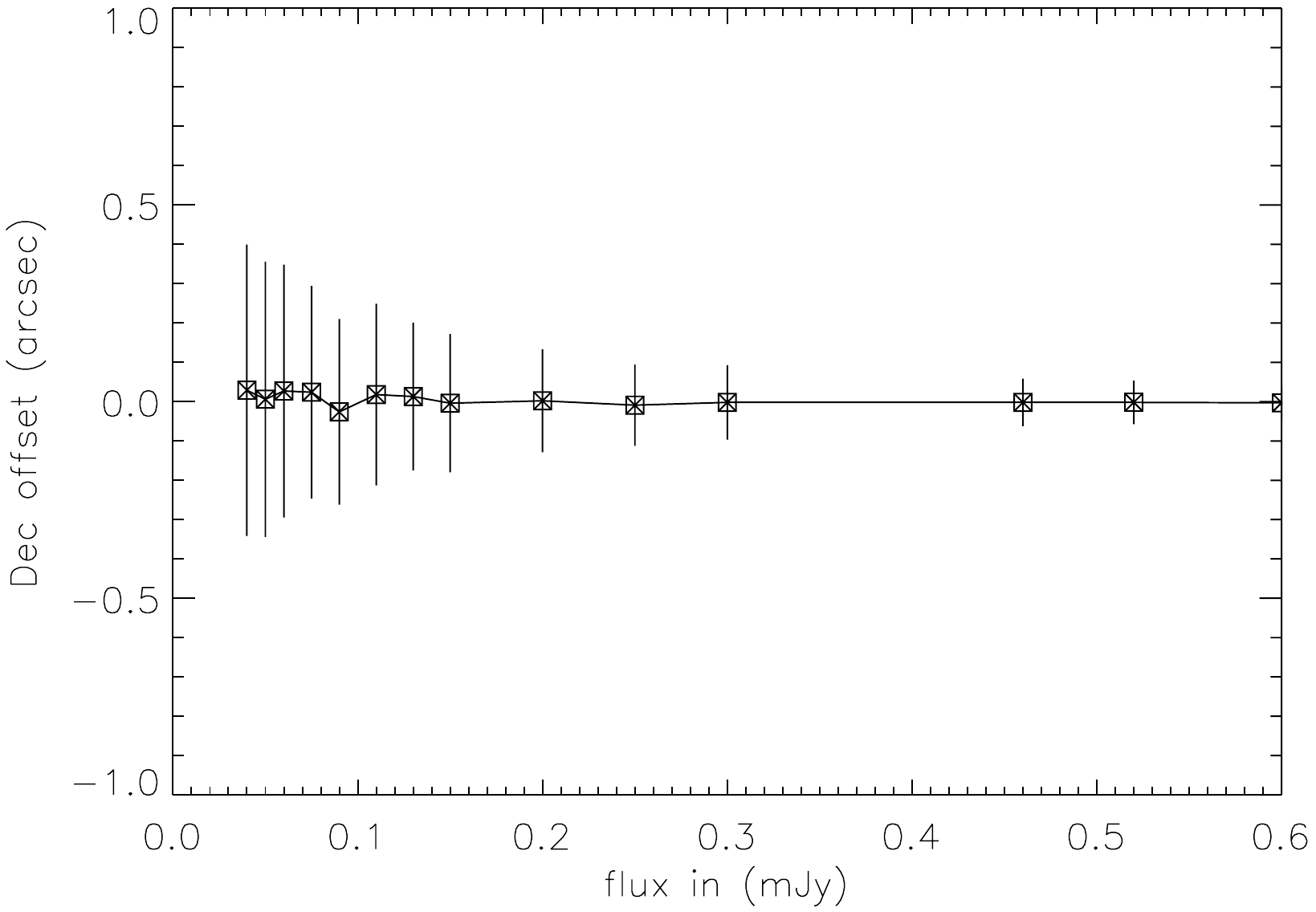}
\caption{LEFT: The offset in RA between the recovered positions of sources in the simulation and the true input positions, as a function of input flux density. The error bars mark the 1 sigma uncertainty in the position as 
a function of input flux density. RIGHT: Same as the left panel, but for offset in Dec.}
\label{fig:delradeldec}
\end{figure*}

Estimates of the positional accuracy of the catalogue can be made by comparing input and output positions.
The median of the RA and Dec offsets as a function of input flux density is shown in Figure \ref{fig:delradeldec}. The positional
accuracy can be estimated from the standard deviation in
the offsets. We find that at the faintest
levels (40 $\mu$Jy) the RA and Dec uncertainties are approximately
0.2 and 0.4 arcsec, respectively. The total positional accuracy is $\sim$0.25 arcsec or better for sources that are brighter than 0.1 mJy.

\subsection{Source Size and Resolution Bias}

Weak and extended radio sources may have peak flux densities that fall below the detection threshold, leading to so-called resolution bias.  
To derive source counts which are complete in terms of total flux density the resolution bias must be determined. 
As in H12 we follow the formalism of  \cite{prandoni2001}  and \cite{huynh2005} in calculating the resolution bias. 

In brief, the maximum size ($\theta_{\rm max}$) a source of total flux density $S_{\rm tot}$ can have is $S_{\rm tot}/\sigma_{\rm det} = \theta_{\rm max}^2 / b_{\rm min} b_{\rm max}\;$, where $b_{\rm min}$ and $b_{\rm max}$ are the synthesized beam FWHM axes and $\sigma_{\rm det}$ is the detection limit. Since the {\em sfind} detection limit varies across the image, we take the 50\% completeness level (52 $\mu$Jy), as determined by the simulations of Section \ref{sec:sims}, to be $\sigma_{\rm det}$.  The minimum angular size ($\theta_{\rm min}$) is estimated from Equation 1, with $\sigma$ equal to the typical noise of the full image (8.4 $\mu$Jy). 

\begin{figure*}
\includegraphics[width=0.85\columnwidth]{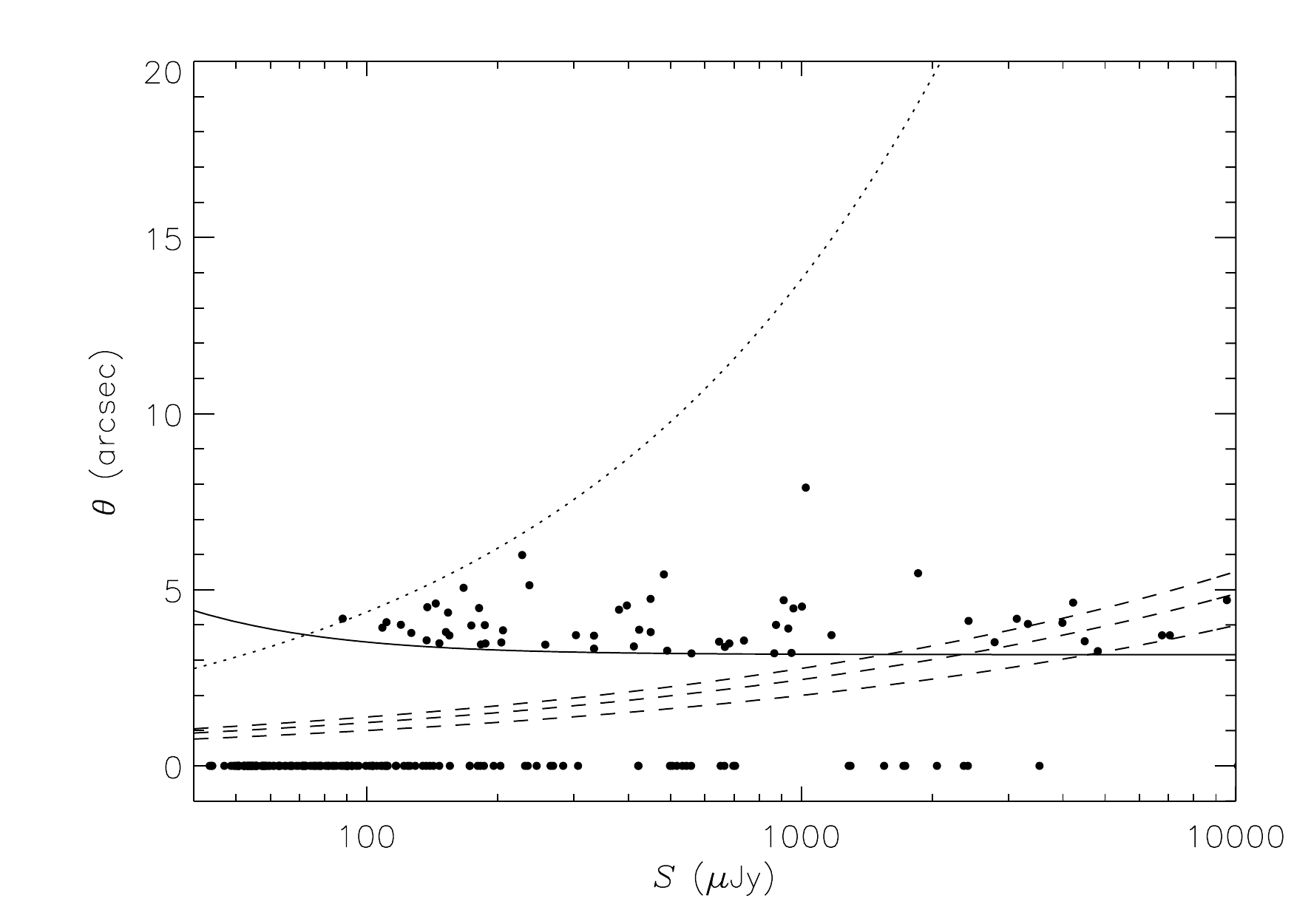}
\hspace{15mm}
\includegraphics[width=0.85\columnwidth]{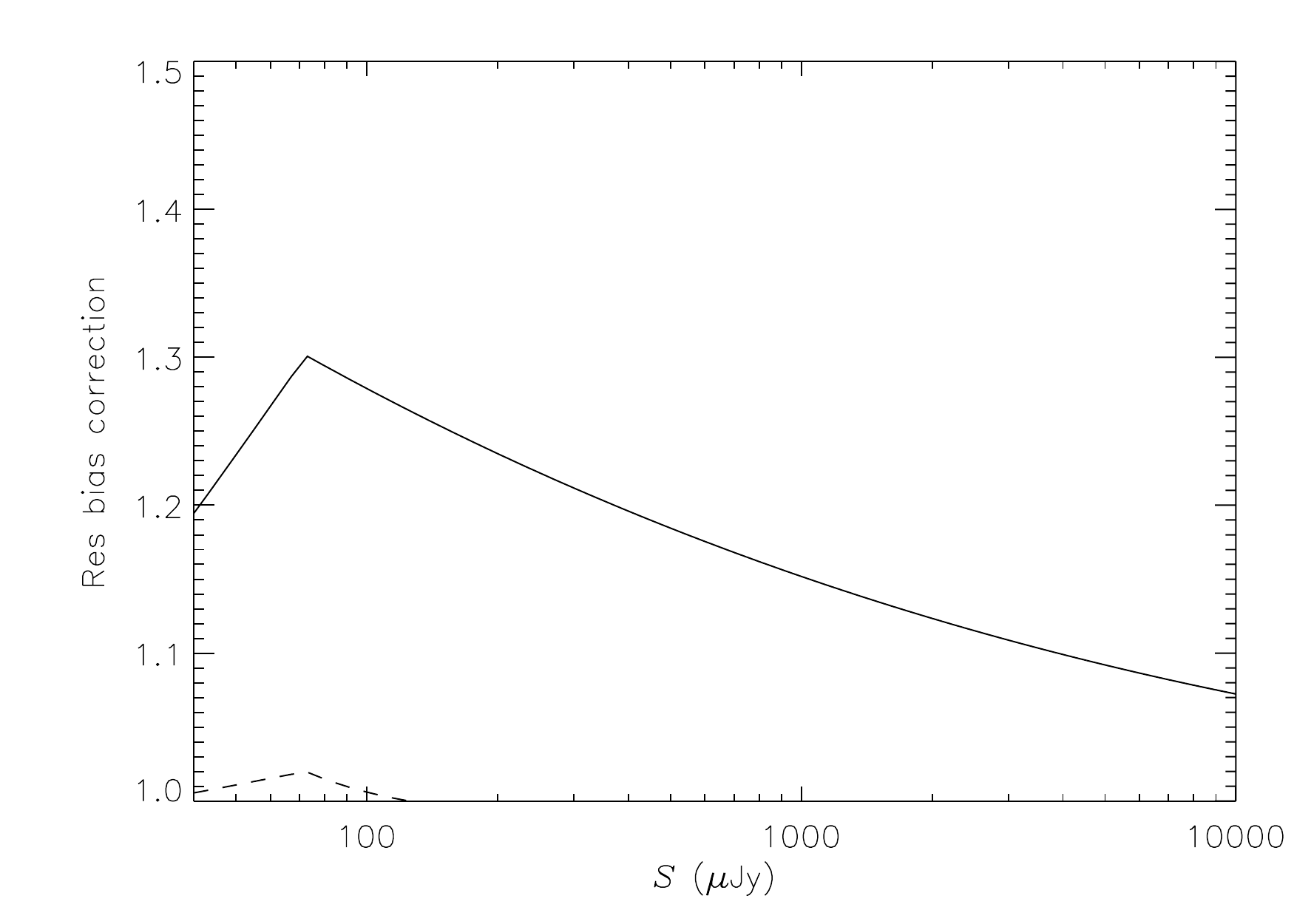} 
\caption{LEFT: The fitted angular size as a function of total flux density. Point sources are shown with an angular size of zero. The solid line indicates the minimum angular size ($\theta_{\rm min}$) of sources in the survey, below which deconvolution is not considered meaningful. The dotted line shows the maximum angular size ($\theta_{\rm max}$) above which the survey becomes incomplete due to resolution bias. The dashed lines indicate the median source sizes expected from the Windhorst et al. (1990) relation, as a function of flux density, for a spectral index of 0, $-$0.5 and $-$0.8 between 1.4 and 5.5 GHz.  RIGHT: The resolution bias correction as a function of flux density, assuming the Windhorst et al. (1990) integral angular source size distribution (solid line) and assuming the Muxlow et al. (2005) size distribution (dashed line).}
\label{fig:rescorr}
\end{figure*}

The angular sizes ($\theta$) of the catalogued sources as a function of total flux density is shown in Figure \ref{fig:rescorr}, where the angular size $\theta$ is defined as the geometric mean of the fitted Gaussian major and minor axes. We find that the largest catalogued sources are in good agreement with the $\theta_{\rm max}$ function. The $\theta_{\rm min}$ constraint is important at low flux density levels, where $\theta_{\rm max}$ becomes unphysical (smaller than a point source). Also shown in Figure \ref{fig:rescorr} (dashed lines) is the expected median angular size obtained from \cite{windhorst1990} for a 1.4 GHz sample, $\theta_{\rm med} = 2\arcsec S_{\rm 1.4 GHz}^{0.30}$, where $S_{\rm 1.4 GHz}$ is in mJy. We extrapolated to 5.5 GHz assuming a spectral index of 0, -0.5 and -0.8 between 1.4 and 5.5 GHz. At the bright end ($S >$ 2 mJy) our source sizes are consistent with the \cite{windhorst1990} relation, however most of the sources are unresolved and therefore we can not draw any conclusions about the full sample. 

The overall angular size limit, $\theta_{\rm lim} = {\rm max}(\theta_{\rm max},\theta_{\rm min})$, and an expected integral size distribution, $h(\theta)$, allows an estimation of the fraction of sources larger than the maximum detectable size, and hence missed by the survey. The resolution bias correction factor is then simply $\frac{1} {1 - h(\theta)}\;.$ The correction factor for \cite{windhorst1990} and \cite{muxlow2005} integral size distributions are shown in Figure \ref{fig:rescorr}.
The resolution bias correction for the \cite{windhorst1990} integral size distribution has a maximum of about 1.3 at a flux density of 70--80 $\mu$Jy. The \cite{windhorst1990} integral size distribution is commonly used to determine resolution bias (e.g. \citealp{prandoni2001}) so we include it in our source count derivation, but we note it is derived from a brighter sample than our work ($S_{\rm 1.4 GHz} > 0.4$ mJy). The \cite{muxlow2005} sample goes to sub-100-$\mu$Jy levels, but it comes from high resolution MERLIN and VLA imaging which may miss low surface brightness galaxies. We note that the resolution bias is negligible if the \cite{muxlow2005} size distribution is assumed.

\section{Source Counts}

The differential radio source counts were constructed from the catalogue of Section 3. Integrated flux densities were used for resolved sources, and components of multiple sources were summed and counted as a single source. The results are summarised in Table \ref{tab:srccount}, where for each bin we report the flux density interval, mean flux density, the number of sources detected ($N$), the number of sources after completeness, flux boosting and resolution bias corrections have been applied ($N_C$), the differential source count  ($dN_{C}/dS$), and the normalised counts ($N_C/N_{\rm exp}$). The counts are normalised to  $N_{\rm exp}$, the number expected in the bin from the standard Euclidean count, for comparison with counts in the literature. 
At 6cm the standard Euclidean integral counts are $N(> S_{\rm 6cm}) = 60 \times S_{\rm 6cm}^{-1.5} \; {\rm sr}^{-1}$, where $S_{\rm 6cm}$ is in Jy \citep{donnelly1987, fomalont1991,ciliegi2003}.
The Poissonian errors in the count are  $C N^{1/2}/N_{\rm exp}$, where $C$ is the total correction factor, $N_C/N$. The estimated total uncertainty in the counts is the Poissonian error with the resolution bias uncertainty (10\%), the flux boosting uncertainty (5 -- 20 \%),  and completeness correction uncertainty (2 -- 4 \%) all added in quadrature.

\begin{figure*}
\includegraphics[width=0.99\columnwidth]{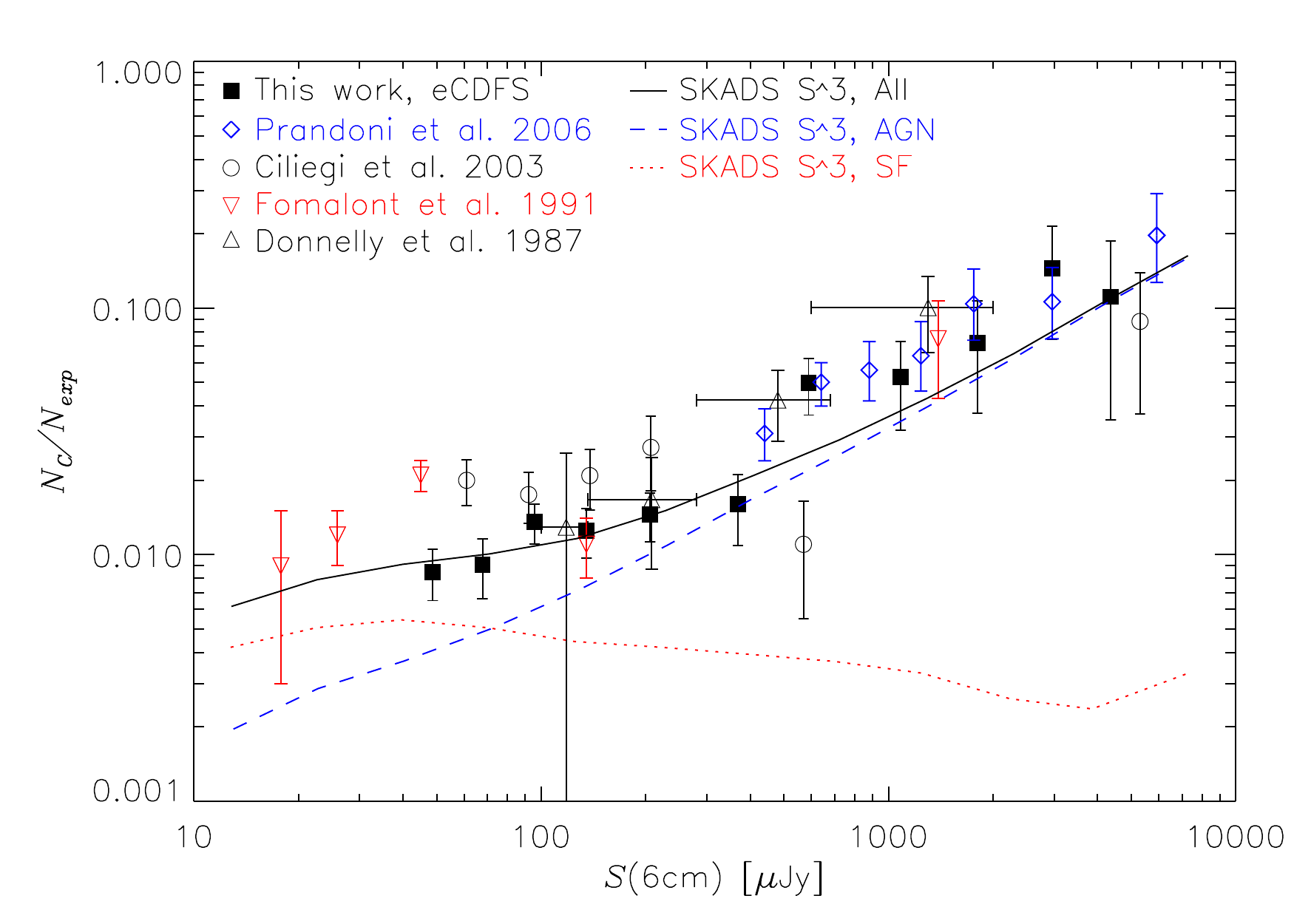}
\hspace{5mm}
\includegraphics[width=0.99\columnwidth]{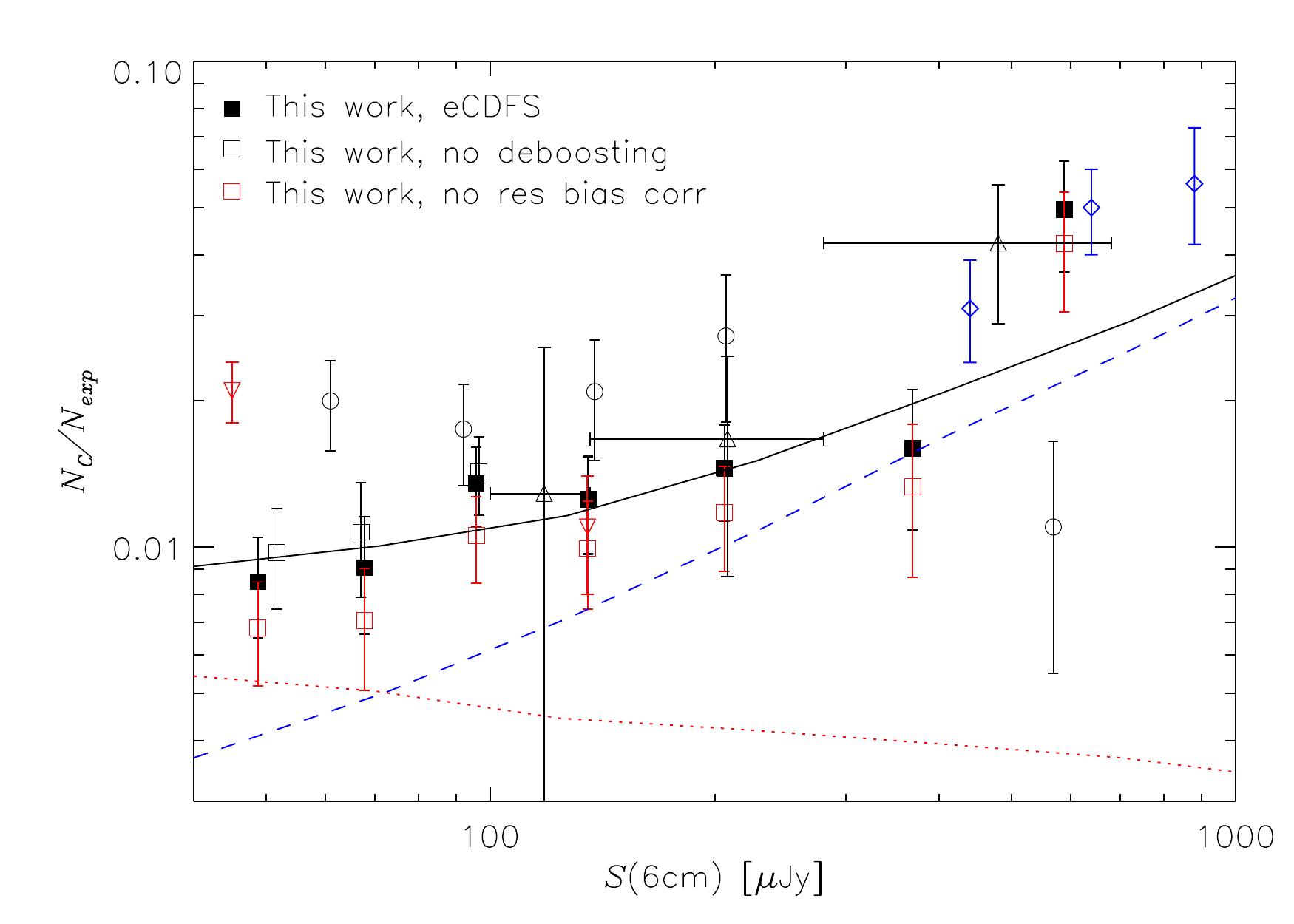} 
\caption{LEFT: Normalized 5.5 GHz differential source counts for different samples: Prandoni et al. (2006) (blue diamonds); Ciliegi et al. (2003) (empty circles); Fomalont et al. (1991) (red upside-down triangles); Donnelly et al. (1987) (empty triangles). The eCDFS 5.5 GHz source counts presented in this work (filled squares) are corrected for completeness, flux boosting and resolution bias as explained in the text. Vertical bars represent Poisson errors on the normalized counts. RIGHT:  A zoom of the source counts in the 0.040 to 1 mJy flux density range, showing the effect of the different corrections. Black empty squares are the counts without the flux boosting correction. Red empty squares are counts without resolution bias correction. Other symbols are as for the left figure.}
\label{fig:sourcecount}
\end{figure*}

Our results are compared with previous work in Figure \ref{fig:sourcecount}. 
Our source counts are consistent with the ATESP 6cm source counts \citep{prandoni2006} for $S_{\rm 6cm} > 0.4$ mJy. 
At fainter flux densities our counts hint at a flattening of the differential source counts, with a slope of $\alpha = 0.32 \pm 0.19$ for $\log(N_C/N_{\rm exp}) \propto \log(S_{\rm 6cm})^\alpha$ at $S_{\rm 6cm} < 0.4$ mJy compared to $\alpha = 0.51 \pm 0.35$ at $S_{\rm 6cm} > 0.4$ mJy, but the difference in $\alpha$ is not statistically significant. 
Our source counts are lower than the \cite{ciliegi2003} counts but consistent within the uncertainties, except for the faintest two bins.  
The counts in our faintest bins ($40 < S < 80$ $\mu$Jy)  are about a factor of two lower than the \cite{ciliegi2003} and \cite{fomalont1991} counts at similar flux densities. 
\cite{fomalont1991} catalogued sources to about 4$\sigma$ in their image, so it is likely that they have spurious sources in their faintest bins. We note that our survey area is 4 times greater than \cite{ciliegi2003} (0.34 deg$^2$ versus 0.087 deg$^2$) and 7 times greater than \cite{fomalont1991} (0.34 deg$^2$ versus 0.05 deg$^2$). Most of the difference in the counts at the faint end can be attributed to cosmic variance and clustering (e.g. H12, \citealp{heywood2013}). The 6cm surveys in the literature have a central frequency of 5 GHz and the difference of 0.5 GHz in the central observing frequency may have an impact on the source count comparison. 
If a spectral index of $-0.8$ is applied to convert our 5.5 GHz flux densities to 5 GHz ones than the source counts change by at most 6\% for $S_{\rm 6cm} < 0.1$ mJy, and hence the different central frequency does not account for the difference in our source counts compared to previous 6cm surveys in the faintest bins.

We also compare the observed source counts to the simulations of \cite{wilman2008,wilman2010} in Figure \ref{fig:sourcecount}. Briefly, this is a semi-empirical extragalactic simulation which uses observed and extrapolated luminosity functions of various radio populations (radio loud AGN split into FRI and FRII classes, radio quiet AGN, `normal' starforming galaxies and starbursts) and places them on top of a dark matter density field with biases to reflect their observed large-scale clustering properties. This simulation (known as SKADS S-cubed\footnote{http://s-cubed.physics.ox.ac.uk}) covers a sky area of 20 $\times$ 20 degrees and comprises 320 million sources to the flux density limit of 10 nJy.  
In general the modelled 4.86 GHz  source counts are in good agreement with the observed counts, and this is remarkable given the level of complexity in the simulation.
However the model counts in the $\sim$0.5--2 mJy flux density range underestimate the number of observed sources by 0.2 to 0.3 dex. FRIs dominate the model count at this flux density level so it is possible that either the FRIs are modelled incorrectly, i.e. jet Lorentz factors or radio lobe ratios used in the models are not correct for sources in this flux density range, or a population is missing from the simulations. A flat spectrum population  detected at higher frequencies ($>$ 10 GHz) but missing from 1.4 GHz surveys was recently identified \citep{whittam2013,franzen2014} and this small excess in the 5.5 GHz counts at $\sim$mJy levels is consistent with this discovery. 

\begin{table}
\centering
\caption{The 5.5 GHz source counts.}
\begin{tabular}{ccrrcc} \hline
$\Delta S$ & $<$S$>$ & $N$ & $N_{C}$ & $dN_{C}/dS$ & $N_{C}/N_{exp}$ \\
($\mu$Jy) & ($\mu$Jy) & & &  (sr$^{-1}$ Jy$^{-1}$) &  ($\times 10^{-2}$)\\ \hline
40 -- 57 & 49 & 33 & 78.8 & $4.61  \times 10^{10}$ & 0.85 $\pm$ 0.20 \\
57 -- 80 & 68 & 30 & 52.3 & $2.16  \times 10^{10}$ & 0.91 $\pm$ 0.25 \\
80 -- 113 & 96 & 33 & 46.3 & $1.36  \times 10^{10}$ & 1.35 $\pm$ 0.25 \\
113 -- 159 & 135 & 20 & 25.6 & $5.31  \times 10^9$ & 1.25 $\pm$ 0.29 \\
159 -- 270 & 206 & 20 & 24.8 & $2.15  \times 10^9$ & 1.46 $\pm$ 0.33 \\
270 -- 459 & 368 & 9 & 10.8 & $5.52  \times 10^8$ & 1.60 $\pm$  0.51 \\
459 -- 780 & 589 & 15 & 17.7 & $5.31 \times 10^8$ & 4.96 $\pm$  1.28 \\
780 -- 1325 & 1084 & 6 & 6.9 & $1.22 \times 10^8$ & 5.25 $\pm$ 2.07 \\
1325 -- 2249 & 1802 & 4 & 4.5 & $4.70 \times 10^7$ & 7.20 $\pm$ 3.47 \\
2249 -- 3820 & 2960 & 4 & 4.4 & $2.73 \times 10^7$ & 14.45 $\pm$ 7.01 \\ 
3820 -- 6487 & 4367 & 2 & 2.2 & $7.93 \times 10^6$ & 11.11 $\pm$ 7.58 \\ \hline
\end{tabular}
\label{tab:srccount}
\end{table}

\section{Radio Spectral Energy Distributions}

\subsection{1.4 to 5.5 GHz Spectral Indices}

\label{sec:spectralindex}

To study the spectral index properties of the faint radio population we matched the 5.5 GHz catalogue to sources in the second data release of the VLA 1.4 GHz survey of the eCDFS \citep{miller2013}. This improves on the initial VLA data with a 0.5 $\mu$Jy/beam rms noise reduction across the VLA mosaic to typical values of 7.4 $\mu$Jy/beam rms (i.e. $\sim$7\%) improvement), and a deeper source catalogue detection limit of 5$\sigma$  versus 7$\sigma$ in the initial release. VLA imaging of the eCDFS has similar coverage to our observations, roughly 34\arcmin $\times$ 34\arcmin. Importantly, the beam of the VLA observations is 2.8$\arcsec$ $\times$ 1.6$\arcsec$ beam, which is only a factor of $\sim$1.5 smaller than our observations. With similar resolution and sensitivity these images have a similar surface brightness sensitivity, and thus the measured flux densities can be used directly for spectral index analyses. 

Multi-component sources were removed from the spectral index analysis as their interpretation is complicated by the core-jet structure, resulting in 177 individual 5.5 GHz sources for investigation. 163/177 (92\%) of the 5.5 GHz sources have a 1.4 GHz match within 2 arcsec (FWHM of the synthesised beam of the VLA observations). The unmatched 5.5 GHz sources were inspected and four had counterparts in the 1.4 GHz image but weren't in the Miller et al. (2013) catalogue. The 1.4 GHz flux density for these sources was measured manually with the MIRIAD task {\em imfit}. In summary 167/177 (94\%) of the 5.5 GHz sources in the 1.4 GHz image area have a 1.4 GHz counterpart, and hence a spectral index measurement (Table \ref{tab:alpha}).  Of the remaining 10 sources, 2 show multi-component source structure in the VLA image and 7 are faint at 5.5 GHz or located at the higher noise edges of the mosaic, indicating they are possibly spurious sources. 

\begin{figure}
\includegraphics[width=0.95\columnwidth]{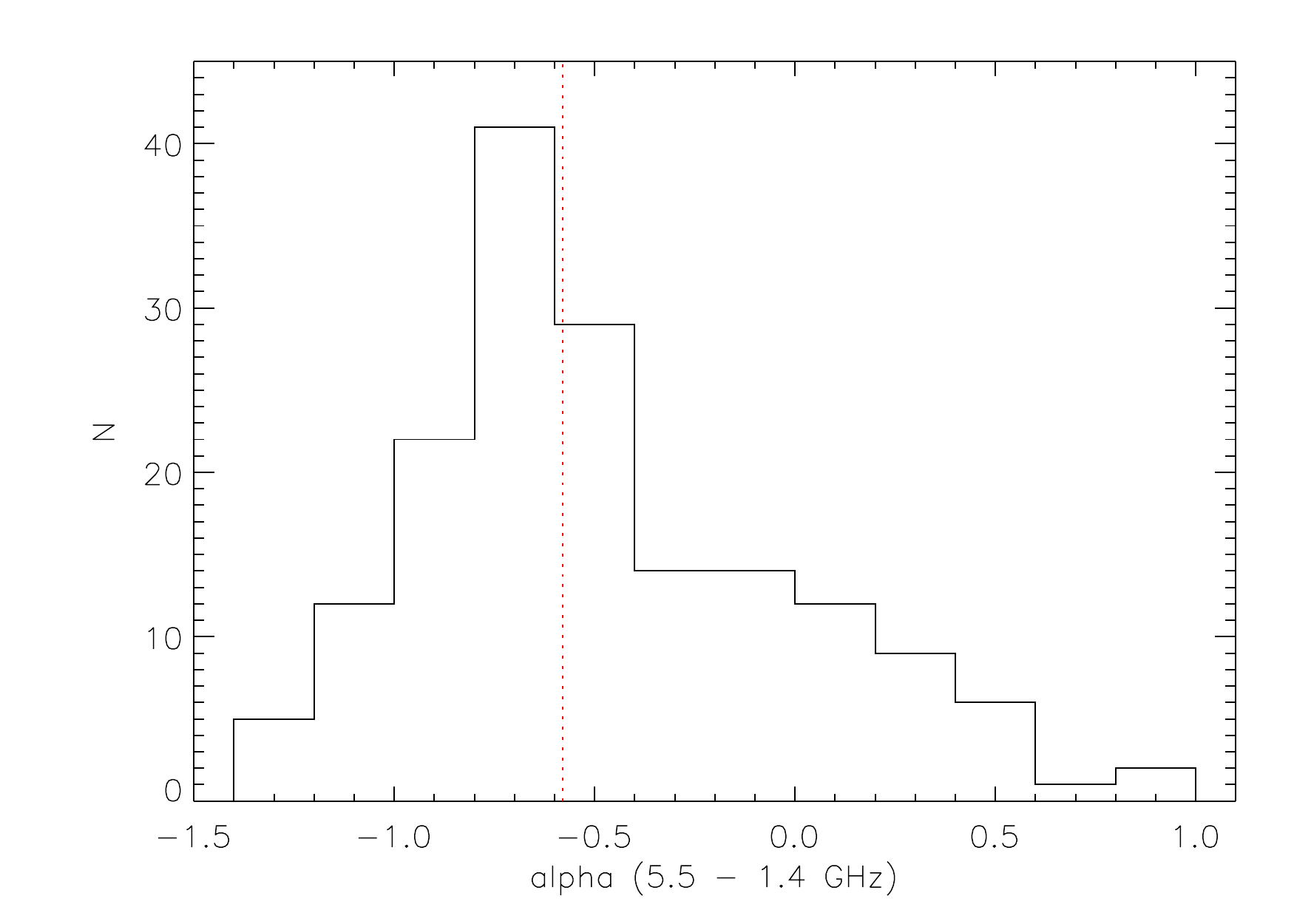}
\caption{Spectral index distribution for sources in the ATCA 6cm sample. The vertical dotted line indicates the median value of the sample ($\alpha_{med} = -0.58$).}
\label{fig:alphahist}
\end{figure}

\begin{figure}
\includegraphics[width=0.95\columnwidth]{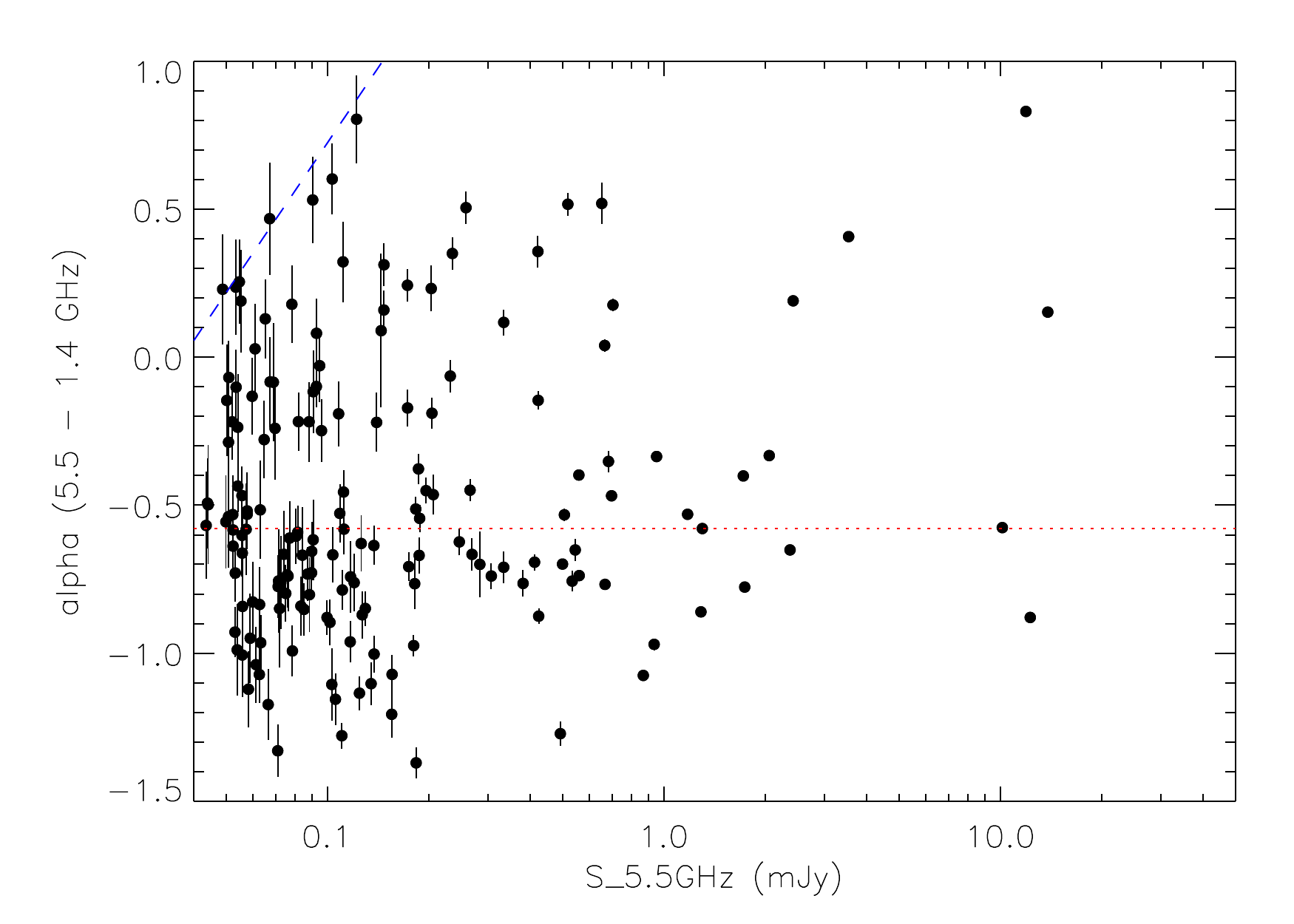}
\caption{The 5.5--1.4  GHz spectral index vs. 5.5 GHz flux density for the ATCA 5.5 GHz selected sample. Only single component sources are shown. The dotted line indicates the median spectral index, $\alpha = -0.58$. The dashed line shows the $5 \sigma$ limit of the VLA 1.4 GHz observations, showing this sample is sensitive to inverted sources at the faintest 5.5 GHz levels.}
\label{fig:alphaflux}
\end{figure}

The median spectral index of this 5.5 GHz selected sample is $\alpha_{\rm med} = -0.58$ (see Figure \ref{fig:alphahist}) and $\alpha_{\rm mean} = -0.47 \pm 0.04$.
This median spectral index is marginally steeper than our previous work which found $\alpha_{\rm med}$  = $-0.40$ (H12). Figure \ref{fig:alphaflux} presents the spectral index as a function of 5.5 GHz flux density, and it shows that a population of steep spectrum sources at $S_{5.5 GHz} < 0.1$ mJy is responsible for the slightly steeper average spectral index compared to earlier work in H12. 
This indicates that these new deeper observations may be starting to probe the star forming population.
However, even at these low flux densities a significant fraction (31/79, 39\%) of the 5.5 GHz sample has a flat or inverted spectral index ($\alpha > -0.5$). 
For $S_{5.5 GHz} > 0.5$ mJy the median spectral index is $\alpha_{\rm med} = -0.47$ and  the average spectral index is $\alpha_{\rm mean} = -0.35 \pm 0.10$, which is consistent with published values for 6cm selected sources of similar flux density. For example Prandoni et al. (2006) found $\alpha_{\rm med}=-0.4$ for $0.4 < S_{5GHz} < 4$ mJy and Donnelly et al. (1987) who found $\alpha_{\rm med} =-0.42$ for $0.4 < S_{5GHz} < 1.2$ mJy.

\begin{table}
\centering
\caption{The 1.4 -- 5.5 GHz spectral index of the ATCA 5.5 GHz sample. }
\begin{tabular}{lllcc} \hline
ID & $S_{5.5 GHz}$  & $S_{1.4 GHz}$  & $\alpha$ & $\delta$$\alpha$  \\
     & ($\mu$Jy) & ($\mu$Jy) & & \\ \hline
     1 &        283 &     729.0 &      -0.70 &      0.11 \\
     2 &        306 &     831.0 &      -0.74 &      0.04 \\
     3 &        544 &    1310.0 &      -0.65 &      0.04 \\
     4 &        155 &     660.0 &      -1.07 &      0.07 \\
     5 &        505 &    1036.0 &      -0.53 &      0.02 \\
     6 &         96 &     133.9 &      -0.25 &      0.11 \\
     7 &         95 &      98.3 &      -0.03 &      0.12 \\
     8 &        421 &     260.0 &       0.36 &      0.05 \\
     9 &        155 &     789.8 &      -1.21 &      0.08 \\
    10 &        231 &     252.0 &      -0.06 &      0.05 \\
    11 &         93 &      83.0 &       0.08 &      0.12 \\
    12 &        265 &     485.6 &      -0.45 &      0.04 \\
    13 &         56 &     136.1 &      -0.66 &      0.13 \\
    14 &         72 &     226.3 &      -0.85 &      0.20 \\
    15 &         76 &     204.6 &      -0.74 &      0.11 \\
    16 &        108 &     139.4 &      -0.19 &      0.11 \\
    17 &        120 &     334.9 &      -0.76 &      0.10 \\
    18 &       1286 &    4108.0 &      -0.86 &      0.01 \\
    19 &        697 &    1312.0 &      -0.47 &      0.01 \\
    21 &       1298 &    2836.0 &      -0.58 &      0.01 \\
    22 &        704 &     554.6 &       0.18 &      0.02 \\
    23 &        246 &     571.1 &      -0.62 &      0.05 \\
    24 &        137 &     323.6 &      -0.64 &      0.07 \\
    25 &         67 &      35.7 &       0.47 &      0.19 \\
    26 &        100 &     326.2 &      -0.88 &      0.06 \\
    27 &      10114 &   22000.0 &      -0.58 &      0.00 \\
    28 &        127 &     410.0 &      -0.87 &      0.08 \\
    29 &        499 &    1281.0 &      -0.70 &      0.01 \\
    30 &         51 &     104.9 &      -0.54 &      0.17 \\
    31 &         63 &     193.7 &      -0.84 &      0.13 \\
    32 &        333 &     284.3 &       0.12 &      0.04 \\
    33 &         61 &     248.8 &      -1.04 &      0.13 \\
    34 &         65 &      54.7 &       0.13 &      0.13 \\
    35 &         49 &      35.7 &       0.23 &      0.19 \\
    36 &         60 &     182.6 &      -0.83 &      0.14 \\
    37 &         63 &     126.4 &      -0.52 &      0.17 \\
    39 &        949 &    1493.0 &      -0.34 &      0.01 \\
    40 &        122 &      41.1 &       0.80 &      0.15 \\
    41 &        683 &    1098.0 &      -0.35 &      0.04 \\
    42 &         52 &     123.8 &      -0.64 &      0.11 \\
    43 &         65 &      94.2 &      -0.28 &      0.13 \\
    44 &        203 &     148.3 &       0.23 &      0.08 \\
    45 &         70 &      96.4 &      -0.24 &      0.17 \\
    47 &        206 &     385.1 &      -0.46 &      0.07 \\
    48 &         59 &     211.9 &      -0.95 &      0.15 \\
    49 &        867 &    3700.0 &      -1.08 &      0.01 \\
    50 &       2366 &    5697.0 &      -0.65 &      0.01 \\
    51 &        109 &     221.4 &      -0.53 &      0.10 \\
    52 &         44 &      93.7 &      -0.57 &      0.18 \\
    53 &         76 &     206.8 &      -0.74 &      0.12 \\
    54 &         50 &     105.5 &      -0.56 &      0.16 \\
    56 &        933 &    3458.0 &      -0.97 &      0.02 \\
    57 &        126 &     294.0 &      -0.63 &      0.09 \\
    58 &         88 &     260.5 &      -0.80 &      0.13 \\
    60 &         74 &     182.1 &      -0.67 &      0.15 \\
    61 &        129 &     406.8 &      -0.85 &      0.06 \\
    62 &        124 &     575.2 &      -1.14 &      0.06 \\
    63 &         53 &     185.7 &      -0.93 &      0.09 \\
    64 &         52 &     107.0 &      -0.53 &      0.13 \\
    65 &         56 &     173.9 &      -0.84 &      0.18 \\
    66 &         56 &     104.5 &      -0.47 &      0.10 \\
    67 &         54 &     204.4 &      -0.99 &      0.16 \\
    68 &        105 &     501.8 &      -1.16 &      0.09 \\ \hline
        \end{tabular}
\label{tab:alpha}
\end{table}

\setcounter{table}{3}
\begin{table}
\caption{continued}  
 \centering
\begin{tabular}{lllcc} \hline
ID & $S_{5.5 GHz}$  & $S_{1.4 GHz}$  & $\alpha$ & $\delta$$\alpha$  \\
     & ($\mu$Jy) & ($\mu$Jy) & &  \\ \hline   
    70 &         67 &      75.4 &      -0.08 &      0.15 \\ 
    71 &        103 &      45.7 &       0.60 &      0.12 \\
    72 &        187 &     461.5 &      -0.67 &      0.06 \\
    73 &        174 &     452.2 &      -0.71 &      0.05 \\
    74 &        381 &    1068.0 &      -0.76 &      0.04 \\
    77 &        204 &     263.3 &      -0.19 &      0.05 \\
    78 &        111 &      71.8 &       0.32 &      0.14 \\
    79 &         57 &     125.4 &      -0.58 &      0.15 \\
    81 &         84 &     207.2 &      -0.67 &      0.16 \\
    82 &        147 &     118.3 &       0.16 &      0.06 \\
    83 &        104 &     254.9 &      -0.67 &      0.09 \\
    84 &         83 &     258.9 &      -0.84 &      0.10 \\
    85 &         53 &      38.7 &       0.24 &      0.16 \\
    86 &        181 &     509.4 &      -0.77 &      0.09 \\
    87 &         55 &      38.7 &       0.25 &      0.14 \\
    88 &         93 &     105.7 &      -0.10 &      0.07 \\
    89 &         87 &     234.2 &      -0.73 &      0.09 \\
    90 &        102 &     340.2 &      -0.90 &      0.08 \\
    92 &         81 &     180.9 &      -0.59 &      0.09 \\
    93 &         77 &     175.7 &      -0.61 &      0.12 \\
    94 &        422 &     513.6 &      -0.15 &      0.03 \\
    95 &       1718 &    2952.0 &      -0.40 &      0.01 \\
    96 &         88 &     118.1 &      -0.22 &      0.14 \\
    97 &        258 &     130.1 &       0.51 &      0.06 \\
    98 &         58 &     117.9 &      -0.53 &      0.11 \\
    99 &         82 &     109.7 &      -0.22 &      0.10 \\
   101 &         90 &      44.0 &       0.53 &      0.15 \\
   102 &         44 &      86.4 &      -0.50 &      0.20 \\
   103 &        424 &    1380.0 &      -0.88 &      0.03 \\
   104 &      11886 &    3871.0 &       0.83 &      0.00 \\
   105 &         90 &     239.3 &      -0.73 &      0.08 \\
   106 &         50 &      61.0 &      -0.15 &      0.19 \\
   107 &       2052 &    3213.0 &      -0.33 &      0.01 \\
   108 &        183 &     364.9 &      -0.51 &      0.04 \\
   109 &        111 &     206.0 &      -0.46 &      0.08 \\
   110 &        183 &    1165.0 &      -1.37 &      0.05 \\
   112 &        517 &     257.2 &       0.52 &      0.04 \\
   113 &       3533 &    2037.0 &       0.41 &      0.01 \\
   114 &         63 &     232.5 &      -0.96 &      0.09 \\
   115 &      13803 &   11230.0 &       0.15 &      0.01 \\
   116 &         73 &     202.7 &      -0.76 &      0.16 \\
   117 &         61 &      58.5 &       0.03 &      0.15 \\
   118 &        111 &     319.5 &      -0.79 &      0.07 \\
   119 &         60 &      71.2 &      -0.13 &      0.13 \\
   120 &         63 &     266.2 &      -1.07 &      0.10 \\
   122 &       2417 &    1868.0 &       0.19 &      0.01 \\
   123 &         81 &     182.1 &      -0.60 &      0.09 \\
   124 &         51 &      55.7 &      -0.07 &      0.13 \\
   126 &         54 &      97.2 &      -0.44 &      0.09 \\
   127 &         75 &     220.3 &      -0.80 &      0.10 \\
   128 &         51 &      74.7 &      -0.29 &      0.15 \\
   129 &        665 &     630.3 &       0.04 &      0.02 \\
   130 &        558 &     954.2 &      -0.40 &      0.02 \\
   131 &         52 &      69.8 &      -0.22 &      0.13 \\
   132 &        110 &     619.0 &      -1.28 &      0.04 \\
   133 &        333 &     868.8 &      -0.71 &      0.05 \\
   134 &        173 &     217.5 &      -0.17 &      0.06 \\
   135 &       1173 &    2402.0 &      -0.53 &      0.01 \\
   136 &         90 &     217.2 &      -0.66 &      0.10 \\
   137 &        147 &      96.3 &       0.31 &      0.07 \\
   138 &        188 &     391.0 &      -0.54 &      0.05 \\
   139 &        186 &     309.5 &      -0.38 &      0.05 \\
   140 &         55 &      42.7 &       0.19 &      0.17 \\
   141 &         79 &     299.8 &      -0.99 &      0.09 \\
   142 &         54 &      74.3 &      -0.24 &      0.18 \\ \hline
         \end{tabular}
\end{table}

\setcounter{table}{3}
\begin{table}
\caption{continued}  
 \centering
\begin{tabular}{lllcc} \hline
ID & $S_{5.5 GHz}$  & $S_{1.4 GHz}$  & $\alpha$ & $\delta$$\alpha$  \\
     & ($\mu$Jy) & ($\mu$Jy) & &  \\ \hline   
   143 &         38 &      75.4 &      -0.52 &      0.15 \\
   144 &         52 &     115.0 &      -0.58 &      0.09 \\
   145 &         91 &     208.1 &      -0.62 &      0.14 \\
   146 &         71 &     202.6 &      -0.77 &      0.11 \\
   147 &         53 &      61.2 &      -0.10 &      0.13 \\
   148 &         71 &     427.6 &      -1.33 &      0.09 \\
   150 &         58 &     116.1 &      -0.52 &      0.13 \\
   151 &         53 &     142.1 &      -0.73 &      0.10 \\
   152 &        268 &     659.7 &      -0.67 &      0.05 \\
   153 &         58 &     264.2 &      -1.12 &      0.13 \\
   155 &        196 &     360.0 &      -0.45 &      0.04 \\
   157 &         56 &     125.3 &      -0.60 &      0.14 \\
   158 &       1735 &    4950.0 &      -0.78 &      0.01 \\
   159 &        180 &     670.8 &      -0.97 &      0.04 \\
   160 &         56 &     216.7 &      -1.01 &      0.14 \\
   161 &        667 &    1880.0 &      -0.77 &      0.01 \\
   162 &        135 &     596.4 &      -1.10 &      0.07 \\
   163 &        173 &     124.2 &       0.24 &      0.06 \\
   164 &        137 &     531.9 &      -1.00 &      0.06 \\
   165 &         85 &     268.5 &      -0.85 &      0.09 \\
   166 &      12243 &   40130.0 &      -0.88 &      0.00 \\
   167 &        103 &     457.5 &      -1.11 &      0.12 \\
   168 &        559 &    1513.0 &      -0.74 &      0.01 \\
   170 &         44 &      85.4 &      -0.49 &      0.15 \\
   171 &         78 &      61.4 &       0.18 &      0.13 \\
   173 &        112 &     244.7 &      -0.58 &      0.08 \\
   174 &         67 &     324.3 &      -1.17 &      0.12 \\
   175 &        117 &     427.1 &      -0.96 &      0.07 \\
   176 &        491 &    2736.0 &      -1.27 &      0.04 \\
   178 &        412 &    1049.0 &      -0.69 &      0.03 \\
   179 &        117 &     318.3 &      -0.74 &      0.12 \\
   180 &         71 &     198.1 &      -0.76 &      0.18 \\
   181 &        533 &    1479.0 &      -0.76 &      0.03 \\
   182 &        235 &     146.1 &       0.35 &      0.06 \\
   183 &         69 &      77.2 &      -0.08 &      0.20 \\
   184 &         91 &     105.9 &      -0.12 &      0.14 \\
   185 &        652 &     325.0 &       0.52 &      0.07 \\
   186 &        144 &     128.0 &       0.09 &      0.26 \\
   187 &        140 &     189.0 &      -0.22 &      0.10 \\ \hline
\end{tabular}
\end{table}

\subsection{Radio Spectral Curvature}

The 4 sub-bands across the full 2 GHz band at 6cm provide us with the opportunity to study in more detail the radio spectral energy distribution (SED) of our sources. 
The radio SED of galaxies can be complex and is not always well described by a power law. Self-synchrotron absorption leads to a turnover at low frequencies ($\nu_{rest} << 1$ GHz) but for young compact AGN the turnover frequency can be on the order of a GHz, and these are known as Gigahertz Peaked Spectrum (GPS) sources (e.g. \citealp{fanti1995,odea1998,randall2011}). The spectral slope can steepen at high frequencies ($\nu_{rest} \gtrsim 10$ GHz)  from inverse-Compton losses (e.g. \citealp{klamer2006}). Alternatively, a restarting AGN could appear to flatten at high frequencies as the high frequency observations are more sensitive to the flat or inverted AGN core while the lower frequency observations detect the steeper old lobes. Furthermore, thermal (free-free) emission becomes increasingly important at  higher frequencies ($\nu_{rest} \gtrsim 10$ GHz) and the relatively flat spectrum of thermal emission can lead to a flattening of the radio SED in starforming galaxies at these frequencies \citep{condon1992}. 

In order to study the spectral curvature of our radio sources we first selected sources which have a S/N $>$ 10 in the 5.5 GHz full-band data, to ensure a good detection in each of the four sub-bands. Furthermore we only examined point sources, to limit any effects from the small differences in the beam-sizes of the sub-band images, resulting in a sample of 59 sources. We supplemented the four 6cm sub-bands and VLA 1.4 GHz detections with data from ATLAS 1.4 GHz Data Release 3 (DR3, Franzen et al. 2015 in press). The ATLAS 1.4 GHz DR3 survey covers the full eCDFS at two sub-bands with central frequencies of 1.4 and 1.7 GHz, reaching a typical sensitivity level of $\sim$ 20$\mu$Jy rms with a beam of 16$\arcsec$ $\times$ 7$\arcsec$ (Franzen et al. 2015 in press). To better explore radio spectral curvature we also include 9 GHz flux densities measured from a 9 GHz image made with CABB 3cm data taken simultaneously with the 6cm data of H12. The 9 GHz image reaches 25 -- 30 $\mu$Jy rms with a beam of  2.9$\arcsec$ $\times$ 1.2$\arcsec$ (Huynh et al. in prep).  The 9 GHz resolution is similar to the VLA 1.4 GHz resolution, however the large beam-size of the ATLAS 1.4 GHz DR3 data could mean discrepant flux densities for sources that are resolved out by the higher resolution images. This should not be a major issue as we are examining only point sources at 5.5 GHz. Furthermore, there is excellent agreement between the VLA 1.4 GHz flux densities and that of ATLAS 1.4 GHz DR3 for $S_{1.4 GHz} > 0.15$ mJy (Franzen et al. 2015 in press). 

We fitted the spectral energy distributions in log space with first and second order polynomials of the form:
\[ \log S = \gamma + \alpha \log \nu + \beta(\log \nu)^2 \;, \]
with units of $S$ in mJy and $\nu$ in GHz. The first order polynomial fit ($\beta = 0$) is the commonly assumed power law $S \propto \nu^\alpha$, with $\alpha$ as the spectral index. We refer to the first order polynomial fit as the log-linear fit and the second order polynomial fit as the log-quadratic fit. 

\begin{figure*}
\includegraphics[trim=0.3 1.75cm 0.3 1.75cm, page=1,clip=true,width=0.99\textwidth]{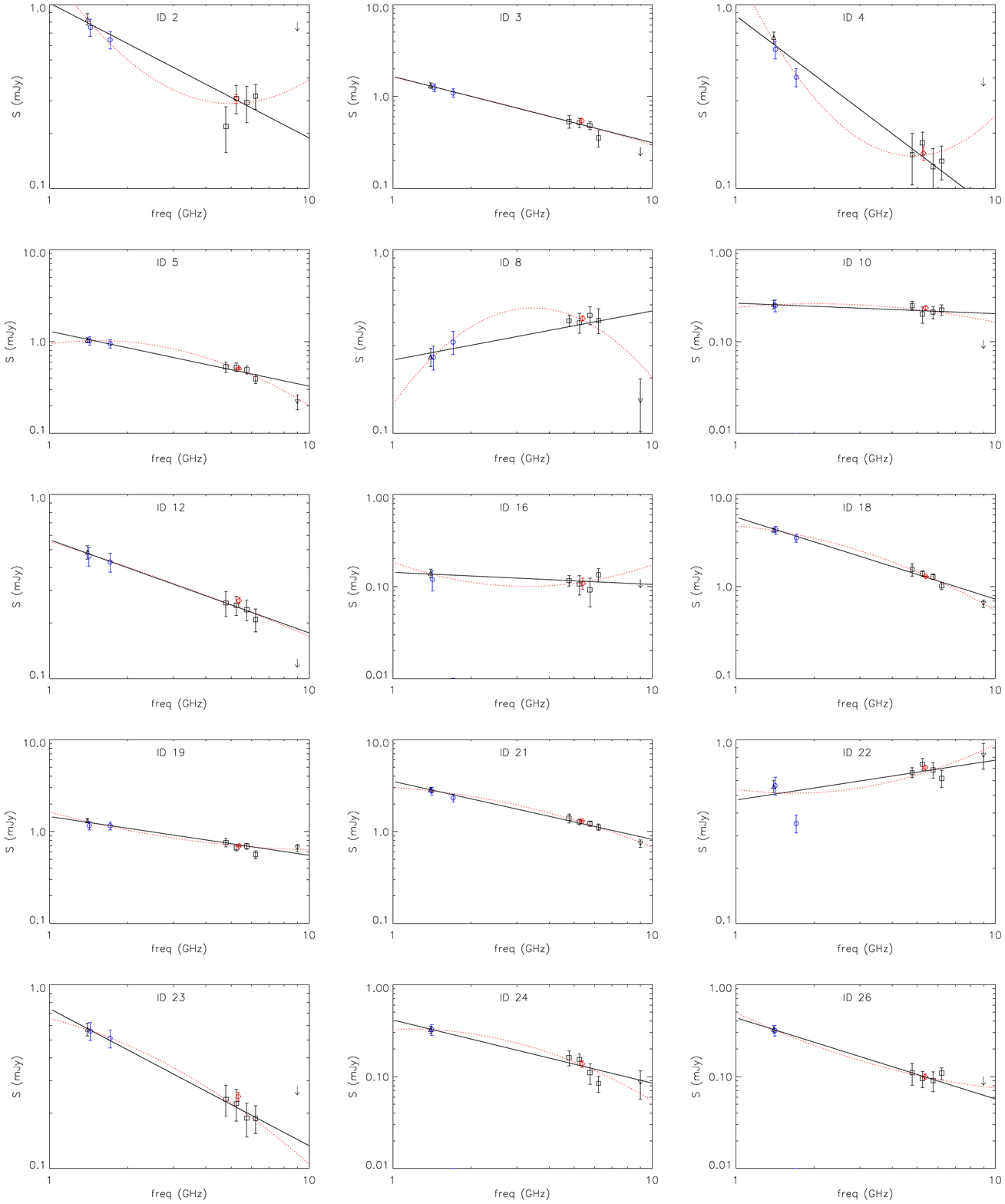}
\caption{The radio spectral energy distribution for all 6cm sources with S/N greater than 10. The datapoints at 1.4 and 1.7 GHz come from Miller et al. (2013) (black triangles) and Franzen et al. (2015, in press) (blue circles). The four measurements across 4.5 to 6.5 GHz (black squares) are from this work. The 9 GHz datapoint (black upside-down triangle, Huynh et al. 2015 in prep) is shown for sources detected at 9 GHz. An arrow is placed at 4$\sigma$ in the case of no detection at 9 GHz. The log-linear fit to this data is shown as a solid black line while the log-quadratic fit is shown as the red dotted line. Only $\sim$10\% of sources show significant curvature. The red diamond indicates the full-band 5.5 GHz flux densities.}
\label{fig:seds}
\end{figure*}

The fitting results for the 59 sources is summarised in Table \ref{tab:sedfit}, and their radio SEDs shown in Figure \ref{fig:seds}. First we compared the spectral index from the log-linear fit to the two point spectral index derived in Section \ref{sec:spectralindex}. We find that the ratio of the two point spectral index to the spectral index from the power law fit to all data, $\alpha_{\rm 5.5GHz -1.4GHz} / \alpha_{\rm fit} $, is 1.10 $\pm$ 0.04, with a median of 1.01. Thus the two measures of the spectral index are consistent at the $\sim$10\% level.  This is reflected in the individual radio SEDs (Figure \ref{fig:seds}), where the measured full-bandwidth 5.5 GHz flux density (shown as a red diamond) is consistent with the log-linear fit (solid black line).

If the log-quadratric fit is accepted only for $|\beta|/\delta \beta > 2$, i.e. $\beta$ is formally greater than zero at better than 2$\sigma$ (95\%) level of confidence, then 13/59 (22\%) of the 5.5 GHz sources are candidates for sources with significant curvature. These are source IDs 5, 8, 18, 27, 29, 50, 94, 104, 108, 112, 113, 139, 158. On examination of these radio sources 8, 94, 108 and 139 are cases where the 9 GHz detection is low S/N and the SED is consistent with a log-linear fit if the 9 GHz datapoint is discarded, so we conservatively exclude these candidates. Sources 104, 112, and 113 have variable flux densities on timescales of months to years \citep{bell2015}, so these are excluded also. Source 158 appears to have a positive curvature and the upturn might be due to the 9 GHz observations picking up the flatter spectrum core of the source. Sources 18, 27, and 50 show negative curvature or steepening spectra. Sources 5 and 29 appear to be GPS sources peaking between 1 to 2 GHz. In summary, the log-quadratic fit is accepted for 6/59 ($\sim$10\%) of the 5.5 GHz sources after examination, with 1 source showing an upturn, 3 sources showing a steepening, and 2 sources exhibiting a GPS SED peaking between 1 to 2 GHz.  

One caveat on these results is that the radio data across 1.4 to 9 GHz were not taken simultaneously and hence source variability can affect the radio SED. \cite{bell2015} have shown that only a few percent of 5.5 GHz sources are variable on the yearly timescale, and these appear to be inverted spectrum sources where variability is intrinsic to the AGN due to changes in the accretion rate, heating of material and reprocessing of energy by the accretion disk. Hence source variability could explain the SEDs of the GPS candidates, IDs 5 and 29, but it is not a likely explanation for the curvature seen in source IDs 18, 27, 50, and 158. 

The fraction of radio sources with significant spectral curvature ranges from almost 100\% in the brightest samples (e.g. \citealp{laing1980}) to 13 -- 49\% for 1 -- 10 mJy level sources \citep{randall2012,ker2012b}.  This is higher than the fraction we observe in our faint 5.5 GHz sample. We also find 2/59 (3\%) sources have a GPS SED, which is lower than the 10\% fraction found in Jy level radio samples \citep{odea1998}. This would imply our low flux density sample exhibits less radio spectral curvature than brighter samples, but there are other effects to consider, such as the signal-to-noise ratio of detections, different frequency coverage (greater frequency coverage makes it easier to detect spectral curvature) and the non-consistent definitions of curvature across the different studies. A homogeneous analysis across a large sample of radio sources is required to draw firm conclusions. 

\begin{table*}
\caption{Summary of radio spectral energy distribution fitting results. }
\begin{tabular}{lcccccccccc} \hline
ID & \multicolumn{4}{c}{Log-linear fit}  & \multicolumn{6}{c}{Log-quadratic fit} \\
     &  $\gamma$ & $\delta\gamma$ & $\alpha$ & $\delta\alpha$  & $\gamma$ & $\delta\gamma$ & $\alpha$ & $\delta\alpha$ & $\beta$ & $\delta\beta$ \\ 
\hline
    2 &    0.01 &    0.03 &   -0.73 &    0.09 &    0.19 &    0.14 &   -2.07 &    1.02 &    1.48 &    1.12 \\
    3 &    0.22 &    0.03 &   -0.72 &    0.07 &    0.21 &    0.12 &   -0.65 &    0.84 &   -0.08 &    0.95 \\
    4 &   -0.06 &    0.04 &   -1.06 &    0.10 &    0.19 &    0.14 &   -2.97 &    1.05 &    2.17 &    1.19 \\
    5 &    0.11 &    0.03 &   -0.60 &    0.05 &   -0.03 &    0.07 &    0.40 &    0.43 &   -1.06 &    0.45 \\
    8 &   -0.60 &    0.05 &    0.27 &    0.08 &   -0.84 &    0.11 &    1.93 &    0.67 &   -1.79 &    0.71 \\
   10 &   -0.58 &    0.05 &   -0.11 &    0.09 &   -0.63 &    0.19 &    0.26 &    1.49 &   -0.42 &    1.67 \\
   12 &   -0.25 &    0.04 &   -0.51 &    0.07 &   -0.26 &    0.13 &   -0.44 &    0.89 &   -0.07 &    0.99 \\
   16 &   -0.85 &    0.06 &   -0.13 &    0.13 &   -0.74 &    0.25 &   -1.03 &    1.98 &    1.00 &    2.20 \\
   18 &    0.75 &    0.03 &   -0.89 &    0.04 &    0.66 &    0.04 &   -0.24 &    0.25 &   -0.68 &    0.26 \\
   19 &    0.16 &    0.03 &   -0.42 &    0.04 &    0.21 &    0.05 &   -0.76 &    0.27 &    0.35 &    0.27 \\
   21 &    0.55 &    0.02 &   -0.63 &    0.04 &    0.48 &    0.04 &   -0.17 &    0.25 &   -0.47 &    0.25 \\
   22 &   -0.33 &    0.03 &    0.22 &    0.05 &   -0.27 &    0.06 &   -0.20 &    0.38 &    0.44 &    0.40 \\
   23 &   -0.13 &    0.04 &   -0.75 &    0.09 &   -0.19 &    0.14 &   -0.35 &    1.01 &   -0.44 &    1.13 \\
   24 &   -0.39 &    0.04 &   -0.69 &    0.09 &   -0.49 &    0.11 &    0.11 &    0.81 &   -0.89 &    0.89 \\
   26 &   -0.36 &    0.04 &   -0.88 &    0.10 &   -0.30 &    0.24 &   -1.34 &    1.90 &    0.52 &    2.15 \\
   27 &    1.45 &    0.02 &   -0.62 &    0.03 &    1.39 &    0.04 &   -0.22 &    0.18 &   -0.40 &    0.17 \\
   29 &    0.23 &    0.03 &   -0.80 &    0.05 &    0.08 &    0.06 &    0.29 &    0.42 &   -1.19 &    0.46 \\
   40 &   -1.47 &    0.15 &    0.71 &    0.21 &   -1.84 &    0.31 &    3.58 &    2.09 &   -3.21 &    2.33 \\
   44 &   -0.83 &    0.07 &    0.14 &    0.11 &   -0.96 &    0.12 &    1.03 &    0.62 &   -0.94 &    0.64 \\
   50 &    0.86 &    0.03 &   -0.70 &    0.04 &    0.79 &    0.04 &   -0.23 &    0.21 &   -0.47 &    0.21 \\
   57 &   -0.43 &    0.05 &   -0.70 &    0.11 &   -0.47 &    0.21 &   -0.33 &    1.67 &   -0.42 &    1.88 \\
   61 &   -0.30 &    0.05 &   -0.79 &    0.11 &   -0.85 &    0.29 &    3.76 &    2.35 &   -5.25 &    2.71 \\
   62 &   -0.09 &    0.03 &   -1.14 &    0.09 &   -0.10 &    0.11 &   -1.06 &    0.74 &   -0.08 &    0.82 \\
   71 &   -1.44 &    0.11 &    0.63 &    0.18 &   -1.07 &    0.34 &   -2.36 &    2.54 &    3.30 &    2.81 \\
   82 &   -0.93 &    0.06 &    0.09 &    0.11 &   -0.89 &    0.24 &   -0.22 &    1.87 &    0.36 &    2.13 \\
   83 &   -0.50 &    0.05 &   -0.76 &    0.13 &   -0.79 &    0.29 &    1.61 &    2.27 &   -2.69 &    2.56 \\
   84 &   -0.51 &    0.06 &   -0.75 &    0.13 &   -0.25 &    0.18 &   -2.75 &    1.38 &    2.15 &    1.47 \\
   90 &   -0.34 &    0.05 &   -0.86 &    0.10 &   -0.30 &    0.25 &   -1.17 &    2.01 &    0.35 &    2.28 \\
   92 &   -0.69 &    0.06 &   -0.43 &    0.12 &   -0.68 &    0.13 &   -0.54 &    0.89 &    0.12 &    0.98 \\
   94 &   -0.26 &    0.03 &   -0.23 &    0.06 &   -0.41 &    0.07 &    0.89 &    0.47 &   -1.20 &    0.50 \\
   95 &    0.54 &    0.02 &   -0.38 &    0.04 &    0.52 &    0.04 &   -0.26 &    0.21 &   -0.12 &    0.21 \\
  101 &   -1.41 &    0.12 &    0.46 &    0.18 &   -1.48 &    0.18 &    0.92 &    1.00 &   -0.48 &    1.04 \\
  104 &    0.53 &    0.02 &    0.68 &    0.03 &    0.34 &    0.04 &    1.86 &    0.18 &   -1.17 &    0.18 \\
  107 &    0.53 &    0.02 &   -0.29 &    0.04 &    0.58 &    0.04 &   -0.56 &    0.21 &    0.28 &    0.21 \\
  108 &   -0.39 &    0.04 &   -0.45 &    0.07 &   -0.23 &    0.08 &   -1.71 &    0.53 &    1.33 &    0.55 \\
  109 &   -0.62 &    0.06 &   -0.47 &    0.10 &   -0.58 &    0.14 &   -0.79 &    1.02 &    0.34 &    1.10 \\
  112 &   -0.62 &    0.05 &    0.42 &    0.07 &   -0.72 &    0.07 &    1.04 &    0.32 &   -0.63 &    0.31 \\
  113 &    0.29 &    0.02 &    0.30 &    0.04 &    0.11 &    0.04 &    1.45 &    0.20 &   -1.14 &    0.19 \\
  115 &    1.02 &    0.02 &    0.17 &    0.04 &    1.00 &    0.04 &    0.26 &    0.19 &   -0.09 &    0.19 \\
  118 &   -0.38 &    0.04 &   -0.84 &    0.09 &   -0.40 &    0.12 &   -0.70 &    0.82 &   -0.15 &    0.89 \\
  122 &    0.25 &    0.03 &    0.16 &    0.04 &    0.20 &    0.04 &    0.47 &    0.22 &   -0.31 &    0.22 \\
  123 &   -0.69 &    0.06 &   -0.45 &    0.12 &   -0.63 &    0.16 &   -0.88 &    1.16 &    0.47 &    1.25 \\
  129 &   -0.18 &    0.03 &    0.02 &    0.05 &   -0.26 &    0.05 &    0.54 &    0.27 &   -0.53 &    0.27 \\
  130 &    0.02 &    0.03 &   -0.37 &    0.05 &    0.08 &    0.06 &   -0.77 &    0.34 &    0.42 &    0.35 \\
  132 &   -0.03 &    0.03 &   -1.26 &    0.08 &   -0.08 &    0.11 &   -0.88 &    0.80 &   -0.42 &    0.88 \\
  134 &   -0.62 &    0.06 &   -0.27 &    0.11 &   -0.66 &    0.13 &    0.01 &    0.89 &   -0.30 &    0.95 \\
  139 &   -0.53 &    0.04 &   -0.18 &    0.07 &   -0.24 &    0.08 &   -2.32 &    0.45 &    2.17 &    0.45 \\
  152 &   -0.12 &    0.04 &   -0.72 &    0.08 &   -0.07 &    0.13 &   -1.03 &    0.93 &    0.35 &    1.04 \\
  155 &   -0.32 &    0.04 &   -0.57 &    0.08 &   -0.50 &    0.11 &    0.73 &    0.76 &   -1.41 &    0.82 \\
  158 &    0.78 &    0.03 &   -0.69 &    0.04 &    0.86 &    0.04 &   -1.18 &    0.24 &    0.50 &    0.24 \\
  159 &   -0.02 &    0.03 &   -1.04 &    0.08 &   -0.07 &    0.09 &   -0.65 &    0.66 &   -0.43 &    0.72 \\
  162 &   -0.07 &    0.04 &   -1.10 &    0.10 &    0.01 &    0.14 &   -1.67 &    1.03 &    0.65 &    1.14 \\
  163 &   -0.93 &    0.06 &    0.19 &    0.10 &   -0.97 &    0.18 &    0.50 &    1.34 &   -0.35 &    1.49 \\
  173 &   -0.55 &    0.06 &   -0.59 &    0.12 &   -0.29 &    0.32 &   -2.66 &    2.50 &    2.28 &    2.76 \\
  175 &   -0.25 &    0.04 &   -0.94 &    0.10 &   -0.49 &    0.25 &    1.02 &    2.02 &   -2.23 &    2.29 \\
  181 &    0.26 &    0.03 &   -0.78 &    0.05 &    0.36 &    0.06 &   -1.47 &    0.38 &    0.73 &    0.40 \\
  182 &   -0.85 &    0.06 &    0.17 &    0.10 &   -1.06 &    0.14 &    1.84 &    1.01 &   -1.87 &    1.12 \\
  185 &   -0.56 &    0.05 &    0.46 &    0.08 &   -0.71 &    0.11 &    1.47 &    0.65 &   -1.10 &    0.70 \\
  187 &   -0.68 &    0.05 &   -0.17 &    0.12 &   -0.68 &    0.20 &   -0.20 &    1.45 &    0.03 &    1.60 \\
\hline
\end{tabular}
\label{tab:sedfit}
\end{table*}

\section{Concluding Remarks}

We have presented new observations at 5.5 GHz of the extended Chandra Deep Field South with the Australia Telescope Compact Array.
Combined with our earlier data, this resulting image of 0.34 deg$^2$ reaches a noise level of $\sim$8.6 $\mu$Jy rms, for a synthesized beam of 5.0 $\times$ 2.0 arcsec. 
This new image is the largest mosaic ever made at this frequency to these depths. 
Using a false-discovery-rate method, we extracted 189 individual radio sources. Twelve sources were resolved multiple sources with AGN core-lobe or lobe-lobe structures and hence fitted as multiple components.

We derived source counts at 5.5 GHz after careful corrections for completeness, flux boosting and resolution bias. These are amongst the deepest source counts ever calculated at 6cm but come from an area 4 to 7 times larger than the previous surveys to these depths. The ATLAS 5.5 GHz counts are consistent with the counts derived from other 5 GHz surveys at brighter flux densities, but are lower than counts in the literature by a factor of two for $S_{5.5GHz} < 0.1$ mJy. Most of this discrepancy is attributed to cosmic variance because of the small effective area of the surveys at faint flux densities. This fluctuation in the 5.5 GHz source counts at the faint end is similar to that seen at 1.4 GHz for $S_{1.4GHz} < 0.1$ mJy (e.g. \citealp{norris2011}). In general there is good agreement between the observed counts and that of semi-empirical simulations of Wilman et al. 2008, but there maybe an excess in observed sources in the $\sim$0.5 -- 2 mJy flux density range, which may be related to flat-spectrum sources detected at sub-mJy levels at higher frequencies ($>$10 GHz, Whittam et al. 2013; Franzen et al. 2014).

The 1.4 -- 5.5 GHz spectral index has also been determined for the 5.5 GHz sample.  We find a median spectral index for the ATCA 5.5 GHz sample of $\alpha_{\rm med} = -0.58$. This is steeper than the median spectral index for sub-mJy samples at 5.5 GHz and steeper than our previous result in Data Release 1 \citep{huynh2012a}. These new deeper observations may be starting to probe the starforming population. However a significant fraction (39\%) of the faintest sources ($0.05 < S_{5.5GHz} < 0.1$ mJy) show a flat or inverted spectral index ($\alpha > -0.5$). 

The radio SEDs of the brighter sources (S/N $>$ 10) in our 5.5 GHz sample were studied in detail by combining 4 flux density measurements in this work, spanning 4.5 to 6.5 GHz,  with literature data at 1.4 and 9 GHz. We fit the radio SEDs with both a log-linear and log-quadratic function to search for significant curvature over 0.8 dex in frequency. The log-quadratic fit is accepted for 10\% of the 5.5 GHz sources, with 1 source showing an upturn, 3 sources showing a steepening, and 2 sources exhibiting a GPS SED peaking between 1 to 2 GHz.  
 
New radio facilities are becoming available such as the upgraded VLA (the Karl G. Jansky VLA) and the Square Kilometre Array pathfinders, ASKAP and MeerKAT. In the next few years deep radio surveys will routinely achieve rms sensitivities of $\sim$1$\mu$Jy at frequencies near 1.4 GHz (e.g. Condon et al. 2012), providing valuable insight into the star formation and AGN activity in galaxies. Higher frequency radio surveys appear to select flat-spectrum populations not present in 1.4 GHz surveys of similar depth. Hence deep observations, at 5 GHz and above, will remain important for a full understanding the faint radio population. 

\section*{Acknowledgements}

The ATCA is part of the Australia Telescope which is funded by the Commonwealth of Australia for operation as a National Facility managed by CSIRO.
NS is a recipient of an ARC Future Fellowship. 

\bibliographystyle{mn2e}
\bibliography{refs}

\begin{thebibliography}{68}
\expandafter\ifx\csname natexlab\endcsname\relax\def\natexlab#1{#1}\fi

\bibitem[{{Aird} {et~al}\mbox{.}(2010){Aird}, {Nandra}, {Laird}, {Georgakakis},
  {Ashby}, {Barmby}, {Coil}, {Huang}, {Koekemoer}, {Steidel}, \&
  {Willmer}}]{aird2010}
{Aird} J. {et~al.}, 2010, MNRAS, 401, 2531

\bibitem[{{Baars} {et~al}\mbox{.}(1977){Baars}, {Genzel}, {Pauliny-Toth}, \&
  {Witzel}}]{Baars1977}
{Baars} J.~W.~M., {Genzel} R., {Pauliny-Toth} I.~I.~K., {Witzel} A., 1977,
  A\&A, 61, 99

\bibitem[{{Bell} {et~al}\mbox{.}(2015){Bell}, {Huynh}, {Hancock}, {Murphy},
  {Gaensler}, {Burlon}, {Trott}, \& {Bannister}}]{bell2015}
{Bell} M.~E., {Huynh} M.~T., {Hancock} P., {Murphy} T., {Gaensler} B.~M.,
  {Burlon} D., {Trott} C., {Bannister} K., 2015, MNRAS, 450, 4221

\bibitem[{{Bertin} \& {Arnouts}(1996)}]{bertin1996}
{Bertin} E., {Arnouts} S., 1996, A\&AS, 117, 393

\bibitem[{{Bonzini} {et~al}\mbox{.}(2013){Bonzini}, {Padovani}, {Mainieri},
  {Kellermann}, {Miller}, {Rosati}, {Tozzi}, \& {Vattakunnel}}]{bonzini2013}
{Bonzini} M., {Padovani} P., {Mainieri} V., {Kellermann} K.~I., {Miller} N.,
  {Rosati} P., {Tozzi} P., {Vattakunnel} S., 2013, MNRAS, 436, 3759

\bibitem[{{Briggs}(1995)}]{briggs1995}
{Briggs} D.~S., 1995, PhD thesis, New Mexico Institute of Mining and Technology

\bibitem[{{Ciliegi} {et~al}\mbox{.}(2003){Ciliegi}, {Zamorani}, {Hasinger},
  {Lehmann}, {Szokoly}, \& {Wilson}}]{ciliegi2003}
{Ciliegi} P., {Zamorani} G., {Hasinger} G., {Lehmann} I., {Szokoly} G.,
  {Wilson} G., 2003, A\&A, 398, 901

\bibitem[{{Condon}(1984{\natexlab{a}})}]{condon1984b}
{Condon} J.~J., 1984{\natexlab{a}}, ApJ, 287, 461

\bibitem[{{Condon}(1984{\natexlab{b}})}]{condon1984a}
{Condon} J.~J., 1984{\natexlab{b}}, ApJ, 284, 44

\bibitem[{{Condon}(1989)}]{condon1989}
{Condon} J.~J., 1989, ApJ, 338, 13

\bibitem[{{Condon}(1992)}]{condon1992}
{Condon} J.~J., 1992, ARA\&A, 30, 575

\bibitem[{{Condon} {et~al}\mbox{.}(2012){Condon}, {Cotton}, {Fomalont},
  {Kellermann}, {Miller}, {Perley}, {Scott}, {Vernstrom}, \&
  {Wall}}]{condon2012}
{Condon} J.~J. {et~al.}, 2012, ApJ, 758, 23

\bibitem[{{Condon} {et~al}\mbox{.}(1998){Condon}, {Cotton}, {Greisen}, {Yin},
  {Perley}, {Taylor}, \& {Broderick}}]{condon1998}
{Condon} J.~J., {Cotton} W.~D., {Greisen} E.~W., {Yin} Q.~F., {Perley} R.~A.,
  {Taylor} G.~B., {Broderick} J.~J., 1998, AJ, 115, 1693

\bibitem[{{Condon} \& {Ledden}(1981)}]{condon1981}
{Condon} J.~J., {Ledden} J.~E., 1981, AJ, 86, 643

\bibitem[{{Cowie} {et~al}\mbox{.}(1996){Cowie}, {Songaila}, {Hu}, \&
  {Cohen}}]{cowie1996}
{Cowie} L.~L., {Songaila} A., {Hu} E.~M., {Cohen} J.~G., 1996, AJ, 112, 839

\bibitem[{{Donnelly} {et~al}\mbox{.}(1987){Donnelly}, {Partridge}, \&
  {Windhorst}}]{donnelly1987}
{Donnelly} R.~H., {Partridge} R.~B., {Windhorst} R.~A., 1987, ApJ, 321, 94

\bibitem[{{Fanti} {et~al}\mbox{.}(1995){Fanti}, {Fanti}, {Dallacasa},
  {Schilizzi}, {Spencer}, \& {Stanghellini}}]{fanti1995}
{Fanti} C., {Fanti} R., {Dallacasa} D., {Schilizzi} R.~T., {Spencer} R.~E.,
  {Stanghellini} C., 1995, A\&A, 302, 317

\bibitem[{{Fomalont} {et~al}\mbox{.}(1991){Fomalont}, {Windhorst}, {Kristian},
  \& {Kellerman}}]{fomalont1991}
{Fomalont} E.~B., {Windhorst} R.~A., {Kristian} J.~A., {Kellerman} K.~I., 1991,
  AJ, 102, 1258

\bibitem[{{Franzen} {et~al}\mbox{.}(2014){Franzen}, {Sadler}, {Chhetri},
  {Ekers}, {Mahony}, {Murphy}, {Norris}, {Waldram}, \& {Whittam}}]{franzen2014}
{Franzen} T.~M.~O. {et~al.}, 2014, MNRAS, 439, 1212

\bibitem[{{Hales} {et~al}\mbox{.}(2012){Hales}, {Murphy}, {Curran},
  {Middelberg}, {Gaensler}, \& {Norris}}]{hales2012}
{Hales} C.~A., {Murphy} T., {Curran} J.~R., {Middelberg} E., {Gaensler} B.~M.,
  {Norris} R.~P., 2012, MNRAS, 425, 979

\bibitem[{{Hancock} {et~al}\mbox{.}(2012){Hancock}, {Murphy}, {Gaensler},
  {Hopkins}, \& {Curran}}]{hancock2012}
{Hancock} P.~J., {Murphy} T., {Gaensler} B.~M., {Hopkins} A., {Curran} J.~R.,
  2012, MNRAS, 422, 1812

\bibitem[{{Hasinger} {et~al}\mbox{.}(2005){Hasinger}, {Miyaji}, \&
  {Schmidt}}]{hasinger2005}
{Hasinger} G., {Miyaji} T., {Schmidt} M., 2005, A\&A, 441, 417

\bibitem[{{Heywood} {et~al}\mbox{.}(2013){Heywood}, {Jarvis}, \&
  {Condon}}]{heywood2013}
{Heywood} I., {Jarvis} M.~J., {Condon} J.~J., 2013, MNRAS, 432, 2625

\bibitem[{{Hopkins} \& {Beacom}(2006)}]{hopkins2006}
{Hopkins} A.~M., {Beacom} J.~F., 2006, ApJ, 651, 142

\bibitem[{{Hopkins} {et~al}\mbox{.}(2002){Hopkins}, {Miller}, {Connolly},
  {Genovese}, {Nichol}, \& {Wasserman}}]{hopkins2002}
{Hopkins} A.~M., {Miller} C.~J., {Connolly} A.~J., {Genovese} C., {Nichol}
  R.~C., {Wasserman} L., 2002, AJ, 123, 1086

\bibitem[{{Hopkins} {et~al}\mbox{.}(1998){Hopkins}, {Mobasher}, {Cram}, \&
  {Rowan-Robinson}}]{hopkins1998}
{Hopkins} A.~M., {Mobasher} B., {Cram} L., {Rowan-Robinson} M., 1998, MNRAS,
  296, 839

\bibitem[{{Huynh} {et~al}\mbox{.}(2012{\natexlab{a}}){Huynh}, {Hopkins},
  {Norris}, {Hancock}, {Murphy}, {Jurek}, \& {Whiting}}]{huynh2012b}
{Huynh} M.~T., {Hopkins} A., {Norris} R., {Hancock} P., {Murphy} T., {Jurek}
  R., {Whiting} M., 2012{\natexlab{a}}, PASA, 29, 229

\bibitem[{{Huynh} {et~al}\mbox{.}(2012{\natexlab{b}}){Huynh}, {Hopkins},
  {Lenc}, {Mao}, {Middelberg}, {Norris}, \& {Randall}}]{huynh2012a}
{Huynh} M.~T., {Hopkins} A.~M., {Lenc} E., {Mao} M.~Y., {Middelberg} E.,
  {Norris} R.~P., {Randall} K.~E., 2012{\natexlab{b}}, MNRAS, 426, 2342

\bibitem[{{Huynh} {et~al}\mbox{.}(2008){Huynh}, {Jackson}, {Norris}, \&
  {Fernandez-Soto}}]{huynh2008}
{Huynh} M.~T., {Jackson} C.~A., {Norris} R.~P., {Fernandez-Soto} A., 2008, AJ,
  135, 2470

\bibitem[{{Huynh} {et~al}\mbox{.}(2005){Huynh}, {Jackson}, {Norris}, \&
  {Prandoni}}]{huynh2005}
{Huynh} M.~T., {Jackson} C.~A., {Norris} R.~P., {Prandoni} I., 2005, AJ, 130,
  1373

\bibitem[{{Ibar} {et~al}\mbox{.}(2009){Ibar}, {Ivison}, {Biggs}, {Lal}, {Best},
  \& {Green}}]{ibar2009}
{Ibar} E., {Ivison} R.~J., {Biggs} A.~D., {Lal} D.~V., {Best} P.~N., {Green}
  D.~A., 2009, MNRAS, 397, 281

\bibitem[{{Jarvis} \& {Rawlings}(2004)}]{jarvis2004}
{Jarvis} M.~J., {Rawlings} S., 2004, NewAR, 48, 1173

\bibitem[{{Juneau} {et~al}\mbox{.}(2005){Juneau}, {Glazebrook}, {Crampton},
  {McCarthy}, {Savaglio}, {Abraham}, {Carlberg}, {Chen}, {Le Borgne}, {Marzke},
  {Roth}, {J{\o}rgensen}, {Hook}, \& {Murowinski}}]{juneau2005}
{Juneau} S. {et~al.}, 2005, ApJL, 619, L135

\bibitem[{{Kellermann} {et~al}\mbox{.}(2008){Kellermann}, {Fomalont},
  {Mainieri}, {Padovani}, {Rosati}, {Shaver}, {Tozzi}, \&
  {Miller}}]{kellermann2008}
{Kellermann} K.~I., {Fomalont} E.~B., {Mainieri} V., {Padovani} P., {Rosati}
  P., {Shaver} P., {Tozzi} P., {Miller} N., 2008, ApJS, 179, 71

\bibitem[{{Ker}(2012)}]{ker2012b}
{Ker} L.~M., 2012, PhD thesis, University of Edinburgh

\bibitem[{{Klamer} {et~al}\mbox{.}(2006){Klamer}, {Ekers}, {Bryant},
  {Hunstead}, {Sadler}, \& {De Breuck}}]{klamer2006}
{Klamer} I.~J., {Ekers} R.~D., {Bryant} J.~J., {Hunstead} R.~W., {Sadler}
  E.~M., {De Breuck} C., 2006, MNRAS, 371, 852

\bibitem[{{Laing} \& {Peacock}(1980)}]{laing1980}
{Laing} R.~A., {Peacock} J.~A., 1980, MNRAS, 190, 903

\bibitem[{{Magorrian} {et~al}\mbox{.}(1998){Magorrian}, {Tremaine},
  {Richstone}, {Bender}, {Bower}, {Dressler}, {Faber}, {Gebhardt}, {Green},
  {Grillmair}, {Kormendy}, \& {Lauer}}]{magorrian1998}
{Magorrian} J. {et~al.}, 1998, AJ, 115, 2285

\bibitem[{{Miller} {et~al}\mbox{.}(2001){Miller}, {Genovese}, {Nichol},
  {Wasserman}, {Connolly}, {Reichart}, {Hopkins}, {Schneider}, \&
  {Moore}}]{miller2001}
{Miller} C.~J. {et~al.}, 2001, AJ, 122, 3492

\bibitem[{{Miller} {et~al}\mbox{.}(2013){Miller}, {Bonzini}, {Fomalont},
  {Kellermann}, {Mainieri}, {Padovani}, {Rosati}, {Tozzi}, \&
  {Vattakunnel}}]{miller2013}
{Miller} N.~A. {et~al.}, 2013, ApJS, 205, 13

\bibitem[{{Mobasher} {et~al}\mbox{.}(2009){Mobasher}, {Dahlen}, {Hopkins},
  {Scoville}, {Capak}, {Rich}, {Sanders}, {Schinnerer}, {Ilbert}, {Salvato}, \&
  {Sheth}}]{mobasher2009}
{Mobasher} B. {et~al.}, 2009, ApJ, 690, 1074

\bibitem[{{Murphy}(2009)}]{murphy2009}
{Murphy} E.~J., 2009, ApJ, 706, 482

\bibitem[{{Murphy} {et~al}\mbox{.}(2013){Murphy}, {Chatterjee}, {Kaplan},
  {Banyer}, {Bell}, {Bignall}, {Bower}, {Cameron}, {Coward}, {Cordes}, {Croft},
  {Curran}, {Djorgovski}, {Farrell}, {Frail}, {Gaensler}, {Galloway}, {Gendre},
  {Green}, {Hancock}, {Johnston}, {Kamble}, {Law}, {Lazio}, {Lo}, {Macquart},
  {Rea}, {Rebbapragada}, {Reynolds}, {Ryder}, {Schmidt}, {Soria}, {Stairs},
  {Tingay}, {Torkelsson}, {Wagstaff}, {Walker}, {Wayth}, \&
  {Williams}}]{murphy2013}
{Murphy} T. {et~al.}, 2013, PASA, 30, 6

\bibitem[{{Muxlow} {et~al}\mbox{.}(2005){Muxlow}, {Richards}, {Garrington},
  {Wilkinson}, {Anderson}, {Richards}, {Axon}, {Fomalont}, {Kellermann},
  {Partridge}, \& {Windhorst}}]{muxlow2005}
{Muxlow} T.~W.~B. {et~al.}, 2005, MNRAS, 358, 1159

\bibitem[{{Norris} {et~al}\mbox{.}(2011){Norris}, {Hopkins}, {Afonso}, {Brown},
  {Condon}, {Dunne}, {Feain}, {Hollow}, {Jarvis}, {Johnston-Hollitt}, {Lenc},
  {Middelberg}, {Padovani}, {Prandoni}, {Rudnick}, {Seymour}, {Umana},
  {Andernach}, {Alexander}, {Appleton}, {Bacon}, {Banfield}, {Becker}, {Brown},
  {Ciliegi}, {Jackson}, {Eales}, {Edge}, {Gaensler}, {Giovannini}, {Hales},
  {Hancock}, {Huynh}, {Ibar}, {Ivison}, {Kennicutt}, {Kimball}, {Koekemoer},
  {Koribalski}, {L{\'o}pez-S{\'a}nchez}, {Mao}, {Murphy}, {Messias},
  {Pimbblet}, {Raccanelli}, {Randall}, {Reiprich}, {Roseboom},
  {R{\"o}ttgering}, {Saikia}, {Sharp}, {Slee}, {Smail}, {Thompson}, {Urquhart},
  {Wall}, \& {Zhao}}]{norris2011}
{Norris} R.~P. {et~al.}, 2011, PASA, 28, 215

\bibitem[{{O'Dea}(1998)}]{odea1998}
{O'Dea} C.~P., 1998, PASP, 110, 493

\bibitem[{{Offringa} {et~al}\mbox{.}(2010){Offringa}, {de Bruyn}, {Biehl},
  {Zaroubi}, {Bernardi}, \& {Pandey}}]{offringa2010}
{Offringa} A.~R., {de Bruyn} A.~G., {Biehl} M., {Zaroubi} S., {Bernardi} G.,
  {Pandey} V.~N., 2010, MNRAS, 405, 155

\bibitem[{{Owen} {et~al}\mbox{.}(1983){Owen}, {Condon}, \& {Ledden}}]{owen1983}
{Owen} F.~N., {Condon} J.~J., {Ledden} J.~E., 1983, AJ, 88, 1

\bibitem[{{Padovani} {et~al}\mbox{.}(2009){Padovani}, {Mainieri}, {Tozzi},
  {Kellermann}, {Fomalont}, {Miller}, {Rosati}, \& {Shaver}}]{padovani2009}
{Padovani} P., {Mainieri} V., {Tozzi} P., {Kellermann} K.~I., {Fomalont} E.~B.,
  {Miller} N., {Rosati} P., {Shaver} P., 2009, ApJ, 694, 235

\bibitem[{{Padovani} {et~al}\mbox{.}(2011){Padovani}, {Miller}, {Kellermann},
  {Mainieri}, {Rosati}, \& {Tozzi}}]{padovani2011}
{Padovani} P., {Miller} N., {Kellermann} K.~I., {Mainieri} V., {Rosati} P.,
  {Tozzi} P., 2011, ApJ, 740, 20

\bibitem[{{Peacock} \& {Wall}(1981)}]{peacock1981}
{Peacock} J.~A., {Wall} J.~V., 1981, MNRAS, 194, 331

\bibitem[{{Prandoni} {et~al}\mbox{.}(2001){Prandoni}, {Gregorini}, {Parma}, {de
  Ruiter}, {Vettolani}, {Wieringa}, \& {Ekers}}]{prandoni2001}
{Prandoni} I., {Gregorini} L., {Parma} P., {de Ruiter} H.~R., {Vettolani} G.,
  {Wieringa} M.~H., {Ekers} R.~D., 2001, A\&A, 365, 392

\bibitem[{{Prandoni} {et~al}\mbox{.}(2006){Prandoni}, {Parma}, {Wieringa}, {de
  Ruiter}, {Gregorini}, {Mignano}, {Vettolani}, \& {Ekers}}]{prandoni2006}
{Prandoni} I., {Parma} P., {Wieringa} M.~H., {de Ruiter} H.~R., {Gregorini} L.,
  {Mignano} A., {Vettolani} G., {Ekers} R.~D., 2006, A\&A, 457, 517

\bibitem[{{Randall} {et~al}\mbox{.}(2011){Randall}, {Hopkins}, {Norris}, \&
  {Edwards}}]{randall2011}
{Randall} K.~E., {Hopkins} A.~M., {Norris} R.~P., {Edwards} P.~G., 2011, MNRAS,
  416, 1135

\bibitem[{{Randall} {et~al}\mbox{.}(2012){Randall}, {Hopkins}, {Norris},
  {Zinn}, {Middelberg}, {Mao}, \& {Sharp}}]{randall2012}
{Randall} K.~E., {Hopkins} A.~M., {Norris} R.~P., {Zinn} P.-C., {Middelberg}
  E., {Mao} M.~Y., {Sharp} R.~G., 2012, MNRAS, 421, 1644

\bibitem[{{Rowan-Robinson} {et~al}\mbox{.}(1993){Rowan-Robinson}, {Benn},
  {Lawrence}, {McMahon}, \& {Broadhurst}}]{rowan-robinson1993}
{Rowan-Robinson} M., {Benn} C.~R., {Lawrence} A., {McMahon} R.~G., {Broadhurst}
  T.~J., 1993, MNRAS, 263, 123

\bibitem[{{Schinnerer} {et~al}\mbox{.}(2010){Schinnerer}, {Sargent}, {Bondi},
  {Smol{\v c}i{\'c}}, {Datta}, {Carilli}, {Bertoldi}, {Blain}, {Ciliegi},
  {Koekemoer}, \& {Scoville}}]{schinnerer2010}
{Schinnerer} E. {et~al.}, 2010, ApJS, 188, 384

\bibitem[{{Schinnerer} {et~al}\mbox{.}(2007){Schinnerer}, {Smol{\v c}i{\'c}},
  {Carilli}, {Bondi}, {Ciliegi}, {Jahnke}, {Scoville}, {Aussel}, {Bertoldi},
  {Blain}, {Impey}, {Koekemoer}, {Le Fevre}, \& {Urry}}]{schinnerer2007}
{Schinnerer} E. {et~al.}, 2007, ApJS, 172, 46

\bibitem[{{Seymour} {et~al}\mbox{.}(2008){Seymour}, {Dwelly}, {Moss},
  {McHardy}, {Zoghbi}, {Rieke}, {Page}, {Hopkins}, \& {Loaring}}]{seymour2008b}
{Seymour} N. {et~al.}, 2008, MNRAS, 386, 1695

\bibitem[{{Smol{\v c}i{\'c}} {et~al}\mbox{.}(2008){Smol{\v c}i{\'c}},
  {Schinnerer}, {Scodeggio}, {Franzetti}, {Aussel}, {Bondi}, {Brusa},
  {Carilli}, {Capak}, {Charlot}, {Ciliegi}, {Ilbert}, {Ivezi{\'c}}, {Jahnke},
  {McCracken}, {Obri{\'c}}, {Salvato}, {Sanders}, {Scoville}, {Trump},
  {Tremonti}, {Tasca}, {Walcher}, \& {Zamorani}}]{smolcic2008}
{Smol{\v c}i{\'c}} V. {et~al.}, 2008, ApJS, 177, 14

\bibitem[{{Whiting}(2012)}]{whiting2012}
{Whiting} M.~T., 2012, MNRAS, 421, 3242

\bibitem[{{Whittam} {et~al}\mbox{.}(2013){Whittam}, {Riley}, {Green}, {Jarvis},
  {Prandoni}, {Guglielmino}, {Morganti}, {R{\"o}ttgering}, \&
  {Garrett}}]{whittam2013}
{Whittam} I.~H. {et~al.}, 2013, MNRAS, 429, 2080

\bibitem[{{Wilman} {et~al}\mbox{.}(2010){Wilman}, {Jarvis}, {Mauch},
  {Rawlings}, \& {Hickey}}]{wilman2010}
{Wilman} R.~J., {Jarvis} M.~J., {Mauch} T., {Rawlings} S., {Hickey} S., 2010,
  MNRAS, 405, 447

\bibitem[{{Wilman} {et~al}\mbox{.}(2008){Wilman}, {Miller}, {Jarvis}, {Mauch},
  {Levrier}, {Abdalla}, {Rawlings}, {Kl{\"o}ckner}, {Obreschkow}, {Olteanu}, \&
  {Young}}]{wilman2008}
{Wilman} R.~J. {et~al.}, 2008, MNRAS, 388, 1335

\bibitem[{{Wilson} {et~al}\mbox{.}(2011){Wilson}, {Ferris}, {Axtens}, {Brown},
  {Davis}, {Hampson}, {Leach}, {Roberts}, {Saunders}, {Koribalski}, {Caswell},
  {Lenc}, {Stevens}, {Voronkov}, {Wieringa}, {Brooks}, {Edwards}, {Ekers},
  {Emonts}, {Hindson}, {Johnston}, {Maddison}, {Mahony}, {Malu}, {Massardi},
  {Mao}, {McConnell}, {Norris}, {Schnitzeler}, {Subrahmanyan}, {Urquhart},
  {Thompson}, \& {Wark}}]{wilson2011}
{Wilson} W.~E. {et~al.}, 2011, MNRAS, 416, 832

\bibitem[{{Windhorst} {et~al}\mbox{.}(1990){Windhorst}, {Mathis}, \&
  {Neuschaefer}}]{windhorst1990}
{Windhorst} R., {Mathis} D., {Neuschaefer} L., 1990, in Astronomical Society of
  the Pacific Conference Series, Vol.~10, Evolution of the Universe of
  Galaxies, {R.~G.~Kron}, ed., pp. 389--403

\bibitem[{{Windhorst} {et~al}\mbox{.}(1985){Windhorst}, {Miley}, {Owen},
  {Kron}, \& {Koo}}]{windhorst1985}
{Windhorst} R.~A., {Miley} G.~K., {Owen} F.~N., {Kron} R.~G., {Koo} D.~C.,
  1985, ApJ, 289, 494

\bibitem[{{Witzel} {et~al}\mbox{.}(1979){Witzel}, {Pauliny-Toth}, {Nauber}, \&
  {Schmidt}}]{witzel1979}
{Witzel} A., {Pauliny-Toth} I.~I.~K., {Nauber} U., {Schmidt} J., 1979, AJ, 84,
  942

\end{thebibliography}

\begin{figure*}
\includegraphics[trim=0.3 1.75cm 0.3 1.75cm, page=2,clip=true,width=0.99\textwidth]{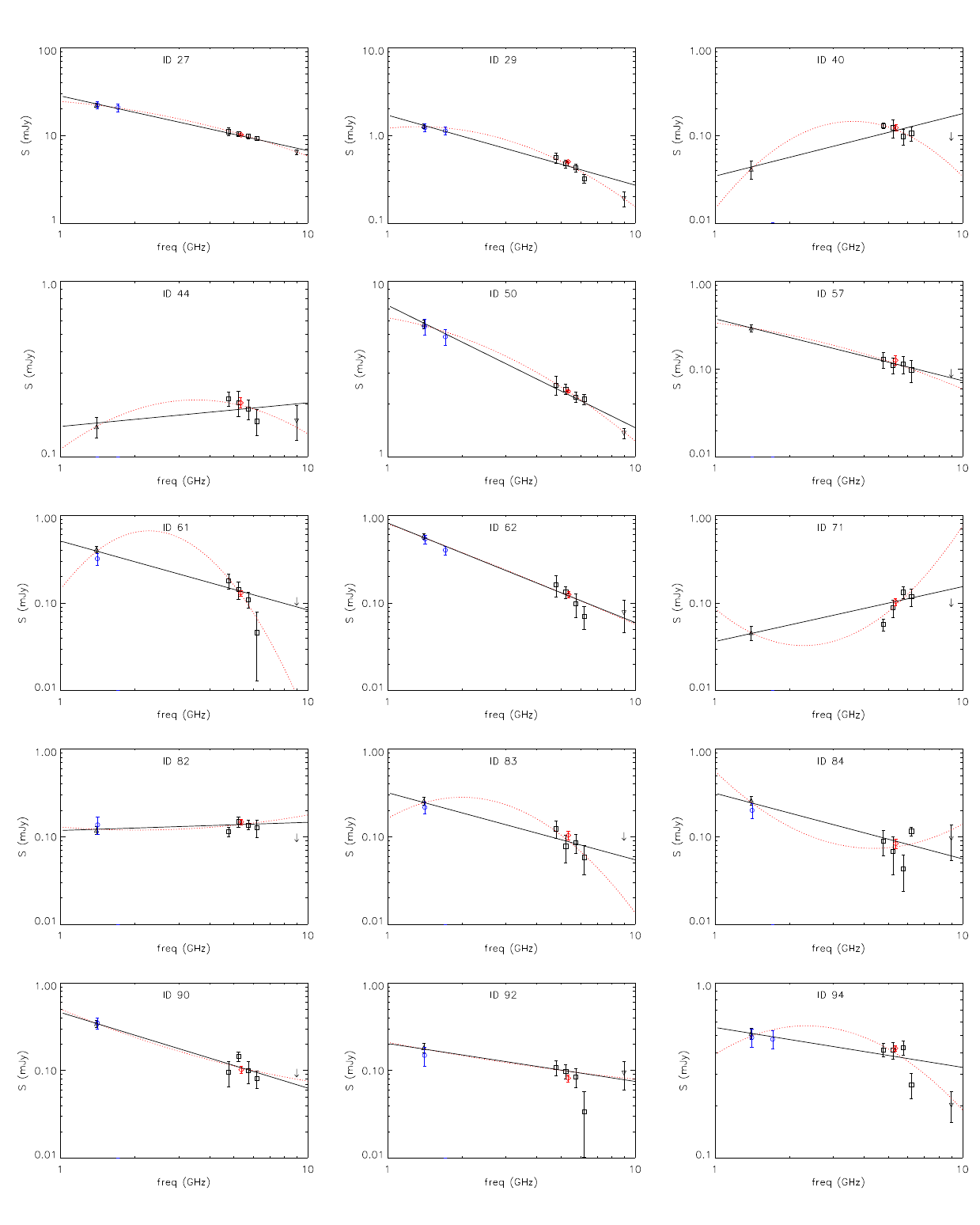}
\begin{center}
{\bf Figure~\ref{fig:seds}} (continued)
\end{center}
\end{figure*}
\begin{figure*}
\includegraphics[trim=0.3 1.75cm 0.3 1.75cm, page=3,clip=true,width=0.99\textwidth]{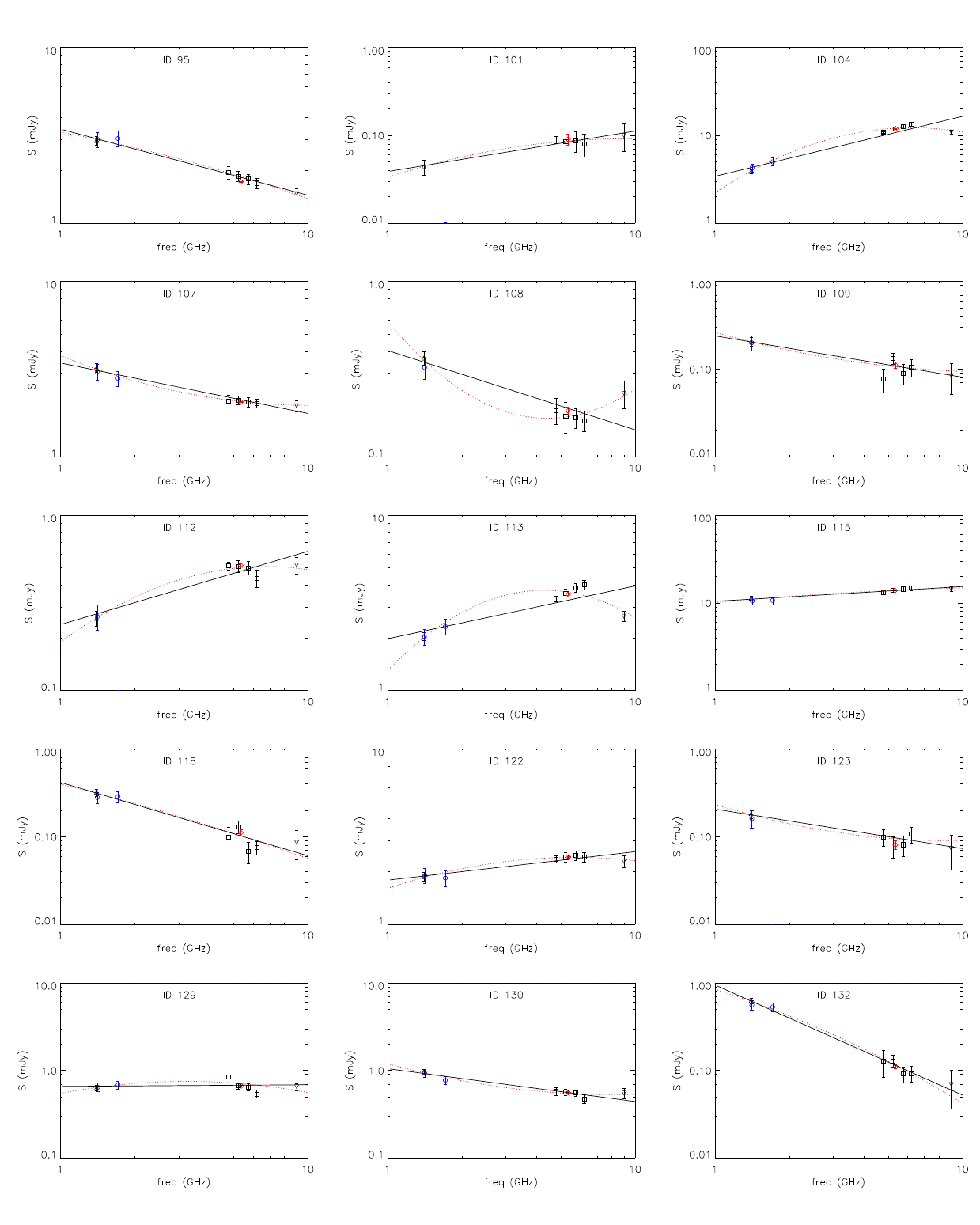}
\begin{center}
{\bf Figure~\ref{fig:seds}} (continued)
\end{center}
\end{figure*}
\begin{figure*}
\includegraphics[trim=0.3 1.75cm 0.3 1.75cm, page=4,clip=true,width=0.99\textwidth]{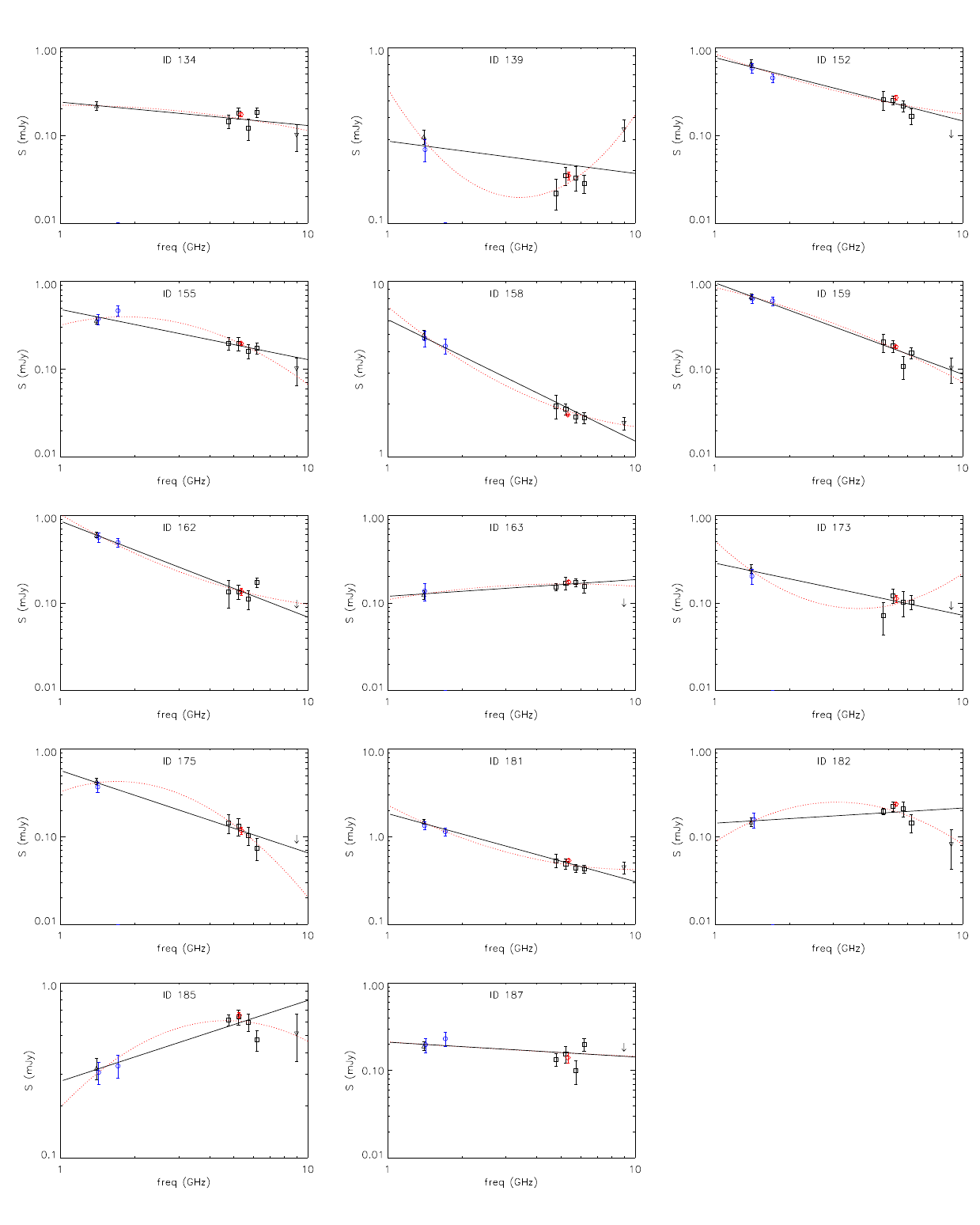}
\begin{center}
{\bf Figure~\ref{fig:seds}} (continued)
\end{center}
\end{figure*}

\label{lastpage}

\end{document}